\documentclass[journal, 12pt, draftcls, a4paper, onecolumn, twoside]{IEEEtran}

\usepackage{cite}
\usepackage[cmex10]{amsmath}
\usepackage{amsthm}
\usepackage{amssymb}

\usepackage{array}

\usepackage[T1]{fontenc}
\usepackage[utf8x]{inputenc}
\usepackage[english]{babel}

\usepackage{tikz}
\usetikzlibrary{positioning,chains,fit,calc,shapes,arrows}

\newtheorem{definition}{Definition}

\newtheorem{theorem}{Theorem}
\newtheorem{lemma}{Lemma}
\newtheorem{corollary}{Corollary}

\newcommand{\F}{\mathbb{F}}

\newcommand{\R}{\mathbb{R}}
\newcommand{\Z}{\mathbb{Z}}

\newcommand{\E}{\mathbb{E}}
\newcommand{\Zn}{\mathbb{Z}^n}
\newcommand{\Rn}{\mathbb{R}^n}

\newcommand{\Fp}{\mathbb{F}_p}
\newcommand{\Fpn}{\mathbb{F}_p^n}
\newcommand{\zero}{\mathbf{0}}
\newcommand{\vor}{\mathcal{V}}
\newcommand{\prob}{\mathcal{P}}

\newcommand{\B}{\mathcal{B}}
\newcommand{\HH}{{\mathbf H}}
\newcommand{\NN}{\mathcal{N}}

\DeclareMathOperator*{\vol}{Vol}
\DeclareMathOperator*{\dHmin}{d_{H_{\min}}}
\DeclareMathOperator*{\dEmin}{d_{E_{\min}}}

\DeclareMathOperator*{\var}{Var}
\DeclareMathOperator*{\supp}{Supp}

\DeclareMathOperator*{\eff}{eff}
\DeclareMathOperator*{\cov}{Cov}
\DeclareMathOperator*{\snr}{SNR}

\DeclareMathOperator*{\dec}{dec}

\DeclareMathOperator*{\pol}{Pol}

\begin{document}
%
% paper title
% Titles are generally capitalized except for words such as a, an, and, as,
% at, but, by, for, in, nor, of, on, or, the, to and up, which are usually
% not capitalized unless they are the first or last word of the title.
% Linebreaks \\ can be used within to get better formatting as desired.
% Do not put math or special symbols in the title.
\title{LDA Lattices Without Dithering Achieve Capacity on the Gaussian Channel}
%
%
% author names and IEEE memberships
% note positions of commas and nonbreaking spaces ( ~ ) LaTeX will not break
% a structure at a ~ so this keeps an author's name from being broken across
% two lines.
% use \thanks{} to gain access to the first footnote area
% a separate \thanks must be used for each paragraph as LaTeX2e's \thanks
% was not built to handle multiple paragraphs
%

\author{Nicola~di~Pietro, %~\IEEEmembership{Member,~IEEE,}
        Gilles~Z\'emor, %~\IEEEmembership{Fellow,~OSA,}
        and~Joseph~J.~Boutros,~\IEEEmembership{Senior~Member,~IEEE}% <-this % stops a space
\thanks{This manuscript was submitted to the IEEE Transactions on Information Theory, paper IT-16-0169, March 2016. The research work on LDA lattices presented here was supported by QNRF, a member of Qatar Foundation, under NPRP project 5-597-2-241. A small part of the contents of this paper was presented at the 2014 Joint Workshop on Coding and Communications and at the 2016 International Zurich Seminar on Communications.}
\thanks{Nicola di Pietro and Joseph J.~Boutros are with the Department
of Electrical and Computer Engineering, Texas A\&M University at Qatar, 
c/o Qatar Foundation, Education City, Doha, Qatar, P.O.\ Box 23874 (e-mail: nicola.ndp@gmail.com; boutros@ieee.org).}% <-this % stops a space
\thanks{Gilles Z\'emor is with the Institut de Math\'ematiques de Bordeaux UMR 5251,
Universit\'e de Bordeaux, 351 cours de la Lib\'eration - F33405, Talence, France (e-mail: zemor@math.u-bordeaux.fr).}% <-this % stops a space
%\thanks{Manuscript received April 19, 2005; revised September 17, 2014.}}
}

% note the % following the last \IEEEmembership and also \thanks - 
% these prevent an unwanted space from occurring between the last author name
% and the end of the author line. i.e., if you had this:
% 
% \author{....lastname \thanks{...} \thanks{...} }
%                     ^------------^------------^----Do not want these spaces!
%
% a space would be appended to the last name and could cause every name on that
% line to be shifted left slightly. This is one of those "LaTeX things". For
% instance, "\textbf{A} \textbf{B}" will typeset as "A B" not "AB". To get
% "AB" then you have to do: "\textbf{A}\textbf{B}"
% \thanks is no different in this regard, so shield the last } of each \thanks
% that ends a line with a % and do not let a space in before the next \thanks.
% Spaces after \IEEEmembership other than the last one are OK (and needed) as
% you are supposed to have spaces between the names. For what it is worth,
% this is a minor point as most people would not even notice if the said evil
% space somehow managed to creep in.

% The paper headers
\markboth{Transactions on Information Theory,~Vol., No., Month~year}%
{di Pietro \MakeLowercase{\textit{et al.}}: LDA Lattices Achieve Capacity}
% The only time the second header will appear is for the odd numbered pages
% after the title page when using the twoside option.
% 
% *** Note that you probably will NOT want to include the author's ***
% *** name in the headers of peer review papers.                   ***
% You can use \ifCLASSOPTIONpeerreview for conditional compilation here if
% you desire.

% If you want to put a publisher's ID mark on the page you can do it like
% this:
%\IEEEpubid{0000--0000/00\$00.00~\copyright~2014 IEEE}
% Remember, if you use this you must call \IEEEpubidadjcol in the second
% column for its text to clear the IEEEpubid mark.

% use for special paper notices
%\IEEEspecialpapernotice{(Invited Paper)}

% make the title area
\maketitle

% As a general rule, do not put math, special symbols or citations
% in the abstract or keywords.
\begin{abstract}
  This paper deals with Low-Density Construction-A (LDA) lattices, which are obtained via Construction A from non-binary Low-Density Parity-Check codes. More precisely, a proof is provided that Voronoi constellations of LDA lattices achieve the capacity of the AWGN channel under lattice encoding and decoding. This is obtained after showing the same result for more general Construction-A lattice constellations. The theoretical analysis is carried out in a way that allows to describe how the prime number underlying Construction A behaves as a function of the lattice dimension. Moreover, no dithering is required in the transmission scheme, simplifying some previous solutions of the problem. Remarkably, capacity is achievable with LDA lattice codes whose parity-check matrices have constant row and column Hamming weights. Some expansion properties of random bipartite graphs constitute an extremely important tool for dealing with sparse matrices and allow to find a lower bound of the minimum Euclidean distance of LDA lattices in our ensemble.
\end{abstract}

% Note that keywords are not normally used for peerreview papers.

% For peer review papers, you can put extra information on the cover
% page as needed:
% \ifCLASSOPTIONpeerreview
% \begin{center} \bfseries EDICS Category: 3-BBND \end{center}
% \fi
%
% For peerreview papers, this IEEEtran command inserts a page break and
% creates the second title. It will be ignored for other modes.
\IEEEpeerreviewmaketitle

\section{Introduction}
\label{sec:intro}
% The very first letter is a 2 line initial drop letter followed
% by the rest of the first word in caps.
% 
% form to use if the first word consists of a single letter:
% \IEEEPARstart{A}{demo} file is ....
% 
% form to use if you need the single drop letter followed by
% normal text (unknown if ever used by IEEE):
% \IEEEPARstart{A}{}demo file is ....
% 
% Some journals put the first two words in caps:
% \IEEEPARstart{T}{his demo} file is ....
% 
% Here we have the typical use of a "T" for an initial drop letter
% and "HIS" in caps to complete the first word.

% You must have at least 2 lines in the paragraph with the drop letter
% (should never be an issue)
\IEEEPARstart{T}{his} paper addresses the problem of communication over the Additive White Gaussian Noise (AWGN) channel with lattice codes. The first notable work on the possibility of sending information with lattices over the AWGN channel with satisfactory performance is due to de Buda and dates back to 1975 \cite{deBuda1975}. He showed how lattice codes whose shaping region is a ball can be reliably decoded at any asymptotic rate up to $\frac{1}{2}\log_2(\snr)$ under \emph{lattice decoding}. This decoding strategy does not take into account the shaping region that defines the constellation. In other words, a lattice decoder simply returns the lattice point closest to the decoder input, regardless of whether it belongs to the constellation or not. As a consequence, the decoding decision regions are all equivalent and coincide with the Voronoi regions of the lattice points. Of course, this method is suboptimal with respect to the maximum likelihood (ML) decoder. Nevertheless, its easier algorithmic nature makes it appealing for both theoretical analysis and practical implementation.

The work by de Buda continued \cite{deBuda1989} and was partially corrected by Linder, Schlegel and Zeger \cite{Linder1993}. They were able to prove that lattice codes can attain the capacity of the AWGN channel under optimal decoding, with shaping determined by ``thin'' spherical shells. This peculiar shaping region actually makes the code lose most of its lattice structure and look similar to a random code on a sphere. Urbanke and Rimoldi \cite{Urbanke1998} completed this work with the proof that lattice codes made up of the intersection between a ball and a lattice are capacity-achieving under optimal nearest-codeword decoding. 

Thus, %the theoretical problem of showing that lattice codes are capacity-achieving was solved. 
it was shown that lattice codes are capacity-achieving. 
Nonetheless, the question of whether this result can be obtained under (\emph{a priori} non-optimal) lattice decoding remained answerless. In 1997, Loeliger \cite{Loeliger1997} proved the achievability of the rate $\frac{1}{2}\log_2(\snr)$ with Construction-A lattices over non-binary alphabets and conjectured that this limit could not be overcome with lattice decoding. It has been necessary to wait for Erez and Zamir's solution to the problem \cite{Erez2004}, based on the Modulo-Lattice Additive Noise (MLAN) channel and Voronoi constellations with Construction-A lattices. More recently, Belfiore and Ling \cite{Ling2014} proposed a solution that involves %a non-uniform distribution on the channel inputs and 
an infinite (but energetically finite) codebook.

Once the theoretical problem of non-constructively achieving capacity with ML decoding was solved, it left the place also to the challenge of designing some constructive families of lattices adapted to iterative decoding with close-to-capacity performance. The intention was, and still is, to translate into concrete evidence the theoretical effort of showing that lattices are adequately suited to block coding in high dimensions for the AWGN channel. Most of the proposed families are inspired by LDPC and turbo codes \cite{Sadeghi2006,Baik2008,Sommer2008,Sakzad2010,Sakzad2012,Sadeghi2013} and an interesting work on lattices based on polar codes exists \cite{Yan2014,Yan2013,Yan2012}; the latter are also shown to be capacity-achieving. 

The authors of this paper have contributed to this field with the introduction of two lattice families: the most recent are the \emph{Generalized Low-Density (GLD) lattices} \cite{Boutros2014, Boutros2015}. They show great performance under iterative decoding and numerical simulations have been run in remarkably high dimensions (up to one million). Moreover, a theoretical analysis about the possibility of achieving the so called Poltyrev capacity with infinite GLD-lattice constellations is provided in~\cite{diPietro2015}.

The second family consists of \emph{Low-Density Construction-A (LDA) lattices}, to which this paper is entirely devoted. LDA lattices put together the strength of Construction A \cite{Leech1971} and LDPC codes (over a non-binary prime field) \cite{Gallager1963}. Their main feature is that their corresponding parity-check matrix is sparse. As one can guess, this is the key idea to redirect their decoding to well-performing, implementable LDPC decoding algorithms. LDA lattices were first envisaged in \cite{Erez2002} and were referred to with this name and reintroduced by di Pietro \textit{et al.}\ \cite{diPietro2012}, together with an efficient iterative algorithm to decode them. A theoretical analysis of the Poltyrev-capacity-achieving qualities of infinite LDA constellations was carried out by the same authors \cite{diPietro2013,diPietro2013bis}, whereas the ``goodness'' properties of LDA lattices are studied in \cite{Vatedka2014,Vatedka2015}. The problem of attaining the real capacity of the AWGN channel with finite LDA constellations was addressed and a solution was developed in the first author's dissertation \cite{diPietro2014}. The main purpose of this work is to give a detailed account of this solution: improvements will also be provided along the way. 

\subsection{Original contributions and main features of this paper}
Defoliated of all technical hypotheses, our main accomplishment can be stated as follows:
\begin{theorem}
  \label{thm:light_version}
  For every $\snr > 1$, there exists a random ensemble of LDA lattices that achieves capacity of the AWGN channel under lattice encoding and decoding.
\end{theorem}
One may question the point of proving this kind of result for lattices that are designed for iterative decoding in high dimensions, knowing well that it will be impractical to implement a lattice decoder. Historically, lattice decoding has been considered conceptually simpler than ML decoding for constellations of points in $\Rn$ with little structure, and therefore as a possible intermediate step towards polynomial-time and more practical decoding algorithms. For us, knowing that the LDA family has the potential to reach capacity justifies and encourages further research into the design and study of practical iterative techniques for this family of lattices or some of its subfamilies. %Well, the answer is that showing that LDA lattices can achieve capacity is the most important theoretical justification to the design and simulation work, to any implementation of LDA lattices even for suboptimal schemes and decoding techniques. 

The more precise version of Theorem \ref{thm:light_version} is Theorem \ref{thm:awgn_capacity_LDA} of Section \ref{sec:main_theorem} and all the other results of this dissertation are intermediate steps to reach its proof. The most relevant of these is Theorem \ref{thm:awgn_capacity} of Section \ref{sec:capacity_with_random_construction_A}, which is the analogue of Theorem \ref{thm:light_version} or \ref{thm:awgn_capacity_LDA} for more general, non-LDA Construction-A finite lattice constellations. The capacity of the AWGN channel was previously shown to be achievable by lattice decoding of lattice code ensembles by Erez and Zamir \cite{Erez2004}, Ordentlich and Erez \cite{Ordentlich2012}, Ling and Belfiore \cite{Ling2014}, and recently for polar decoding by Yan \textit{et al.}\ \cite{Yan2014}. The additional insight provided by our proof techniques includes the following: %There are some reasons for which it is worthwhile to provide Theorem \ref{thm:awgn_capacity} in the way we do: 
\begin{itemize}
\item We are able to prove the capacity-achieving properties of Construction-A lattices without using the theoretical tool of the MLAN channel \cite{Erez2004,Ordentlich2012}; in particular, we do not assume that the sender and the receiver share the common randomness known as \emph{dither}, even if we apply Minimum Mean Square Error (MMSE) estimation of the channel output. This solves a problem raised by Forney \cite{Forney2004} who points out that in this context avoiding the use of a dither has to be possible, but no proof had ever been provided, to the best of our knowledge. 
\item We still rely on Voronoi lattice constellations and do not need to introduce Gaussian coding \cite{Ling2014,Yan2014}.
\item We follow the work of \cite{Erez2004} and \cite{Ordentlich2012} in that we use Construction A together with randomly chosen $p$-ary codes. As before, $p$ has to be a growing function of the lattice dimension $n$: however, we are able to reduce its growth rate.
\item Last, but not least, this proof technique adapts to the case of LDA lattices, whereas how to adapt previous proofs to the LDA case is to us very much unclear. %we have been motivated in this redemonstration work because our arguments seem to be the most suitable to be adapted to LDA lattices.
\end{itemize} 

Among the main aspects that characterise our work, it is important to remark that the row and column Hamming weights of the parity-check matrices of the non-binary LDPC codes that underlie our construction are reasonably small constants and do not need to tend to infinity with the lattice dimension. This is an appreciable feature, because the complexity of the LDA decoding algorithm is directly proportional to those numbers. The minimum value of the constant row weight as a function of the parameters of the construction is explicitly given in Theorem \ref{thm:awgn_capacity_LDA} (compare also with \cite{diPietro2013bis}). Notice that for binary LDPC codes to achieve the capacity of any memoryless binary symmetric channel or of the binary erasure channel, asymptotically infinite row weights are mandatorily required \cite{Gallager1963,MacKay1999,Sason2003}. Some graph-based, capacity-achieving binary codes with bounded decoding complexity in spite of their unbounded maximum row weight are instead Pfister \textit{et al.}'s IRA codes \cite{Pfister2005}.
  
  Our LDA ensemble is based on random bipartite graphs. These graphs are known to have some particular \emph{expansion properties} that, qualitatively speaking, say that all ``small enough'' sets of nodes have ``large enough'' neighborhoods. We exploit intensively these properties, formally made explicit in Lemma \ref{lem:good_graphs} and Corollary \ref{cor:good_graphs} of Section \ref{sec:graph_theoretical_tools}; they turn out to be two of the most important theoretical pillars of our analysis. Lemma \ref{lem:min_H_dist} and Corollary \ref{cor:fundamental_gain_LDA} of Section \ref{sec:random_LDA_ensemble} consist of a lower-bound of the minimum Euclidean distance and fundamental gain of our LDA ensemble and are an example of how expansion properties are used in our setting.
  
 As a final comment, notice that our capacity-achieving result for LDA lattices does not hold for $\snr \leq 1$. Nevertheless, this is not a very constraining restriction: for very small $\snr$ there is no need for using lattice constellations for communications over the AWGN channel and classical coded binary modulations are already known to work in a more than satisfactory way \cite{Richardson2008}.
  
\subsection{Structure of the paper}
Our paper is structured as follows: Section \ref{sec:lattices} contains a list of definitions about lattices and lattice constellations. In Section \ref{sec:lemmas}, we state four useful lemmas, which will be often employed in the following. Section \ref{sec:capacity_for_random_construction_A} recalls the main features of Theorem \ref{thm:awgn_capacity}, which shows how and under what conditions random Construction-A lattice constellations achieve the capacity of the AWGN channel. Section \ref{sec:the_random_ensemble} and Section \ref{sec:encoding_and_decoding} provide a formal definition of those constellations and of the information transmission scheme that we consider. In Section \ref{sec:main_ideas}, we give a general description of the main ideas that lead to the proof of Theorem \ref{thm:awgn_capacity} and Theorem \ref{thm:awgn_capacity_LDA}. The complete detailed proof of Theorem \ref{thm:awgn_capacity} is provided in Section \ref{sec:detailed_proof}. Section \ref{sec:graph_theoretical_tools} is an independent section which presents the expansion properties of bipartite graphs. Section \ref{sec:capacity_for_LDA} is an introduction to the LDA setting, to which the second part of the paper is entirely devoted. Our random LDA-lattice constellations are presented in Section \ref{sec:random_LDA_ensemble}, which contains also a result on the minimum distance of their underlying LDPC codes and Hermite constants. The detailed proof of Theorem \ref{thm:awgn_capacity_LDA} on the capacity-achieving properties of LDA lattices is provided in Section \ref{sec:LDA_capacity_detailed_proof}. Section \ref{sec:conclusion} recalls the main results of this paper and contains some concluding remarks. Finally, the appendices contain the proofs of most of the lemmas which are not treated in detail in the other sections.
  
\subsection{Notation}
Throughout the whole paper we will very often use asymptotic relations between functions of the lattice dimension $n$. As usual, the symbol $\sim$ indicates the ``asymptotic equality'': $f(n) \sim g(n)$ if $\lim_{n \to \infty} f(n)/g(n) = 1$. The notation $f(n) \lesssim g(n)$ indicates that $f(n) \sim s(n) \leq g(n)$ for some $s(n)$; or, equivalently, that $f(n) \leq t(n) \sim g(n)$, for some $t(n)$. With analogous meaning, we can write $g(n) \gtrsim f(n)$. The symbols $o(\cdot)$ and $O(\cdot)$ refer to the standard Bachmann-Landau notation in the variable $n$.

We say that a function $f(n)$ grows subexponentially fast in $n$ if $f(n) = O(\exp(n^\beta))$ for some $0 < \beta< 1$. Observe that $n^{n^\gamma}$ is subexponential for every $0 < \gamma< 1$. 

We will very often deal with balls and spheres and we will denote $B_{\mathbf{c},n}(\rho) \subseteq \Rn$ the $n$-dimensional ball centered at $\mathbf{c}$ with radius $\rho$.

A crucial parameter of our analysis is the prime number $p$ that underlies Construction A (cf. Definition \ref{def:constructionA}). It needs to tend to infinity when the lattice dimension $n$ grows and we are interested in describing the growth of $p$ as a function of $n$. For this reason, $p$ is defined as $p = n^\lambda$ for some positive constant $\lambda$. It is clear that if $n$ changes and $\lambda$ is fixed, then in general $n^\lambda$ is not a prime number. It would be more precise to say that $p(\lambda)$ is the closest prime number to $n^\lambda$, or that $p = n^{\lambda(n)}$ for some $\lambda(n)$ assuming values in an interval properly centered at our fixed value $\lambda$. Nevertheless, it is possible to show that this variation of $\lambda(n)$ concerns a range which is narrow enough not to impact any of the asymptotic estimations that we compute letting $n$ tend to infinity. In other words, there always exists a prime number $p$ close enough to $n^\lambda$ to make accurate the approximation $p = n^\lambda$ (for example, we can apply Bertrand's Postulate \cite{Erdos1934}). Despite the slight abuse of notation, we prefer to keep it that way from now on, in order to write the proofs in the clearest way possible and avoid the overabundance of symbols.

\section{Lattices and lattice codes for the AWGN channel}
\label{sec:lattices}
We assume that the reader is already familiar with lattices as mathematical objects and constellations for the transmission of information; excellent references are \cite{Conway1999,Ebeling2013,Zamir2014}. We recall here some definitions that we will need in the following, mainly with the purpose of fixing our notation.

In this paper we exclusively deal with real lattices, i.e., discrete additive subgroups of the Euclidean vector space $\Rn$. They are always full-rank and the letter $n$ indicates the lattice rank and the dimension of the Euclidean space as well. 
\begin{definition}[Voronoi region]
  We call \emph{Voronoi region} of a lattice point $\mathbf{x} \in \Lambda$ the set
  \begin{equation*}
    \mathcal{V}(\mathbf{x}) = \{ \mathbf{y} \in \Rn : \Vert \mathbf{y}-\mathbf{x} \Vert \leq \Vert \mathbf{y}-\mathbf{z} \Vert,\ \forall \mathbf{z} \in \Lambda \smallsetminus \{\mathbf{x}\}\}.
  \end{equation*}
  We call Voronoi region of the lattice, and denote it $\mathcal{V}(\Lambda)$, the Voronoi region of $\zero$.
\end{definition}
\begin{definition}[Effective radius]
  \label{def:effective_radius}
  The \emph{effective radius} of a lattice $\Lambda$ is the radius of the ball whose volume is equal to the volume of $\mathcal{V}(\Lambda)$. 
\end{definition}
\begin{definition}[Volume of a lattice]
  The \emph{volume} of a lattice $\Lambda$ is defined as
  \begin{equation*}
    \vol(\Lambda) = \vol \left( \mathcal{V}(\Lambda) \right).
  \end{equation*}
\end{definition}
\begin{definition}[Minimum Euclidean distance and fundamental gain]
  The \emph{minimum Euclidean distance} of a lattice $\Lambda \subseteq \Rn$ is defined as
  \begin{equation*}
    %\label{eq:min_E_dist}
    \dEmin(\Lambda) = \min_{\substack{\mathbf{x},\mathbf{y} \in \Lambda \\ \mathbf{x} \neq \mathbf{y}}} \Vert\mathbf{x} - \mathbf{y}\Vert = \min_{\mathbf{x} \in \Lambda \smallsetminus \{\zero\} } \Vert\mathbf{x}\Vert.
  \end{equation*}
  The \emph{fundamental gain} of $\Lambda$ is
  \begin{equation}
    \label{eq:fundamental_gain}
    \gamma(\Lambda) = \frac{\dEmin(\Lambda)^2}{\vol(\Lambda)^{\frac{2}{n}}}.
  \end{equation}
  It is also known as the \emph{Hermite constant} of the lattice.
\end{definition}
\begin{definition}[Voronoi constellation]
  \label{def:voronoi_constellation}
  Consider two lattices $\Lambda$ and $\Lambda_f$; we say that they are \emph{nested} if $\Lambda \subseteq \Lambda_f$. We call \emph{Voronoi constellation \cite{Conway1983,Forney1989}} of two nested lattices the lattice code
  \begin{equation*}
    \mathcal{C} = \Lambda_f \cap \mathcal{V}(\Lambda).
  \end{equation*}
  In this context, $\Lambda$ is often called the \emph{shaping lattice} and $\Lambda_f$ the \emph{fine lattice}.
  \end{definition}
We can deduce from the previous definition that the Voronoi constellation has cardinality $\vol(\Lambda)/\vol(\Lambda_f)$ and its elements are the representatives of the congruence classes of $\Lambda_f/\Lambda$ with minimum norm. More precisely, if some points of $\Lambda_f$ lie exactly on the boundary of $\mathcal{V}(\Lambda)$, we are implicitely assuming that only one of them is taken for each congruence class. Equivalently, we can modify the definition of the Voronoi region to design its boundary in such a way that the lattice code consists precisely of one representative for each congruence class. 
\begin{definition}[Construction A \cite{Leech1971}]
  \label{def:constructionA}
  Let $C=C[n,k]_p$ be a $p$-ary linear code of length $n$ and dimension $k$ and let us embed $C$ into $\Zn$ via $\Fp^n \hookrightarrow \{0,1,\ldots,p-1\}^n$. We say that the lattice $\Lambda \subseteq \Rn$ is built with Construction A from $C$ when
  \begin{equation*}
    \Lambda = C + p\Zn = \{\mathbf{x} \in \Rn : \mathbf{x} = \mathbf{c} + p\mathbf{z},\ \exists \mathbf{c} \in C,\mathbf{z}\in \Zn\} \subseteq \Zn.
  \end{equation*}
  If $H$ is a parity-check matrix of $C$, we call it also the \emph{parity-check matrix} of $\Lambda = C+p\Zn$, because
  \begin{equation*}
    \Lambda = \{\mathbf{x} \in \Zn : H \mathbf{x}^T \equiv \zero^T \bmod p \}.
  \end{equation*}
\end{definition}
Notice that the definition of Construction A could be made more general \cite{Conway1999}, but we stick here to the one that will give rise to our lattice code ensembles in the following sections.
\begin{definition}[Capacity-achieving family]
  Let $\mathbf{C}$ be the capacity of our channel. We say that a family of lattice codes is \emph{capacity-achieving} under some decoding procedure if for every $\delta > 0$ and for every $\varepsilon > 0$ there exists a lattice code in the family with rate at least $\mathbf{C}-\delta$ and decoding error probability at most $\varepsilon$.
\end{definition}
\begin{definition}[Wiener coefficient]
  Let $\mathbf{x}$ be the random variable that represents the AWGN channel input and let $\mathbf{y} = \mathbf{x} + \mathbf{w}$ be its random output, then the \emph{Wiener coefficient \cite[Chap.~2]{Haykin2013}} is
  \begin{equation*}
    \alpha = \arg \min_{\beta \in \R} \E[\Vert\mathbf{x} - \beta\mathbf{y}\Vert^2].
  \end{equation*}
  The minimum in the previous formula is usually called \emph{Minimum Mean Squared Error} and the Wiener coefficient is also called \emph{MMSE coefficient}.
\end{definition}
It is well known that, if $\E[\Vert\mathbf{x}\Vert^2] = nP$ and $w_i \sim \mathcal{N}(0,\sigma^2)$ for every $i$, then \cite[Lemma~4.1]{diPietro2014} 
\begin{equation*}
  \alpha = \frac{P}{P + \sigma^2}.
\end{equation*}
\begin{definition}[Lattice quantizer]
  We denote $Q_{\Lambda}(\cdot)$ the \emph{quantizer} of a lattice $\Lambda$ associated with $\vor(\Lambda)$:
  \begin{equation*}
    Q_{\Lambda}(\mathbf{y}) = \arg \min_{\mathbf{x} \in \Lambda} \Vert \mathbf{y} - \mathbf{x} \Vert.
  \end{equation*}
\end{definition}
Notice that the quantizer is \emph{a priori} not defined for the points of the boundary of $\vor(\Lambda)$; this will never be a problem for us, basically because those points belong to a region of the space of measure $0$. If needed, the previous definition can be made more formal with little effort to avoid any kind of ambiguity.
\begin{definition}[MMSE lattice decoder]
  Let $\mathbf{x}$ be the AWGN channel input and $\mathbf{y}$ its random output. We call \emph{MMSE lattice decoder} the decoder that proposes the point $\hat{\mathbf{x}} = Q_{\Lambda}(\alpha \mathbf{y})$ as the channel input guess, where $\alpha$ is the Wiener coefficient. 
\end{definition}
Notice that multiplication by $\alpha$ is essential for achieving capacity with a lattice decoder, as it was for Erez and Zamir \cite{Erez2004,Forney2004}. We will give a geometrical explanation of this in Section \ref{sec:geometric_description}.

\section{Some useful lemmas} %-------------------------------- LEMMAS ---------------------------------
\label{sec:lemmas}

This section contains some lemmas that deal with probability theory, combinatorics and geometry. They are quite classical and will be often applied in the sequel, sometimes even implicitly, when the context will be clear enough. The first one describes the ``typical'' norm of a random additive white Gaussian noise vector in very high dimension. For constant standard deviation $\sigma$, the statement is simply the weak law of large numbers; in Appendix \ref{sec:proof_typical_noise} we give a proof that works also for $\sigma = \sigma(n)$.
\begin{lemma}[Typical norm of the AWG noise]
  \label{lem:typical_noise}
  Consider $n$ i.i.d.\ random variables $X_1, \ldots, X_n$, each of them following a Gaussian distribution
  of mean $0$ and variance $\sigma^2$. Let $\rho = \sqrt{\sum_{i=1}^{n}X_i^2}$. Then, for every $\varepsilon > 0$,
  \begin{equation*}
    \lim_{n \to \infty} \prob \left\{\sigma\sqrt{n} \left( 1 - \varepsilon \right) \leq \rho \leq \sigma\sqrt{n} \left( 1 + \varepsilon \right) \right\} = 1.
  \end{equation*}
\end{lemma}

In the next chapters, we will often need to count the number of integer points inside a sphere of a given radius. For this purpose, we will use the following lemma, whose proof is in Appendix \ref{sec:proof_sphere_points}.
\begin{lemma}[Integer points inside a sphere]
  \label{lem:integer_sphere_points}
  Let $B_{\mathbf{c},n}(\rho) = \{ \mathbf{x} \in \Rn : \Vert\mathbf{x}-\mathbf{c}\Vert^2 \leq \rho^2\}$ be the ball centered at $\mathbf{c}$ of
  radius $\rho$. Let $N = |\Zn \cap B_{\mathbf{c},n}(\rho)|$. Then
  \begin{equation*}
    \vol \left( B_{\mathbf{c},n} \left( \rho \right) \right) \left( \max \left\{ 1 - \frac{\sqrt{n}}{2 \rho}, 0 \right\} \right)^n \leq 
    N \leq \vol \left( B_{\mathbf{c},n} \left( \rho \right) \right) \left(1 + \frac{\sqrt{n}}{2 \rho}\right)^n.
  \end{equation*}
\end{lemma}

\begin{lemma}[Asymptotic volume of a ball]
  \label{lem:volume_sphere}
  Stirling's formula yields
  \begin{equation*}
    \vol(B_{\mathbf{c},n}(\rho)) = \frac{(\sqrt{\pi}\rho)^n}{\Gamma\left(\frac{n}{2} + 1\right)} \sim \frac{1}{\sqrt{\pi n}} \left(\frac{\sqrt{2 \pi e }\rho}{\sqrt{n}}\right)^n,
  \end{equation*}
  where $\Gamma(\cdot)$ is Euler's Gamma function.
\end{lemma}

\begin{lemma}[Bounds of the binomial coefficient]
  \label{lem:binomial_coeff}
  Let $n$ be a natural number and let $0 < \theta < 1$ be any rational number such that $\theta n$ is natural, too. If $h(n)$ is the binary entropy function, then:
  \begin{equation}
    \label{eq:binom_coeff_bounds}
    \frac{1}{\sqrt{8n\theta(1-\theta)}} 2^{nh(\theta)} \leq \binom{n}{\theta n} \leq \frac{1}{\sqrt{2 \pi n \theta (1-\theta)}} 2^{nh(\theta)}.
  \end{equation}
  For $k\in \mathbb{N}$ smaller than $n$, another classical upper bound of the binomial coefficient is
  \begin{equation*}
    \binom{n}{k} \leq \min \left\{n^k,n^{n-k},2^n\right\}.
  \end{equation*}
\end{lemma}
The proof of \eqref{eq:binom_coeff_bounds}, as it is proposed in \cite[p.~309]{MacWilliams1977}, is nothing more than a direct computation that employs Stirling's inequality to approximate the factorial functions in the binomial coefficient.

\section{Random Construction-A lattices achieve capacity}
\label{sec:capacity_for_random_construction_A}
The first main result of this paper is Theorem \ref{thm:awgn_capacity} of Section \ref{sec:capacity_with_random_construction_A}, which consists of a new proof that there exists a random ensemble of Construction-A lattices that achieves capacity under MMSE lattice decoding when $\snr > 1$. Our work preserves the main advantages of the already known results on Construction-A lattices, while overcoming some of their less attractive aspects. The main features of our proof are:
\begin{itemize}
\item Our family is similar to the one proposed by Ordentlich and Erez \cite{Ordentlich2012}, but our approach is dual with respect to theirs: our Construction-A ensemble is defined via a set of parity-check matrices (cf. Section \ref{sec:the_random_ensemble}), whereas they employ generator matrices.
\item We still adopt the technique of Voronoi constellations for shaping.
\item We do not need dithering anymore. This meets the purpose of Ling and Belfiore \cite{Ling2014} of avoiding the unpractical sharing of common randomness between the sender and the receiver. However, they pay the price of a non-constructive encoder. Our proof instead does not need lattice Gaussian distribution and we still have an \emph{a priori} uniform distribution over the lattice constellation. Moreover, an explicit bijection exists that maps messages to constellation points (cf. \eqref{eq:phi}). This is desirable when we think of practical implementations of our encoding and decoding scheme. Our transmission scheme is summarized in Fig.~\ref{fig:enc_dec_scheme} and treated in detail in Section \ref{sec:encoding_and_decoding}.
\item We still rely on the idea of scaling the AWGN channel output by the Wiener coefficient, before performing lattice decoding. This enhances the strength of the decoder.
\item We restrict our construction to the case $\snr > 1$. The reasons of this choice will be explained in Section \ref{sec:geometric_description}.
\item With respect to Ordentlich and Erez's construction, we decrease the size of the prime number needed for Construction A as a function of $n$, still attaining capacity (recall that they have $p \approx n^{3/2}$). Again, this has practical advantages.
\end{itemize}

Most of the previously listed features will concern Theorem \ref{thm:awgn_capacity_LDA}, too. Sections from \ref{sec:capacity_for_random_construction_A} to \ref{sec:graph_theoretical_tools}, although they are self-contained and relevant on their own, can be also considered as an essential and detailed introduction to the proof of Theorem \ref{thm:awgn_capacity_LDA}, which restricts the random Construction-A ensemble to a Low-Density Construction-A (LDA) ensemble. Presenting first Theorem \ref{thm:awgn_capacity} allows the reader to understand the strategy and the tools required to show how capacity is achieved independently from the problems that arise from other less general constructions. Consequently, when moving to the proof of Theorem \ref{thm:awgn_capacity_LDA}, we will be able to focus more on those technicalities that strictly belong to the low-density structure associated with LDA lattices.

\begin{figure}[!t]
  \centering
  \fbox{
    \begin{minipage}{0.9\linewidth}
      {\footnotesize \ }
      \begin{enumerate}
      \item {\bf Generation of the random lattice.} Choose with uniform distribution over $\Fp$ a parity-check matrix $H$ of dimension $(\ell + r) \times n$, with $\ell = n(R_f-R)$ and $r = n(1-R_f)$; see \eqref{eq:parity_check_matrix}.
      \item {\bf Encoding of a message $\mathbf{m} \in \Fp^\ell$.} Find a vector $\mathbf{x} \in \Zn$ of smallest norm such that $H\mathbf{x}^T \equiv (\mathbf{m}\ |\ \zero)^T \in \Fp^\ell \times \Fp^r$. The messages are supposed to be uniformly chosen.
      \item {\bf Decoding of the received vector $\mathbf{y}$.} MMSE lattice decoding of the channel output $\mathbf{y}$: $\hat{\mathbf{x}} = Q_{\Lambda_f}(\alpha \mathbf{y})$, where $\alpha$ is the Wiener coefficient.
      \end{enumerate}
          {\footnotesize \ }
    \end{minipage}
  }
  \caption[Our encoding and decoding scheme.]{Our encoding and decoding scheme.
    \label{fig:enc_dec_scheme}}
\end{figure}

\section{The random Construction-A ensemble}%------------------------- SUBSECTION - THE RANDOM ENSEMBLE ---------------------
\label{sec:the_random_ensemble}

Our random Construction-A ensemble is simply given by a random parity-check matrix, whose entries are independent random variables uniformly distributed over $\{0,1,\ldots,p-1\} \simeq \Fp$. In particular, let $H$ be this matrix, of dimension $n(1-R) \times n$ for some $0 < R < 1$ and let $H_f$ be its lower submatrix formed by the last $n(1-R_f)$ rows of $H$ for some $R < R_f < 1$:
\begin{equation}
  \label{eq:parity_check_matrix}
  H = \left(
  \begin{array}{c}
    H' \\ \hline
    H_f
  \end{array}
  \right).
\end{equation} 
The submatrix $H_f$ defines a linear code $C_f$ over $\Fp$ and the whole matrix $H$ defines a subcode $C$ of $C_f$.
The two lattices $\Lambda$ and $\Lambda_f$, obtained with Construction A respectively from $C$ and $C_f$ are nested:
\begin{equation*}
  \Lambda = \{ \mathbf{x} \in \Zn : H\mathbf{x}^T \equiv \zero^T \bmod p\} \subseteq  \{ \mathbf{x} \in \Zn : H_f\mathbf{x}^T \equiv \zero^T \bmod p\} = \Lambda_f.
\end{equation*}
The Voronoi constellation that we consider is then given by $\Lambda_f \cap \vor(\Lambda)$ (see Definition \ref{def:voronoi_constellation}). %Notice that we will let the dimension $n$ change and tend to infinity all along our proofs, so it would be more proper to call the lattices of the random ensemble $\Lambda^{(n)}$ and $\Lambda_f^{(n)}$; nevertheless, since this will not cause any ambiguity, we will keep the lightened notation $\Lambda$ and $\Lambda_f$.
If we suppose that all the rows of $H$ are linearly independent, then $R$ and $R_f$ are the real rates of the codes $C$ and $C_f$ respectively. It is known that
\begin{equation*}
  \vol(\Lambda) = p^{n(1-R)} \text{ and } \vol(\Lambda_f) = p^{n(1-R_f)},
\end{equation*}
from which we deduce that the cardinality $M$ of the lattice constellation is
\begin{equation}
  \label{eq:constellation_size}
  M = |\Lambda_f/\Lambda| = \frac{\vol(\Lambda)}{\vol(\Lambda_f)} = p^{n(R_f-R)}.
\end{equation}
Notice that the probability that the rank of $H$ is strictly smaller than $n(1-R)$ can be shown to decrease to $0$ very fast when $n$ tends to infinity; hence we will work as if $H$ always had full rank. %In other terms, an $H$ whose rows are not independent has to be considered a ``bad'' choice and discarded; the contribution of these ``bad'' matrices to the average arguments that we carry out asymptotically vanishes and is completely negligible.

\section{Encoding and decoding}%------------------------- SUBSECTION - ENCODING AND DECODING ---------------------
\label{sec:encoding_and_decoding}
The points of the constellation (or equivalently the cosets of $\Lambda_f/\Lambda$) are indexed by the $p^{n(R_f-R)}$ different syndromes of the form $(s_1, s_2, \ldots, s_{n(R_f-R)},0,\ldots,0)$ associated with the matrix $H$, where all the $s_i \in \Fp$. More explicitly, let $\ell = n(R_f-R)$ and let $\Fp^\ell$ be (in $1$-$1$ correspondence with) the set of the messages; the bijection
\begin{equation}
  \label{eq:phi}
  \begin{aligned}
    \varphi \colon & \Lambda_f \cap \vor(\Lambda) \to \Fp^\ell \\
    & \mathbf{x} \mapsto H'\mathbf{x}^T \bmod p
  \end{aligned}
\end{equation}
makes a constructive encoding possible (recall that $H'$ is the upper submatrix of $H$). Our transmission scheme works as follows:
\begin{enumerate}
\item The sender pairs up a message and a syndrome and transmits $\mathbf{x}$, the corresponding constellation point obtained via $\varphi^{-1}$, over the AWGN channel.
\item The receiver gets the channel output $\mathbf{y} = \mathbf{x} + \mathbf{w}$ and multiplies it by the Wiener coefficient $\alpha$.
\item Then, he performs lattice decoding of $\alpha \mathbf{y}$ and gets $\hat{\mathbf{x}} = Q_{\Lambda_f}(\alpha \mathbf{y})$.
\item The decoded message will be the one associated with $\varphi(\hat{\mathbf{x}})$.
\end{enumerate}
A final remark on the bijection $\varphi$: for every $\mathbf{s}' \in \Fp^\ell$, let $\mathbf{x} \in \Lambda_f$ be any solution of the linear system $H'\mathbf{x}^T \equiv \mathbf{s}'^T \bmod p$. Then
\begin{equation*}
  \varphi^{-1}(\mathbf{s}') = \mathbf{x} - Q_{\Lambda}(\mathbf{x})
\end{equation*}
and the encoding operation can be substantially performed thanks to a lattice decoder, too.

\section{How to achieve capacity - Overview and discussion on our proof}%------------------------- SUBSECTION - MAIN IDEAS ---------------------
\label{sec:main_ideas}

We will now give a general description of our proof, by the means of a heuristic argument that does not take into account all the probabilistic and asymptotic aspects of the rigorous demonstration.

\subsection{Geometric description}%----------------------------------- SUBSECTION - GEOMETRIC DESCRIPTION -------------------------------
\label{sec:geometric_description}

Our result is based on the following facts:
\begin{itemize}
\item The points of the constellation typically have the same norm and lie very close to the surface of a sphere of a given radius (cf. Lemma \ref{lem:typical_norm}).
\item The AWG noise is typically almost orthogonal to the sent vector, in the sense that, if $\mathbf{x}$ is our transmitted constellation point and $\mathbf{w}$ is the noise, then the scalar product $\mathbf{x} \mathbf{w}^T$ has a ``small enough'' absolute value (cf. Lemma \ref{lem:orthogonal_noise}).
\item The effective noise due to MMSE scaling and the sent point are not decorrelated. Consequently, it is not possible to show that MMSE lattice decoding works with very high probability independently of the sent point. Nevertheless, Theorem \ref{thm:awgn_capacity} is based on the fact that the number of points for which this does not happen is not big enough to perturb the average error probability of the family.
\item For a certain MMSE-scaled channel output, we look for lattice points inside a sphere centered at it and with a typical radius to be specified later. Basically, there will be no decoding error if the only lattice point in this \emph{decoding sphere} is the transmitted one. In a few particular cases, we will need to show explicitly that even if there is more than one lattice point in the decoding sphere, the decoder output will still be the channel input.
\end{itemize}

\begin{figure}[!t]
  %\captionsetup{singlelinecheck=off} % only with package ``caption''
  \begin{center}
    \begin{tikzpicture}[scale=4.0,cap=round,>=latex]
      
      \draw[thick] (0cm,0cm) circle(1cm);
      
      % x:
      \draw[gray, ->] (0cm,0cm) -- (60:1cm);
      % w:
      \draw[gray, ->] (60:1cm) -- (1.1cm, 0.52cm);
      % y:
      \draw[gray, ->] (0cm,0cm) -- (1.1cm, 0.52cm);      % h:
      \draw[gray, ->] (60:1cm) -- (0.7433cm,0.351373cm);    
      
      \draw (1.2cm, 0.53cm) node[fill=white] {$\mathbf{y}$};
      \draw (0.83cm, 0.78cm) node[fill=white] {$\mathbf{w}$};
      \draw (0.82cm,0.28cm) node[fill=white] {$\alpha \mathbf{y}$};
      \draw (-0.06,-0.05cm) node[fill=white] {$\zero$};
      \draw (60:1.1cm) node[fill=white] {$\mathbf{x}$};
      \draw (0.563cm,0.52cm) node[fill=white] {$\mathbf{h}$};

      \draw[rotate=60, gray] (1cm,0cm) +(-2pt,-2pt) rectangle +(0pt,0pt);
      \draw[rotate=25.3, gray] (0.82189cm,0cm) +(0pt,0pt) rectangle +(2pt,2pt);
      
      \filldraw[black] (0cm,0cm) circle(0.4pt);
      \filldraw[black] (1.1cm, 0.52cm) circle(0.4pt);
      \filldraw[black] (60:1cm) circle(0.4pt);
      \filldraw[black] (0.7433cm,0.351373cm) circle(0.4pt);
    \end{tikzpicture}
  \end{center}
  \caption[Geometric interpretation of MMSE scaling.]{
    \label{fig:geometric_interpretation}
    Geometric interpretation: % Package ``caption'' needed for using \itemize here
    %\begin{itemize}
    %\item 
    $\mathbf{x}$ is the transmitted constellation point; $\Vert\mathbf{x}\Vert^2 = nP$.
    %\item 
    The AWG noise vector is $\mathbf{w}$, with $\Vert\mathbf{w}\Vert^2 = n\sigma^2$.
    %\item 
    The AWGN channel output is $\mathbf{y} = \mathbf{x} + \mathbf{w}$.
    %\item 
    The Wiener coefficient is $\alpha = \frac{P}{P+\sigma^2}$
    %\item 
    and $\alpha \mathbf{y}$ is the lattice decoder input.
    %\item 
    $\mathbf{h}$ is the effective noise corresponding to MMSE scaling.
    %\end{itemize}
  }
\end{figure}
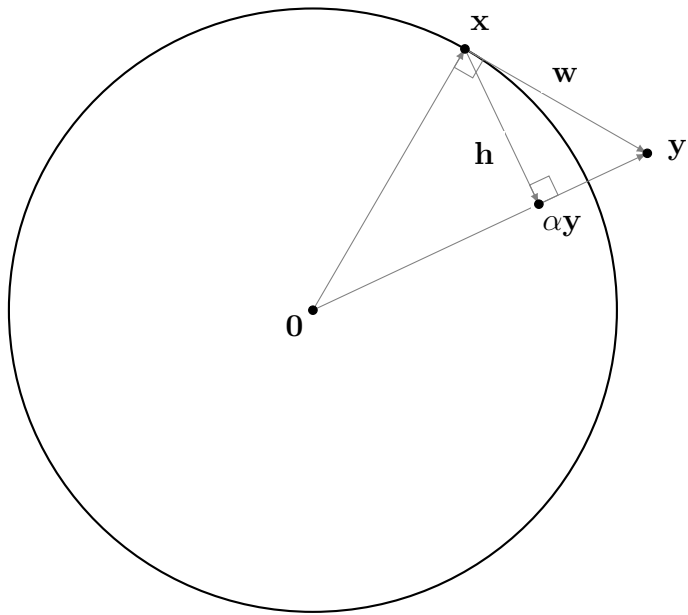

Consider that when we use the adverb ``typically'', we mean ``with probability tending to $1$ when $n$ tends to infinity''. The accurate proof will be treated in all detail in the sequel, but let us try to understand the geometric sense of the elements that we have just listed. So, suppose that the channel input is a point $\mathbf{x}$ whose norm is fixed to be $\Vert\mathbf{x}\Vert = \sqrt{nP}$, for some $P > 0$, which will turn out to be the average (and asymptotically maximum) power of the constellation. Suppose also that $\mathbf{x}\mathbf{w}^T = 0$ (this is a stronger hypothesis than the statement of Lemma \ref{lem:orthogonal_noise}, but it helps to understand the more general scenario); if $\mathbf{y} = \mathbf{x} + \mathbf{w}$ is the channel output, then $\Vert\mathbf{y}\Vert^2 = \Vert\mathbf{x}\Vert^2 + \Vert\mathbf{w}\Vert^2$. Now, let us multiply $\mathbf{y}$ by the scalar value $\alpha$ that minimizes the distance between $\mathbf{x}$ and $\alpha \mathbf{y}$. If $\sigma^2$ is the AWG noise variance per dimension, basic Euclidean geometry (see Fig.~\ref{fig:geometric_interpretation}) tells us that if $\Vert\mathbf{w}\Vert^2 = n\sigma^2$, then $\alpha = P/(\sigma^2 + P)$ is precisely the Wiener coefficient. This lets us guess that MMSE scaling helps in bringing the decoder input closer to the sent point. 

The receiver passes $\alpha \mathbf{y}$ to the lattice decoder and there will be no decoding error if there is no other lattice point closer to $\alpha \mathbf{y}$ than $\mathbf{x}$. We will show that this typically happens when
\begin{enumerate}
\item $\snr > 1$.
\item $P \approx p^{2(1-R)}/2 \pi e$;
\item $\Vert\alpha \mathbf{y} - \mathbf{x}\Vert^2 < np^{2(1-R_f)}/2 \pi e$.
\end{enumerate}
Notice that the latter bound concretely means that 
our constellation tolerates an ``effective'' noise after MMSE scaling whose variance per dimension is less than
\begin{equation*}
  \sigma_{\pol}^2 = \frac{p^{2(1-R_f)}}{2 \pi e}.
\end{equation*}
This value is far from being fortuitous: it is precisely the so called \emph{Poltyrev limit} or \emph{Poltyrev capacity} of the random infinite constellation $\Lambda_f$ \cite[Definition 2.19]{diPietro2014},\cite{Poltyrev1994,Loeliger1997}. We intuitively understand that this is the good condition on the maximum bearable noise, admitting that no problem comes from the fact that the ``effective'' noise and the sent point $\mathbf{x}$ are not decorrelated (incidentally, this would be the case if we used dithering).

The condition on the signal-to-noise ratio can be simply understood with the following argument: let us call $\mathbf{h} = \alpha \mathbf{y} - \mathbf{x}$ and suppose that it takes the maximum value according to the third condition above here, $\Vert\mathbf{h}\Vert^2 = np^{2(1-R_f)}/2 \pi e = n\sigma^2_{\dec}$. We drop the index ``$\pol$'' and use ``$\dec$'' instead, to indicate that the quantity corresponds to the (upper bound of the) reliably decodable effective noise and to the \emph{decoding sphere} defined in the proof of Theorem \ref{thm:awgn_capacity}. If we want good decoding, we need $\alpha \mathbf{y}$ to be closer to $\mathbf{x}$ than to $\zero$, because the latter deterministically belongs to any Voronoi constellation; in other terms, it is necessary that $\Vert\alpha \mathbf{y}\Vert^2 > \Vert\mathbf{h}\Vert^2$. Again, a Euclidean geometry argument based on Fig.~\ref{fig:geometric_interpretation} shows that (always supposing that $\mathbf{x}\mathbf{w}^T = 0$)
\begin{equation}
  \label{eq:alpha_noise}
  n\sigma_{\dec}^2 = \Vert\mathbf{h}\Vert^2 = \frac{\Vert\mathbf{x}\Vert^2 \Vert\mathbf{w}\Vert^2}{\Vert\mathbf{y}\Vert^2} = \frac{n^2 P\sigma^2}{nP + n\sigma^2} = \frac{n P\sigma^2}{P + \sigma^2},
\end{equation}
whereas
\begin{equation*}
  \Vert\alpha \mathbf{y}\Vert^2 = \frac{P^2(nP + n\sigma^2)}{(P+\sigma^2)^2} = \frac{nP^2}{P+\sigma^2}.
\end{equation*}
Then, $\Vert\alpha \mathbf{y}\Vert^2 > \Vert\mathbf{h}\Vert^2$ becomes
\begin{equation*}
  \frac{nP^2}{P+\sigma^2} > \frac{n P\sigma^2}{P + \sigma^2},
\end{equation*}
that is $P > \sigma^2$ or, equivalently, $\snr > 1$. This gives a first explanation why we do not treat the case $\snr \leq 1$.

Taking $\Vert\mathbf{h}\Vert^2 = n\sigma^2_{\dec}$ corresponds to a maximum rate for the constellation that equals capacity, as can be understood from the following calculation: from \eqref{eq:alpha_noise} we can derive that
\begin{equation*}
  \sigma^2 = \frac{P\sigma^2_{\dec}}{P - \sigma^2_{\dec}}.
\end{equation*}
This implies that
\begin{equation*}
  \snr = \frac{P}{\sigma^2} = \frac{P}{\sigma^2_{\dec}}  - 1.
\end{equation*}
Observe that the previous formula shows how decoding $\alpha \mathbf{y}$ enhances the strength of the constellation, as if we had an ``effective'' signal-to-noise ratio $\snr_{\eff} = P/\sigma^2_{\dec} = \snr + 1$. This heuristically explains how we manage to gain the ``plus 1'' in the formula $\frac{1}{2}\log_2(\snr)$, which was the conjectured maximum achievable rate in this context, before the introduction of MMSE scaling \cite{Loeliger1997}. The same argument was pointed out in Erez and Zamir's work \cite{Erez2004}. To conclude, recall that we make the hypothesis that $P \approx p^{2(1-R)}/2 \pi e$; this and \eqref{eq:constellation_size} can be used to show that the AWGN capacity is
\begin{align*}
  %\label{eq:snr_1}
  \frac{1}{2} \log_2( 1 + \snr) & =  \frac{1}{2}\log_2 \left( \frac{P}{\sigma^2_{\dec}} \right) \\
  & \approx \frac{1}{2} \log_2( p^{2(R_f-R)}) \\
  %\label{eq:snr_2}
  & = \frac{1}{n} \log_2( p^{n(R_f-R)}),
\end{align*}
which is exactly the rate of our constellation. A stronger rate would go beyond capacity, the ``effective'' noise would make $\Vert\mathbf{h}\Vert^2$ exceed $n\sigma^2_{\dec}$ and no reliable decoding could be guaranteed.

\subsection{Originality of our proof and lattice decoding of $\alpha \mathbf{y}$}%------------ SUBSECTION - DECODING OF aY ----------
\label{sec:originality_proof}

What we have explained till now gives an intuitive description of the typical geometry that characterises the AWG noise and the random Voronoi constellations of Construction-A nested lattices. Nevertheless, it does not directly drop a hint on the original idea behind our proof that allows to avoid dithering. It is worth the effort of spending some words about that now, before moving on to the detailed proof.

The main argument is the following: if $\alpha \mathbf{y}$ is the real point that the receiver passes to the lattice decoder, we fix as our working environment the sphere $B_{\alpha \mathbf{y},n}(\sqrt{n}\sigma_{\dec})$, which we call the \emph{decoding sphere}. After ensuring that the sent point $\mathbf{x}$ lies in it, our general strategy aims to prove that it is the only lattice point inside the \emph{decoding sphere}. This would imply that lattice decoding does not fail, but unfortunately this does not happen for every instance of the AWG noise and may not happen for every point of the constellation. Hence, we apply an averaging argument that leads among other things to the estimation of (a more elaborate version of) the following sum:
\begin{equation*}
  %\label{eq:sum_prototype}
  \sum_{\mathbf{z} \in B_{\alpha \mathbf{y},n}(\sqrt{n}\sigma_{\dec}) \smallsetminus \{\mathbf{x}\}}\prob\{\mathbf{z} \in \Lambda_f\ |\ \mathbf{x} \in \Lambda_f\}.
\end{equation*}
Showing that this sum vanishes when $n$ tends to infinity will be our main goal. It will be clear later that the best situation possible is when the two events $\{\mathbf{z} \in \Lambda_f\}$ and $\{\mathbf{x} \in \Lambda_f\}$ are independent; but, in principle, they may not be, also because the multiplication by $\alpha$ adds some correlation between $\mathbf{x}$ and the ``effective'' noise $\alpha \mathbf{y} - \mathbf{x}$. One can interpret Erez and Zamir's dithering technique as a method of eliminating this correlation. We do not use dither and consequently there will be \emph{a priori} some $\mathbf{x}$ for which the probability in the previous sum takes a ``big'' value, while at the same time we need to show that the whole sum is ``small''. The originality of our analysis consists of deducing that the proportion of this kind of points in the constellation is very small and the total error decoding probability still goes to $0$ when $n$ tends to infinity (see Lemma \ref{lem:good_points} and its application to \eqref{eq:cardinality_of_S'} in the proof of Theorem \ref{thm:awgn_capacity}).

\section{The detailed proof}%------------------------- SUBSECTION - DETAILED PROOF ---------------------
\label{sec:detailed_proof}

From now on, we will go into all the technical aspects of our proof that there exists a random capacity-achieving Construction-A lattice family. This result will be formally stated and proved in Theorem \ref{thm:awgn_capacity}. For the sake of clearness, we have taken out of its proof a certain number of lemmas, that we present below here. Except for Lemma \ref{lem:typical_norm}, their proofs are in the appendices, because they do not rely on coding or information-theoretical techniques.

\subsection{The typical norm of a constellation point}

We now evaluate precisely the typical norm of a constellation point. Let $\rho_{\eff}^{(n)}$ be the effective radius of the n-dimensional shaping lattice $\Lambda$ (see also Definition \ref{def:effective_radius} and \ref{def:voronoi_constellation}). It is the radius of the ball which has the same volume as $\vor(\Lambda)$, the Voronoi region of the shaping lattice: $\vol(\vor(\Lambda)) = \vol\left(B_{\zero,n}\left(\rho_{\eff}^{(n)}\right)\right)$. Hence,
\begin{equation*}
  \rho_{\eff}^{(n)} = p^{(1-R)}\vol(B_{\zero,n}(1))^{-1/n} \sim \frac{\sqrt{n} p^{(1-R)}}{\sqrt{2 \pi e}},
\end{equation*}
by Lemma \ref{lem:volume_sphere}. We denote the asymptotic value
\begin{equation}
  \label{eq:effective_radius}
  \rho_{\eff} = \frac{\sqrt{n} p^{(1-R)}}{\sqrt{2 \pi e}}.
\end{equation}
We claim that for $n$ large enough almost all the points of the constellation lie very close to the surface of the ball $B_{\zero,n}(\rho_{\eff})$. Before formally proving this, we need the following lemma, whose proof is in Appendix \ref{sec:proof_equivalent_points}.
\begin{lemma}
  \label{lem:equivalent_points}
  Let $\B = B_{\mathbf{c},n}(\rho)$ and let $\mathbf{x}$ be any point of $\B \cap \Zn$. If $p$ is a prime number and $\mu \in \Fp$, then
  \begin{equation*}
    |\{\mathbf{z} \in \B \cap \Zn : \mathbf{z} \equiv \mu \mathbf{x} \bmod p\}| \leq 1 + \frac{4\rho^2}{p^2}\left( \frac{8n\rho^2}{p^2} \right)^{4\rho^2/p^2}.
  \end{equation*}
\end{lemma}

We are ready to state and demonstrate the lemma about the typical norm of a constellation point. The constellation we consider is the one presented in Section \ref{sec:the_random_ensemble}:
\begin{lemma}[Typical norm of a constellation point]
  \label{lem:typical_norm}
  Let $\mathbf{s} = (s_1, s_2, \ldots, s_{n(R_f-R)},0,\ldots,0) \in \Fp^{n(1-R)} \smallsetminus \{\zero\}$ be any non-zero syndrome in the notation of Section \ref{sec:encoding_and_decoding}. Suppose that $p=n^{\lambda}$ for some $\lambda > 0$ and let $\omega$ be a constant such that 
  \begin{equation}
    \label{eq:omega_lambda}
    0 < \omega < \min \{\lambda(1-R), 2\lambda R,1\}.
  \end{equation}
  If $\mathbf{x}$ is the random (over the choice of the matrix $H$) constellation point associated with the syndrome $\mathbf{s}$ via $\varphi^{-1}$ as in \eqref{eq:phi}, then
  \begin{equation}
    \label{eq:typical_norm}
    \lim_{n\to \infty} \prob\left\{\rho_{\eff}\left(1-\frac{1}{n^\omega}\right) \leq \Vert\mathbf{x}\Vert \leq \rho_{\eff}\left(1+\frac{1}{n^\omega}\right)\right\} = 1.
  \end{equation}
\end{lemma}
\begin{IEEEproof}
  Let $X_\rho$ be the random variable that counts the number of points with syndrome $\mathbf{s}$ in the $n$-dimensional ball $B_{\zero, n}(\rho)$ centered at $\zero$ with radius $\rho$. For any $\mathbf{x} \in \Zn \cap B_{\zero,n}(\rho)$, we define the random variable
  \begin{equation*}
    X_{\mathbf{x}} =
    \begin{cases}
      1, & \text{if } H\mathbf{x}^T \equiv \mathbf{s}^T \bmod p \\ 
      0, & \text{otherwise} \\
    \end{cases}
  \end{equation*}
  that depends on the random choice of $H$. In particular,
  \begin{equation*}
    \prob\{ X_{\mathbf{x}} = 1 \}=
    \begin{cases}
      \left(\frac{1}{p}\right)^{n(1-R)}, & \text{if } \mathbf{x} \in \Zn \cap B_{\zero,n}(\rho) \smallsetminus p\Zn \\ 
      0, & \text{if } \mathbf{x} \in p\Zn \cap B_{\zero,n}(\rho)\\
    \end{cases}
  \end{equation*}
  (recall that $\mathbf{s} \neq \zero$) and clearly
  \begin{equation*}
    X_{\rho} = \sum_{\mathbf{x} \in \Zn \cap B_{\zero,n}(\rho)}X_{\mathbf{x}} = \sum_{\mathbf{x} \in \Zn \cap B_{\zero,n}(\rho) \smallsetminus p\Zn}X_{\mathbf{x}}.
  \end{equation*}

  We will split the proof into two parts. First of all, we will argue that
  \begin{equation}
    \label{eq:first_part}
    \lim_{n\to \infty} \prob\{X_{\rho_{\eff}\left(1-\frac{1}{n^\omega}\right)} > 0\} = 0.
  \end{equation}
  Later, we will show that
  \begin{equation}
    \label{eq:second_part}
    \lim_{n\to \infty} \prob\{X_{\rho_{\eff}\left(1+\frac{1}{n^\omega}\right)} = 0\} = 0.
  \end{equation}
  These two results together imply \eqref{eq:typical_norm}.

  {\bf Proof of \eqref{eq:first_part}.} When $\rho = \rho_{\eff}\left(1-1/n^\omega\right)$,
  \begin{align}
    \E[X_{\rho}]  & = \sum_{\mathbf{x} \in \Zn \cap B_{\zero,n}(\rho)} \prob\{X_{\mathbf{x}}=1\} \nonumber \\
    \label{eq:standard_volume_power_pf_p_ratio}
    & \leq |\Zn \cap B_{\zero,n}(\rho)|\left(\frac{1}{p}\right)^{n(1-R)} \\
    %La disuguaglianza qui sopra viene dal fatto che i punti di pZn hanno probabilità 0 e non 1/p^(n(1-R)) di avere sindrome s diversa da zero.
    & \leq \vol\left(B_{\zero,n}\left(\rho + \frac{\sqrt{n}}{2}\right)\right) \left(\frac{1}{p}\right)^{n(1-R)} \nonumber \\
    \label{eq:standard_computation_2}
    & = \vol\left(B_{\zero,n}(1)\right) \rho_{\eff}^n \left(1 - \frac{1}{n^\omega}\right)^n \left(1 + \frac{\sqrt{n}}{2\rho}\right)^n \left(\frac{1}{p}\right)^{n(1-R)} \\
    \label{eq:effective_radius_volume}
    & \sim \exp \left(-n^{1-\omega} + \sqrt{\frac{\pi e}{2}} n^{1-\lambda(1-R)}\right)
  \end{align}
  where in \eqref{eq:effective_radius_volume} we have used the fact that $\vol\left(B_{\zero,n}(1)\right) \rho_{\eff}^n \sim p^{n(1-R)}$ by definition of effective radius and Lemma \ref{lem:volume_sphere}. The whole quantity tends to $0$, since $1-\omega > 1-\lambda(1-R)$ by \eqref{eq:omega_lambda} and the argument of the exponential function goes to $-\infty$; considering the fact that $\prob\{X_{\rho}>0\} \leq \E[X_{\rho}]$, we also have
  \begin{equation*}
    \lim_{n\to \infty} \prob\left\{X_{\rho_{\eff}\left(1-\frac{1}{n^\omega}\right)} > 0\right\} = 0.
  \end{equation*}

  {\bf Proof of \eqref{eq:second_part}.} Now, let $\rho = \rho_{\eff}\left(1+1/n^\omega\right)$. Taking into account the fact that $|\Zn \cap B_{\zero,n}(\rho) \smallsetminus p\Zn| \sim |\Zn \cap B_{\zero,n}(\rho)|$, we have
  \begin{align}
    %\label{eq:average_first_step}
    \E[X_{\rho}]  & = \sum_{\mathbf{x} \in \Zn \cap B_{\zero,n}(\rho) \smallsetminus p\Zn} \prob\{X_{\mathbf{x}}=1\} \nonumber \\
    \label{eq:general_term_average}
    & = |\Zn \cap B_{\zero,n}(\rho) \smallsetminus p\Zn|\left(\frac{1}{p}\right)^{n(1-R)} \\
    & \sim |\Zn \cap B_{\zero,n}(\rho)| \left(\frac{1}{p}\right)^{n(1-R)} \nonumber \\
    & \geq \vol\left(B_{\zero,n}\left(\rho - \frac{\sqrt{n}}{2}\right)\right) \left(\frac{1}{p}\right)^{n(1-R)} \nonumber \\
    & = \vol\left(B_{\zero,n}(1)\right) \rho_{\eff}^n \left(1 + \frac{1}{n^\omega}\right)^n \left(1 - \frac{\sqrt{n}}{2\rho}\right)^n \left(\frac{1}{p}\right)^{n(1-R)} \nonumber \\
    & \sim \exp \left(n^{1-\omega} - \sqrt{\frac{\pi e}{2}} n^{1-\lambda(1-R)}\right)
    \label{eq:lower_bound_mean}
    %& = \frac{1}{\sqrt{\pi n}} \exp\left( n^{1-\omega} - \sqrt{\frac{\pi e}{2}} \left(1 + \frac{1}{n^\omega}\right)^{-1} \cdot n^{-\lambda(1-R)+1}\right),
  \end{align}
  which tends to infinity, again thanks to \eqref{eq:omega_lambda}. Hence,
  \begin{equation*}
    \lim_{n \to \infty}\E[X_{\rho}] = +\infty.
  \end{equation*}

  Suppose now for a moment that $\var(X_{\rho}) \leq f(n)\E[X_{\rho}]$ for some $f(n) = o(\E[X_{\rho}])$; we would have
  \begin{align}
    \prob\{X_{\rho} = 0\} & \leq \prob\{|X_{\rho} - \E[X_{\rho}]| \geq \E[X_{\rho}]\} \nonumber \\
    \label{eq:cheb_ineq}
    & \leq \frac{\var(X_{\rho})}{\E[X_{\rho}]^2} \\
    & \leq \frac{f(n)}{\E[X_{\rho}]} \longrightarrow 0, \nonumber
  \end{align}
  where we have applied Chebyshev's inequality 
  %(Lemma \ref{lem:chebyshev_inequality}) 
  to obtain \eqref{eq:cheb_ineq}. This would be enough to prove \eqref{eq:second_part} and conclude. For this reason, let us show that $\var(X_{\rho}) \leq  f(n) \E[X_{\rho}]$; to do this, we investigate the quantity 
  \begin{equation*}
    \cov(X_{\mathbf{x}},X_{\mathbf{z}}) = \E[X_{\mathbf{x}}X_{\mathbf{z}}] - \E[X_{\mathbf{x}}]\E[X_{\mathbf{z}}],
  \end{equation*}
  for $\mathbf{x},\mathbf{z} \in \Zn \cap B_{\zero,n}(\rho)$. Observe that, by the definition of the two random variables, if $\mathbf{h}_i$ is the $i$-th row of $H$,
  \begin{align*}
    \E[X_{\mathbf{x}}X_{\mathbf{z}}] & = \prob \{ X_{\mathbf{x}}X_{\mathbf{z}} = 1\} \\
    & = \prob \{X_{\mathbf{x}}=1, X_{\mathbf{z}} = 1 \} \\
    & = \prod_{i=1}^{n(1-R)} \prob\{ \mathbf{h}_i\mathbf{x}^T \equiv s_i \bmod p, \mathbf{h}_i\mathbf{z}^T \equiv s_i \bmod p\}.
  \end{align*}
  There are three possibilities:
  \begin{enumerate}
  \item If $\mathbf{x} \not \equiv a\mathbf{z} \bmod p$ for all $a \in \Fp$, 
    then $X_{\mathbf{x}}$ and $X_{\mathbf{z}}$ are independent and $\cov(X_{\mathbf{x}},X_{\mathbf{z}}) = 0$.
  \item If $\mathbf{x} \equiv a\mathbf{z} \bmod p$ for some $a \in \Fp \smallsetminus \{1\}$, let $i$ be an index such that $s_i \neq 0$ (there always exists, since $\mathbf{s} \neq \zero$). Hence, either $a\mathbf{h}_i\mathbf{z}^T \equiv s_i \bmod p$ or $\mathbf{h}_i\mathbf{z}^T \equiv s_i \bmod p$, with no chance that the two events happen together. Then
    \begin{equation*}
      \prob\{ \mathbf{h}_i\mathbf{x}^T \equiv s_i \bmod p, \mathbf{h}_i\mathbf{z}^T \equiv s_i \bmod p\} = 0,
    \end{equation*}
    $\E[X_{\mathbf{x}}X_{\mathbf{z}}] = 0$ and $\cov(X_{\mathbf{x}},X_{\mathbf{z}}) \leq 0$.
  \item Finally, if  $\mathbf{x} \equiv \mathbf{z} \bmod p$, then $X_{\mathbf{x}}X_{\mathbf{z}} = X_{\mathbf{x}}^2 = X_{\mathbf{x}}$ and $\E[X_{\mathbf{x}}X_{\mathbf{z}}] = \E[X_{\mathbf{x}}]$. That is, $\cov(X_{\mathbf{x}},X_{\mathbf{z}}) \leq \E[X_{\mathbf{x}}]$.
  \end{enumerate}
  Putting all of this together, we have
  \begin{align}
    \var(X_{\rho}) = & \var\left(\sum_{\mathbf{x} \in \Zn \cap B_{\zero,n}(\rho)} X_{\mathbf{x}}\right) \nonumber \\
    = & \sum_{\mathbf{x},\mathbf{z} \in \Zn \cap B_{\zero,n}(\rho)} \cov(X_{\mathbf{x}},X_{\mathbf{z}}) \nonumber \\
    = & \sum_{\substack{\mathbf{x},\mathbf{z} \in \Zn \cap B_{\zero,n}(\rho) \\ \mathbf{x} \not \equiv a\mathbf{z}}} \cov(X_{\mathbf{x}},X_{\mathbf{z}}) + \sum_{\substack{\mathbf{x},\mathbf{z} \in \Zn \cap B_{\zero,n}(\rho) \\ \mathbf{x} \equiv a\mathbf{z},\ a\neq 1}} \cov(X_{\mathbf{x}},X_{\mathbf{z}}) \nonumber \\
    & + \sum_{\substack{\mathbf{x},\mathbf{z} \in \Zn \cap B_{\zero,n}(\rho) \\ \mathbf{x} \equiv \mathbf{z}}} \cov(X_{\mathbf{x}},X_{\mathbf{z}}) \nonumber \\
    \leq & \sum_{\mathbf{x} \in \Zn \cap B_{\zero,n}(\rho)}\sum_{\substack{\mathbf{z} \in \Zn \cap B_{\zero,n}(\rho) \\ \mathbf{x} \equiv \mathbf{z}}} \E[X_{\mathbf{x}}] \nonumber \\
    \label{eq:equivalent_points_estimation}
    \leq & \sum_{\mathbf{x} \in \Zn \cap B_{\zero,n}(\rho)}  \left( 1 + \frac{4\rho^2}{p^2}\left( \frac{8n\rho^2}{p^2} \right)^{4\rho^2/p^2}\right) \E[X_{\mathbf{x}}] \\
    = & \left(1+\frac{4\rho^2}{p^2}\left( \frac{8n\rho^2}{p^2} \right)^{4\rho^2/p^2}\right) \E[X_\rho], \nonumber 
  \end{align}
  where \eqref{eq:equivalent_points_estimation} is a consequence of Lemma \ref{lem:equivalent_points}. The last thing we need to conclude is that
  \begin{equation*}
    \lim_{n\to \infty} \frac{f(n)}{\E[X_\rho]} = \lim_{n\to \infty} \frac{1 + 4\rho^2/p^2 \left( 8n\rho^2/p^2 \right)^{4\rho^2/p^2}}{\E[X_\rho]} = 0.
  \end{equation*}
  Taking into account that $\rho = \sqrt{n}p^{(1-R)}(1+1/n^\omega)/2\pi e$ and $p=n^\lambda$, one can compute that the dominating term (up to some multiplicative constants in the exponent) of the numerator is $n^{n^{1-2\lambda R}} = \exp(n^{1-2\lambda R} \ln n)$. On the other hand, \eqref{eq:lower_bound_mean} and \eqref{eq:omega_lambda} tell that the dominating term in the asymptotic lower bound of the denominator is $\exp(n^{1-\omega})$. Hence, the limit is $0$ if
  \begin{equation*}
    1 - 2\lambda R < 1 - \omega,
  \end{equation*}
  which is true, again by \eqref{eq:omega_lambda}.  
\end{IEEEproof}

\begin{definition}[Shaping sphere]
  \label{def:shaping_sphere}
  We have just proven that almost all the points of the constellation lie very close to the surface of the ball $\B_{\eff} = B_{\zero,n}(\rho_{\eff}(1 + 1/n^\omega))$. For this reason, from now on, we will call the latter the \emph{shaping sphere}.
\end{definition}

\subsection{A property of the Gaussian noise}%------------------------- SUBSECTION - PROPERTY OF THE NOISE ---------------------

The following lemma formally explains in what probabilistic, asymptotic sense the typical AWG noise vector is almost orthogonal to constellation points (see also the comments in Section \ref{sec:geometric_description}). Explicitly, we bound their scalar product by a quantity that in the proof of Theorem \ref{thm:awgn_capacity} turns out to be negligible with respect to their squared norms. Hence, $\Vert\mathbf{x} + \mathbf{w}\Vert^2$ can be accurately enough approximated by $\Vert\mathbf{x}\Vert^2 + \Vert\mathbf{w}\Vert^2$. The proof of the lemma is written in Appendix \ref{sec:proof_orthogonal_noise}.
\begin{lemma}[Orthogonal noise]
  \label{lem:orthogonal_noise}
  Let $\mathbf{x} \in \Rn$ and let $\mathbf{w} = (w_1,w_2,\ldots,w_n)$ be a random AWG noise vector with i.i.d.\ components: $w_i \sim \mathcal{N}(0,\sigma^2)$. Then, for every function $f(n)$ such that $\lim_{n \to \infty}f(n) = +\infty$, we have
  \begin{equation*}
    \lim_{n \to \infty} \prob \{ |\mathbf{x}\mathbf{w}^T| \leq f(n)\sigma \Vert\mathbf{x}\Vert\} = 1.
  \end{equation*}
\end{lemma}

\subsection{Multiple points modulo $p$ in the decoding sphere}%------------------------- SUBSECTION - GOOD POINTS ---------------------
\label{subsec:good_points}

Lemma \ref{lem:equivalent_points} consists of an upper bound of the number of points of the same class modulo $p$ inside a certain ball $\B$. Instead, the following lemma, whose proof is in Appendix \ref{sec:proof_good_points}, counts for how many points of the shaping sphere the previous number is not $0$, when we choose $\B$ to be a particular ball that will appear in the proof of Theorem \ref{thm:awgn_capacity}. 
\begin{lemma}
  \label{lem:good_points}
  Consider the shaping sphere $\B_{\eff} = B_{\zero,n}(\rho_{\eff}(1 + 1/n^\omega))$ and let 
  \begin{equation*}
    \rho = \frac{p^{1-R_f}\sqrt{n}(1+\varepsilon)}{\sqrt{2 \pi e}},
  \end{equation*}
  where $\rho_{\eff} = \sqrt{n}p^{(1-R)}/\sqrt{2 \pi e}$, $p = n^\lambda$ for some constant $\lambda$, $\omega$ is chosen as in Lemma \ref{lem:typical_norm}, and $R$ and $R_f$ are defined in Section \ref{sec:the_random_ensemble}). Moreover, suppose that 
  \begin{equation}
    \label{eq:rates_conditions}
    R > 1/2 \text{\ \ \ and\ \ \ } p^{R_f - R} = \Omega,
  \end{equation}
  for some constant $\Omega > 1$.
  Let $\mu \in \{-(p-1)/2,-(p-3)/2,\ldots,(p-1)/2\} \smallsetminus \{0,1,2\}$. We define 
  \begin{equation*}
    N(\mu) = |\{\mathbf{x} \in \Zn \cap \B_{\eff} : \exists \mathbf{z} \in \Zn \cap B_{\mathbf{x},n}(2\rho) \text{ for which } \mathbf{z} \equiv \mu \mathbf{x} \bmod p\}|. 
  \end{equation*}
  Then, for every function $t(n)$ which grows at most subexponentially fast in $n$,
  \begin{equation}
    \label{eq:big_N}
    N = \sum_{\mu \in \Fp\smallsetminus \{0,1,2\}} N(\mu) = o\left(\frac{p^{n(1-R)}}{t(n)}\right).
  \end{equation}
\end{lemma}

\subsection{The proof that capacity is achieved}
\label{sec:capacity_with_random_construction_A}
We are now ready to state and prove the main result of this section:

\begin{theorem}
  \label{thm:awgn_capacity}
  The random ensemble of nested Construction-A lattices introduced in Section \ref{sec:the_random_ensemble} achieves capacity of the AWGN channel under MMSE lattice decoding, when $\snr > 1$, $R > 1/2$ and $p = n^\lambda$ for some constant $\lambda > (1+R)^{-1}$.
\end{theorem}

\begin{IEEEproof}
  The AWGN channel is defined by the $\snr = P/\sigma^2 > 1$, for some AWG noise variance per dimension $\sigma^2$ and some power constraint $P$. The capacity is then known to be
  \begin{equation*}
    \mathbf{C} = \frac{1}{2} \log_2(1 + \snr).
  \end{equation*}
  Let us call $M = 2^{nR_{\mathcal{C}}}$ the cardinality of our Voronoi constellation; we would like to show that for every fixed rate $R_{\mathcal{C}}$ smaller than capacity, the random ensemble of Section \ref{sec:the_random_ensemble} corresponding to that rate can be reliably decoded. Namely, suppose that $R_{\mathcal{C}} = \gamma \mathbf{C}$, for some constant $0 < \gamma < 1$. Then, we fix the rates of the $\Fp$-linear codes generating the nested lattice ensemble: $1/2 < R < R_f < 1$, such that the constellation $\mathcal{C} = \Lambda_f \cap \vor(\Lambda)$, whose cardinality is $p^{n(R_f-R)}$, has rate
  \begin{equation*}
    \frac{\gamma}{2} \log_2 (1+\snr) = \gamma \mathbf{C} =  R_{\mathcal{C}} = \frac{\log_2p^{n(R_f-R)}}{n} = \log_2p^{R_f-R}, 
  \end{equation*}
  which implies:
  \begin{equation*}
    p^{R_f-R} = (1+\snr)^{\frac{\gamma}{2}}
  \end{equation*}
  (incidentally, notice that \eqref{eq:rates_conditions} is satisfied).
  Now, Lemma \ref{lem:typical_norm} and \eqref{eq:effective_radius} asymptotically imply that the power constraint is
  \begin{equation}
    \label{eq:asymptotic_power}
    P \sim \frac{\rho_{\eff}^2}{n} = \frac{p^{2(1-R)}}{2 \pi e}.
  \end{equation}
  The inequality $R_{\mathcal{C}} = \log_2|\mathcal{C}|/n < \mathbf{C}$ is equivalent to
  \begin{equation}
    \label{eq:new_sigma_max}
    \sigma^2 < \frac{P}{|\mathcal{C}|^{2/n}-1} = \frac{p^{2(1-R)}}{2 \pi e (p^{2(R_f-R)} -1)} = \sigma_{\max}^2.
  \end{equation}
  We have called $\sigma_{\max}^2$ this upper bound because achieving capacity in this setting is equivalent to prove that, for fixed $R_f, R,$ and $\snr$, a random lattice in our ensemble can be reliably decoded (in big enough dimension) for every AWG noise variance value $\sigma^2 = \sigma_{\max}^2(1-\delta')^2$ with $0 < \delta' < 1$. The rest of the proof will be devoted to deriving the latter statement.

  The transmission scheme stays the same as outlined in Fig.~\ref{fig:enc_dec_scheme}. Hence, let us fix a syndrome $\mathbf{s} = (s_1,s_2,\ldots,s_{n(R_f-R)},0,\ldots,0) \in \Fpn$ that represents a message. We recall that the messages are supposed to be \emph{a priori} equiprobable. Let $\mathbf{x}$ be the random coded point associated with $\mathbf{s}$ for some random constellation in the family. If $\mathbf{w}$ is the channel noise (with coordinate-wise variance $\sigma^2$) and $\alpha = P/(P + \sigma^2) = (1+\snr^{-1})^{-1}$ is the Wiener coefficient, we claim that for every $\varepsilon > 0$,
  \begin{equation*}
    \lim_{n \to \infty}\prob\{ \Vert\alpha \mathbf{y} - \mathbf{x}\Vert^2 \leq \alpha n \sigma^2 (1+\varepsilon)^2\} = 1.
  \end{equation*}

  If $\mathbf{s} = \zero$, then $\mathbf{x} = \zero$ and $\mathbf{y} = \mathbf{w}$. The claim is a straightforward consequence of Lemma \ref{lem:typical_noise} (the fact that $\alpha < 1$ is also used). If instead $\mathbf{s} \neq \zero$, let $\varepsilon' < \varepsilon$ be a positive constant, let $f(n)$ be a function such that $\lim_{n \to \infty} f(n) = + \infty$ (to be specified later) and let $\mathcal{E}_1$ be the event
  \begin{equation}
    \label{eq:event_E1}
    \mathcal{E}_1 = \{ \Vert\mathbf{x}\Vert^2 \leq nP(1 + \varepsilon')^2 \} \cap \{ \Vert\mathbf{w}\Vert^2 \leq n\sigma^2(1+\varepsilon')^2 \} \cap \{ |\mathbf{x}\mathbf{w}^T| \leq f(n)\sigma \Vert\mathbf{x}\Vert\}.
  \end{equation}
  Note that, provided that $\varepsilon'$ is small enough, the event $\mathcal{E}_1$ is (asymptotically) contained in the event $\{ \Vert\alpha \mathbf{y} - \mathbf{x}\Vert^2 \leq \alpha n \sigma^2 (1+\varepsilon)^2\}$: indeed, $\mathcal{E}_1$ implies
  \begin{align}
    \Vert\alpha \mathbf{y} - \mathbf{x}\Vert^2 & = (\alpha - 1)^2\Vert\mathbf{x}\Vert^2 + \alpha^2\Vert\mathbf{w}\Vert^2 + 2\alpha(\alpha-1)\mathbf{x}\mathbf{w}^T \nonumber \\
    & \leq \frac{\sigma^4}{(P + \sigma^2)^2}\Vert\mathbf{x}\Vert^2 + \frac{P^2}{(P + \sigma^2)^2}\Vert\mathbf{w}\Vert^2 + \frac{2\sigma^2P}{(P+\sigma^2)^2}|\mathbf{x}\mathbf{w}^T| \nonumber \\
    & \leq \frac{\sigma^4nP(1+\varepsilon')^2}{(P + \sigma^2)^2} + \frac{P^2n\sigma^2(1+\varepsilon')^2}{(P + \sigma^2)^2} + \frac{2\sigma^2Pf(n)\sigma \Vert\mathbf{x}\Vert}{(P+\sigma^2)^2} \nonumber \\
    \label{eq:intermediate_computation}
    & \leq \frac{nP\sigma^2}{P + \sigma^2} \left((1+\varepsilon')^2 + \frac{2f(n)\sigma\sqrt{P}(1+\varepsilon')}{\sqrt{n}(P+\sigma^2)} \right)
  \end{align}
  and
  \begin{align*}
    \lim_{n\to\infty} \frac{2f(n)\sigma\sqrt{P}(1+\varepsilon')}{\sqrt{n}(P+\sigma^2)} & \leq \lim_{n\to\infty} \frac{2f(n) \max \{\sigma^2, P\}(1+\varepsilon')}{\sqrt{n}(P+\sigma^2)}\\
    & \leq \lim_{n\to\infty} \frac{2f(n)(1+\varepsilon')}{\sqrt{n}} = 0,
  \end{align*}
  taking $f(n) = o(\sqrt{n})$. Thus, we can go back to \eqref{eq:intermediate_computation} and obtain (for $n$ big enough and $\varepsilon'$ small enough with respect to $\varepsilon$) that
  \begin{equation*}
    \eqref{eq:intermediate_computation} \leq \frac{nP\sigma^2}{P + \sigma^2}(1+\varepsilon)^2 = \alpha n \sigma^2(1+\varepsilon)^2.
  \end{equation*}
  We are done, because
  \begin{equation}
    \label{eq:x_close_to_alphay}
    \prob\{ \Vert\alpha \mathbf{y} - \mathbf{x}\Vert^2 \leq \alpha n \sigma^2 (1+\varepsilon)^2\} \geq \prob \{\mathcal{E}_1\} \to 1,
  \end{equation}
  by Lemma \ref{lem:typical_norm}, Lemma \ref{lem:typical_noise} and Lemma \ref{lem:orthogonal_noise}.
  Notice also that taking into account \eqref{eq:asymptotic_power} and \eqref{eq:new_sigma_max}, a very simple computation implies that, for any given $\delta'$, there exists $\delta$ (still constant between $0$ and $1$) such that 
  \begin{equation}
    \label{eq:alpha_sigma_max}
    \alpha \sigma^2 < p^{2(1-R_f)}(1-\delta)^2/2 \pi e.
  \end{equation}
  Hence,
  \begin{equation}
    \label{eq:decoding_radius}
    \prob\left\{ \Vert\alpha \mathbf{y} - \mathbf{x}\Vert^2 \leq  \frac{np^{2(1-R_f)}(1-\delta)^2(1+\varepsilon)^2}{2 \pi e}\right\} \geq \prob\{ \Vert\alpha \mathbf{y} - \mathbf{x}\Vert^2 \leq \alpha n \sigma^2 (1+\varepsilon)^2\} \to 1.
  \end{equation}
  We have just shown that with very high probability when $n$ is big enough, the sent point $\mathbf{x}$ lies inside a sphere of radius $\rho_{\dec} = \sqrt{n}p^{(1-R_f)}(1-\delta)(1+\varepsilon)/\sqrt{2\pi e}$ centered at $\alpha \mathbf{y}$. We call this sphere the \emph{decoding sphere} $\B = B_{\alpha \mathbf{y},n}(\rho_{\dec})$ and no decoding error occurs if the only point of $\Lambda_f \cap \B$ is $\mathbf{x}$ (see the related comments in Section \ref{sec:originality_proof}). 

  Let us call the ``good decoding'' event $\mathcal{E}_2 = \{\Lambda_f \cap \B = \{\mathbf{x}\} \}$ and $\mathcal{E}_2^c$ its complement. To prove the theorem, we will show that for every syndrome $\mathbf{s}$, the probability that $\alpha \mathbf{y}$ is not well decoded tends to $0$ for a randomly chosen lattice constellation in the ensemble. Let us call $P_e(\mathbf{s})$ this probability and let $X_{\mathbf{s}}$ be the random variable that represents the constellation point associated with $\mathbf{s}$; $X_{\mathbf{s}}$ takes \emph{a priori} a different value $\mathbf{x} \in \Zn$ for every different choice of a random constellation.

  Let us start with $\mathbf{s} = \zero$. In this case, $\prob \{ X_{\mathbf{s}} = \zero\} = 1$.
  %(one can also deduce it from \eqref{eq:alpha_noise}); 
  To begin, we claim that for every $\mathbf{z} \in p\Zn \smallsetminus \{\zero\}$,
  \begin{equation*}
    \lim_{n \to \infty} \prob\{\Vert\mathbf{w}\Vert^2 \geq \Vert\mathbf{w}-\mathbf{z}\Vert^2\} = 0.
  \end{equation*}
  In other words, the random noise produces a channel output which is typically closer to $\zero$ (the channel input in this case) than to any other point of $p\Zn$. From the point of view of the lattice decoder, this means that the points of $p\Zn$ do not typically induce any decoding errors. Let us prove the claim: since $\mathbf{z}$ belongs to $p\Zn$, a necessary condition when $\Vert\mathbf{w}\Vert^2 \geq \Vert\mathbf{w}-\mathbf{z}\Vert^2$ is that at least one of the coordinates of $\mathbf{w}$ is bigger than $p/2$ in absolute value. Hence
  \begin{align}
    \prob\{\Vert\mathbf{w}\Vert^2 & \geq \Vert\mathbf{w}-\mathbf{z}\Vert^2\} \leq \prob\{|w_i| \geq p/2, \exists i \in \{1,2,\ldots,n\}\} \nonumber \\
    \label{eq:sum_of_prob}
    & \leq \sum_{i=1}^n \prob\{|w_i| \geq p/2\}.
  \end{align}
  Now, $w_i \sim \mathcal{N}(0,\sigma^2)$ for every $i \in \{1,2,\ldots,n\}$ and the probabilities in the previous sum are all identical and independent from $i$. 

  Consider the function $Q(\cdot)$, the tail probability of the standard normal distribution:
  \begin{equation*}
    Q(y) = \frac{1}{\sqrt{2 \pi}} \int_{y}^{\infty} \exp \left(-\frac{u^2}{2}\right) \, \mathrm{d}u.
  \end{equation*}
  For positive $y$, the Chernoff bound states that
  \begin{equation*}
    Q(y) \leq \frac{1}{2}e^{-\frac{y^2}{2}}.
  \end{equation*}
  %Moreover, since $\mathbf{z} \in p\Zn \smallsetminus \{\zero\}$, a lower bound for the norm of $\mathbf{z}$ is $\Vert\mathbf{z}\Vert^2 \geq p^2$.
  Hence, we can go back to \eqref{eq:sum_of_prob} and write (using \eqref{eq:alpha_sigma_max} for the last inequality)
  \begin{align*}
    \sum_{i=1}^n \prob\{|w_i| \geq p/2\} & \leq n \prob\{|w_1| \geq p/2\} \\
    & = 2n Q\left(\frac{p}{2\sigma}\right) \\
    & \leq n \exp\left(-\frac{p^2}{8\sigma^2}\right) \\
    & = n \exp\left(-\frac{\alpha \pi e p^{2R_f}}{4(1-\delta)^2}\right),
  \end{align*}
  which decreases to $0$ because $p = n^\lambda$.

  The claim is proved and we are implicitely saying that with probability tending to $1$ no point of $p\Zn$ different from $\zero$ inside $\B$ can lead to bad decoding. Hence we will restrict our error probability analysis only to points not belonging to $p\Zn$ and, with the help of Lemma \ref{lem:integer_sphere_points} and \ref{lem:volume_sphere}, we obtain
  \begin{align}
    P_e(\mathbf{\zero}) & \sim \prob\{\exists \mathbf{z} \in \Lambda_f \cap \B \smallsetminus p\Zn \} \nonumber \\
    & \leq \sum_{\mathbf{z} \in (\Zn \cap \B) \smallsetminus p\Zn} \prob\{ \mathbf{z} \in \Lambda_f \} \nonumber \\
    & = \sum_{\mathbf{z} \in (\Zn \cap \B) \smallsetminus p\Zn} \left(\frac{1}{p} \right)^{n(1-R_f)} \nonumber \\
    %\label{eq:same_term}
    & \leq |\Zn \cap \B| \left(\frac{1}{p} \right)^{n(1-R_f)} \nonumber \\
    \label{eq:same_term_bis}
    & \leq \vol(\B) \left(1+\frac{\sqrt{n}}{2\rho_{\dec}} \right)^n \left(\frac{1}{p} \right)^{n(1-R_f)}  \\
    & \lesssim \left((1-\delta)(1+\varepsilon)\right)^n g(n), \nonumber
  \end{align}
  where $g(n)$ is a subexponential function. Thus, the dominating term is $\left((1-\delta)(1+\varepsilon)\right)^n$, which tends to $0$ because $(1-\delta)(1+\varepsilon)<1$ (notice that for every fixed $\delta$, we can choose $\varepsilon$ as small as needed).

  Now, let us pass to the case $\mathbf{s} \neq \zero$. Notice that, choosing $\omega$ as in \eqref{eq:omega_lambda}, Lemma \ref{lem:typical_norm} implies that $X_{\mathbf{s}}$ lies inside the shaping sphere 
  %\begin{equation*}
    $\B_{\eff}=B_{\zero,n}(\rho_{\eff}(1+1/n^\omega))$
    %\end{equation*}
  with probability tending to $1$. Therefore,
  \begin{align*}
    P_e(\mathbf{s}) & = \prob \{ \text{ decoding error }|\ X_{\mathbf{s}} \in \B_{\eff}\} \prob\{X_{\mathbf{s}} \in \B_{\eff}\} + \\
    & \ \ \ \ \prob\{\text{ decoding error }|\ X_{\mathbf{s}} \not \in \B_{\eff}\} \prob\{X_{\mathbf{s}} \not \in \B_{\eff}\} \\
    & \sim \prob \{ \text{ decoding error }|\ X_{\mathbf{s}} \in \B_{\eff}\}\prob\{X_{\mathbf{s}} \in \B_{\eff}\}\\
    & = \prob \{X_{\mathbf{s}} \in \B_{\eff}, \text{ decoding error}\}.
  \end{align*}
  For this reason, observing that no point of $p\Zn$ can be the codeword associated with $\mathbf{s}$, we have
  \begin{align}
    P_e(\mathbf{s}) & \sim \prob \{X_{\mathbf{s}} \in \B_{\eff}, \text{ decoding error}\} \nonumber \\
    & \leq \sum_{\mathbf{x} \in \Zn \cap \B_{\eff}} \prob \{ X_{\mathbf{s}} = \mathbf{x}, \mathcal{E}_2^c \} \nonumber \\
    & = \sum_{\mathbf{x} \in (\Zn \cap \B_{\eff}) \smallsetminus p\Zn} \prob \{ X_{\mathbf{s}} = \mathbf{x}, \mathcal{E}_2^c \} \nonumber \\
    \label{eq:P1}
    & \leq \sum_{\mathbf{x} \in (\Zn \cap \B_{\eff}) \smallsetminus p\Zn} \prob \{ X_{\mathbf{s}} = \mathbf{x}, \mathbf{x} \not \in \B\} \\
    \label{eq:P2}
    & \ \ \ \ + \sum_{\mathbf{x} \in (\Zn \cap \B_{\eff}) \smallsetminus p\Zn} \prob \{ X_{\mathbf{s}} = \mathbf{x}, \exists \mathbf{z} \in \Lambda_f \cap \B \smallsetminus \{\mathbf{x}\}\}.
  \end{align}
  We will separately show that \eqref{eq:P1} and \eqref{eq:P2} tend to $0$ when $n$ tends to infinity, which is enough to conclude.

  {\bf Estimation of \eqref{eq:P1}.} By the definition of conditional probability,
  \begin{equation*}
    \eqref{eq:P1} = \sum_{\mathbf{x} \in (\Zn \cap \B_{\eff}) \smallsetminus p\Zn} \prob \{ \mathbf{x} \not \in \B\ |\ X_{\mathbf{s}} = \mathbf{x}\} \prob \{X_{\mathbf{s}} = \mathbf{x}\}.
  \end{equation*}
  \eqref{eq:decoding_radius} tells us that the term $\prob \{ \mathbf{x} \not \in \B\ |\ X_{\mathbf{s}} = \mathbf{x}\}$ is a vanishing term $T(n)$, independently of $\mathbf{x}$. Hence,
  \begin{equation*}
    \eqref{eq:P1} = T(n) \sum_{\mathbf{x} \in (\Zn \cap \B_{\eff}) \smallsetminus p\Zn} \prob \{X_{\mathbf{s}} = \mathbf{x}\} \leq T(n) \to 0.
  \end{equation*}

  {\bf Estimation of \eqref{eq:P2}.} To conclude the proof we only need to show that
  \begin{equation}
    \label{eq:sum_over_all_x}
   \lim_{n \to \infty} \sum_{\mathbf{x} \in (\Zn \cap \B_{\eff})\smallsetminus p\Zn} \prob \{ X_{\mathbf{s}} = \mathbf{x}, \exists \mathbf{z} \in \Lambda_f \cap \B \text{ and } \mathbf{z} \neq \mathbf{x}\} = 0.
  \end{equation}

  Before going on, let us start by making some considerations in a number of particular cases about the error probability, the existence of some $\mathbf{z}$ as in \eqref{eq:sum_over_all_x}, and the corresponding $\mathbf{x}$:
  \begin{enumerate}
  \item First of all, does the point $\mathbf{z} = \zero \in \Lambda_f$ typically induce a decoding error? Actually not, since we claim that
    \begin{equation*}
      \lim_{n \to \infty} \prob\{ \Vert\alpha \mathbf{y}\Vert^2 > \alpha n \sigma^2 (1+\varepsilon)^2\} = 1.
    \end{equation*}
    This and \eqref{eq:x_close_to_alphay} mean that, given any non-zero point $\mathbf{x}$ of the constellation,
    \begin{equation*}
    \lim_{n \to \infty} \prob\{ \Vert\alpha \mathbf{y} - \mathbf{x}\Vert^2 \leq \Vert\alpha \mathbf{y}\Vert^2\} = 1.
    \end{equation*}
    Thus, $\mathbf{x}$ is asymptotically closer to $\alpha \mathbf{y}$ than $\zero$ and the lattice decoder cannot give $\zero$ as an output. Now, let us prove the claim: the condition $\Vert\alpha \mathbf{y}\Vert^2 > \alpha n \sigma^2 (1+\varepsilon)^2$ is equivalent to
    \begin{equation*}
      \Vert\mathbf{y}\Vert^2 > \frac{ n \sigma^2 (1+\varepsilon)^2}{\alpha} = \frac{n \sigma^2 (P+\sigma^2)(1+\varepsilon)^2}{P} = \frac{n(P+\sigma^2)(1+\varepsilon)^2}{\snr}.
    \end{equation*}
    At the same time, Lemma \ref{lem:typical_norm}, Lemma \ref{lem:typical_noise} and Lemma \ref{lem:orthogonal_noise} imply that with probability tending to $1$ as $n$ tends to infinity, the event 
    \begin{equation}
      \label{eq:event_E1prime}
      \mathcal{E}_1' = \{ \Vert\mathbf{x}\Vert^2 \geq nP(1 - \varepsilon')^2 \} \cap \{ \Vert\mathbf{w}\Vert^2 \geq n\sigma^2(1-\varepsilon')^2 \} \cap \{ |\mathbf{x}\mathbf{w}^T| \leq f(n)\sigma \Vert\mathbf{x}\Vert\}
    \end{equation}
    occurs and
    \begin{align*}
      \Vert\mathbf{y}\Vert^2 & = \Vert\mathbf{x}\Vert^2 + \Vert\mathbf{w}\Vert^2 + 2\mathbf{x}\mathbf{w}^T \\
      & \geq nP(1 - \varepsilon')^2 + n\sigma^2(1-\varepsilon')^2 - 2f(n)\sigma \Vert\mathbf{x}\Vert \\
      & \geq n(P+\sigma^2)(1 - \varepsilon')^2 - 2f(n)\sigma\sqrt{nP}(1 + \varepsilon')\\
      & = n(P+\sigma^2)\left((1 - \varepsilon')^2 - \frac{2f(n)\sigma\sqrt{P}(1 + \varepsilon')}{\sqrt{n}(P+\sigma^2)}\right)\\
      & \sim n(P+\sigma^2)(1 - \varepsilon')^2,
    \end{align*}
    where the last asymptotic equality can be derived with the same observations pointed out for \eqref{eq:intermediate_computation}. Thus, it is sufficient to show that
    \begin{equation*}
      \frac{n(P+\sigma^2)(1+\varepsilon)^2}{\snr} < n(P+\sigma^2)(1 - \varepsilon')^2,
    \end{equation*}
    which is true because $\snr$ is bigger than $1$ by hypothesis and $(1+\varepsilon)^2/(1 - \varepsilon')^2$ can be taken to be as close to $1$ as wanted, then \emph{a fortiori} smaller than $\snr$.
  \item The previous argument states that $\zero$ asymptotically almost never causes a decoding error. We would like to treat now the case of all the other points $\mathbf{z} \in p\Zn$. Notice that one of these points can be the lattice decoder output only if it is closer to $\alpha \mathbf{y}$ than $\zero$ itself. That is, dangerous points $\mathbf{z} \in p\Zn\smallsetminus\{\zero\}$ are such that $\Vert\alpha \mathbf{y} - \mathbf{z}\Vert \leq \Vert\alpha \mathbf{y}\Vert$. This implies that there exists $i \in \{1,2,\ldots,n\}$ such that $|\alpha y_i - z_i| \leq |\alpha y_i|$ and $z_i \neq 0$; moreover, the fact that  $\mathbf{z} \in p\Zn$ means that $|z_i| \geq p$. Consequently, $|\alpha y_i|$ has to be bigger than $p/2$ and, \emph{a fortiori}, $|y_i| > p/2$, too, because $\alpha < 1$. Now, $y_i = x_i + w_i$ and a necessary condition for having $|x_i + w_i| > p/2$ is that at least one between $|x_i|$ and $|w_i|$ is bigger than $p/4$. The probability that $|w_i|>p/4$ can be shown to decrease to $0$ when $n$ tends to infinity with the same argument used to treat \eqref{eq:sum_of_prob}. Hence, asymptotically speaking, there can be a decoding error due to points $\mathbf{z} \in p\Zn\smallsetminus\{\zero\}$ only for the $\mathbf{x}$ such that $|x_i| > p/4$ for some $i$. Let us show that also this case does not represent a real problem: recall that $H$ is the random parity-check matrix of the shaping lattice $\Lambda$ and consider the sum
    \begin{align}
      \label{eq:partial_sum_particular_case}
      \sum_{\substack{\mathbf{x} \in (\Zn \cap \B_{\eff})\smallsetminus p\Zn\\|x_i| > p/4,\ \exists i}} & \prob \{ X_{\mathbf{s}} = \mathbf{x}, \exists \mathbf{z} \in \Lambda_f \cap \B \text{ and } \mathbf{z} \neq \mathbf{x}\} \\
      & \leq \sum_{\substack{\mathbf{x} \in (\Zn \cap \B_{\eff})\smallsetminus p\Zn\\|x_i| > p/4,\ \exists i}} \prob \{ X_{\mathbf{s}} = \mathbf{x}\} \nonumber \\
      & \leq \sum_{\substack{\mathbf{x} \in (\Zn \cap \B_{\eff})\smallsetminus p\Zn\\|x_i| > p/4,\ \exists i}} \prob \{ H\mathbf{x}^T \equiv \mathbf{s} \bmod p\} \nonumber \\
      \label{eq:few_points}
      & = |\{\mathbf{x} \in (\Zn \cap \B_{\eff})\smallsetminus p\Zn : |x_i| > p/4, \exists i\}| \left(\frac{1}{p}\right)^{n(1-R)}.
    \end{align}
    Now, if $\rho_{\eff}^2(1+1/n^\omega)^2 < p^2/16$ (i.e., asymptotically, if $\lambda \geq (2R)^{-1}$), the previous quantity is trivially equal to $0$. Then, we suppose $\lambda < (2R)^{-1}$ and go on with the computation: if we call $r = \sqrt{\rho_{\eff}^2(1+1/n^\omega)^2-p^2/16}$, we have
    \begin{align*}
      \eqref{eq:few_points} & \leq n \left|\Z^{n-1} \cap B_{\zero,n-1}(r)\right| \left(\frac{1}{p}\right)^{n(1-R)}\\
      & \leq n \vol \left(B_{\zero,n-1}(r)\right) \left(1+\frac{\sqrt{n-1}}{2r}\right)^{n-1} \left(\frac{1}{p}\right)^{n(1-R)}\\
      & = n \vol \left(B_{\zero,n-1}\left(\rho_{\eff}\left(1+\frac{1}{n^{\omega}}\right)\right)\right) \cdot \\
      & \ \ \ \ \ \ \ \ \cdot \frac{r^{n-1}}{\rho_{\eff}^{n-1}\left(1+1/n^\omega\right)^{n-1}} \left(1+\frac{\sqrt{n-1}}{2r}\right)^{n-1} \left(\frac{1}{p}\right)^{n(1-R)}.
    \end{align*}
    Let us call
    \begin{align*}
      \Psi & = n \vol \left(B_{\zero,n-1}\left(\rho_{\eff}\left(1+\frac{1}{n^{\omega}}\right)\right)\right) \left(\frac{1}{p}\right)^{n(1-R)},\\
      \Phi & = \frac{r^{n-1}}{\rho_{\eff}^{n-1}\left(1+1/n^\omega\right)^{n-1}}, \\
      \Theta & = \left(1+\frac{\sqrt{n-1}}{2r}\right)^{n-1}.
    \end{align*}
    Some simple computations show that the product $\Psi \Theta$ is very similar to \eqref{eq:standard_computation_2} and \eqref{eq:effective_radius_volume} (up to a slight modification of a sign in $\rho$) and it goes to infinity as $O(\exp(n^{1-\omega}))$. On the other hand, $\Phi$ can be shown to be $O(\exp(-Dn^{2\lambda R}))$ for some constant $D$. Hence the whole product tends to $0$ as $n$ grows to infinity when $2\lambda R > 1-\omega$, that is $\omega > 1-2\lambda R$. The hypotheses $R > 1/2$ and $\lambda > (1+R)^{-1} > 1/2$ guarantee that we can take $\omega$ to satisfy the previous condition without contradicting \eqref{eq:omega_lambda}. Thus, we can state that \eqref{eq:partial_sum_particular_case} tends to $0$ when $n$ goes to infinity.
  \item We separately treat also the case of the $\mathbf{z}$ such that $\mathbf{z} \equiv 2\mathbf{x} \bmod p$. Does this kind of $\mathbf{z}$ induce any decoding error? For what $\mathbf{x}$? The strategy to answer these questions is the same that we have adopted in the previous two points. Let us start by considering $\mathbf{z} = 2\mathbf{x}$. There is no decoding error due to $\mathbf{z}$ if $\Vert\alpha \mathbf{y} - 2\mathbf{x}\Vert^2 > \Vert\alpha \mathbf{y} - \mathbf{x}\Vert^2$. Recalling that $\mathbf{y}=\mathbf{x} + \mathbf{w}$, this is equivalent to $3\Vert\mathbf{x}\Vert^2-2\alpha\mathbf{x} \mathbf{y}^T = (3 - 2\alpha)\Vert\mathbf{x}\Vert^2 - 2\alpha\mathbf{x} \mathbf{w}^T > 0$. Since $\alpha < 1$, in order to show that $\mathbf{z} = 2\mathbf{x}$ does not induce any error, it is thus sufficient to show that $\Vert\mathbf{x}\Vert^2 - 2 |\mathbf{x} \mathbf{w}^T| > 0$ with probability tending to $1$ when $n$ tends to infinity. If \eqref{eq:event_E1} and \eqref{eq:event_E1prime} occur,
    \begin{align*}
      \Vert\mathbf{x}\Vert^2 - 2 |\mathbf{x} \mathbf{w}^T| & \geq nP(1-\varepsilon')^2 - 2f(n)\sigma\Vert\mathbf{x}\Vert \\
      & \geq nP(1-\varepsilon')^2 - 2f(n)\sigma\sqrt{nP}(1+\varepsilon') \\
      & = nP(1-\varepsilon')^2 \left(1 - \frac{2f(n)\sigma(1+\varepsilon')}{\sqrt{nP}(1-\varepsilon')^2}\right) \\
      & > nP(1-\varepsilon')^2 \left(1 - \frac{2f(n)(1+\varepsilon')}{\sqrt{n}(1-\varepsilon')^2}\right)
    \end{align*}
    where the last inequality is due to the fact that $\snr = P/\sigma^2 > 1$; taking $f(n)=o(\sqrt{n})$, the lower bound is clearly asymptotically positive and we are done.

    We have proved that $\mathbf{z} = 2\mathbf{x}$ typically does not induce any error. Can we say the same for all the other $\mathbf{z} \equiv 2\mathbf{x} \bmod p$? The only case that could lead to bad decoding is the one of $\mathbf{z} \equiv 2\mathbf{x} \bmod p$ such that $\Vert\alpha \mathbf{y} - \mathbf{z}\Vert^2 < \Vert\alpha \mathbf{y} - 2\mathbf{x}\Vert^2$ (otherwise, the previous computation concerning $\mathbf{z}=2\mathbf{x}$ is sufficient). Let $\mathbf{z} = 2\mathbf{x} + p\mathbf{k}$ for some $\mathbf{k} \in \Zn \smallsetminus \{\zero\}$. Then $\mathbf{z}$ can be closer to $\alpha \mathbf{y}$ than $2\mathbf{x}$ only if there exists $i$ such that
    \begin{equation*}
      |\alpha y_i - 2x_i - pk_i| < |\alpha y_i - 2x_i|,
    \end{equation*}
    for some $k_i \geq 1$. This is possible only if $|\alpha w_i -(2-\alpha)x_i|=|\alpha y_i - 2x_i|>p/2$, which in turn implies that at least one between $|\alpha w_i|$ and $(2-\alpha)|x_i|$ has to be greater than $p/4$. Now, one can use basically the same argument as the one applied for the $\mathbf{z}$ of $p\Zn$ above, and conclude that $\prob\{|\alpha w_i|>p/4\}$ tends to $0$, as does this sum: 
    \begin{equation*}
      \sum_{\substack{\mathbf{x} \in (\Zn \cap \B_{\eff})\smallsetminus p\Zn\\(2-\alpha)|x_i| > p/4,\ \exists i}} \prob \{ X_{\mathbf{s}} = \mathbf{x}, \exists \mathbf{z} \in \Lambda_f \cap \B \text{ and } \mathbf{z} \neq \mathbf{x}\}.
    \end{equation*}
  \item Finally, what about the $\mathbf{z}$ such that $\mathbf{z} \equiv \mathbf{x} \bmod p$? Even if a $\mathbf{z}$ of this kind is closer than $\mathbf{x}$ to $\alpha \mathbf{y}$, its syndrome $H\mathbf{z}^T$ is equal to $\mathbf{s}$, the syndrome of $\mathbf{x}$, and this does not give a decoding error. For this reason, we can actually omit these $\mathbf{z}$ from the total sum and not consider them.
  \end{enumerate}

  Concretely, with the previous four points we have shown that 
  \begin{equation*}
    \lim_{n \to \infty}\sum_{\mathbf{x} \in (\Zn \cap \B_{\eff})\smallsetminus p\Zn} \prob \{ X_{\mathbf{s}} = \mathbf{x}, \exists \mathbf{z}\in \Lambda_f \cap \B \text{ inducing an error}, \mathbf{z} \equiv \mu \mathbf{x},\ \exists \mu \in \{0,1,2\}\} = 0.
  \end{equation*}
  Hence, we can restrict the sum in \eqref{eq:sum_over_all_x} to the set
  \begin{equation}
    \label{eq:S}
    S = \{\mathbf{x} \in (\B_{\eff}\cap \Zn) \smallsetminus p\Zn : \mathbf{z} \equiv \mu\mathbf{x} \bmod p \text{ produces no error}, \forall \mu \in \{0,1,2\}\}. 
  \end{equation}

Recall that $H = [(H')^T\ |\ (H_f)^T]^T$ is the random parity-check matrix of $\Lambda$, whereas $H_f$ is the random submatrix of $H$ that defines $\Lambda_f$. Hence, if $\mathbf{s} = (\mathbf{m}\ |\ \zero) \in \Fp^{n(R_f-R)} \times \Fp^{n(1-R_f)}$, then the sum that we need to estimate is less than
\begin{align*}
    \sum_{\mathbf{x} \in S} & \sum_{\substack{\mathbf{z} \in \Zn \\ \mathbf{z} \not \equiv \mu \mathbf{x},\ \mu=0,1,2}} \prob \{X_{\mathbf{s}} = \mathbf{x},\mathbf{z} \in (\Lambda_f \cap \B)\}  \\
    & \leq \sum_{\mathbf{x} \in S} \sum_{\substack{\mathbf{z} \in \Zn \\ \mathbf{z} \not \equiv \mu \mathbf{x},\ \mu=0,1,2}} \prob \{H\mathbf{x}^T \equiv \mathbf{s}^T \bmod p, H_f\mathbf{z}^T \equiv \zero^T \bmod p, \mathbf{z} \in \B\}  \\
    & \stackrel{(a)}{=} \sum_{\mathbf{x} \in S} \prob \{H'\mathbf{x}^T \equiv \mathbf{m}^T \bmod p\} \\
    & \ \ \ \ \ \ \ \ \ \ \ \ \sum_{\substack{\mathbf{z} \in \Zn \\ \mathbf{z} \not \equiv \mu \mathbf{x},\ \mu=0,1,2}} \prob \{H_f\mathbf{x}^T \equiv \zero^T \bmod p, H_f\mathbf{z}^T \equiv \zero^T \bmod p, \mathbf{z} \in \B\}  \\
    & = \sum_{\mathbf{x} \in S} \left( \frac{1}{p} \right)^{n(R_f-R)} \\
    & \ \ \ \ \ \ \ \ \ \ \ \ \sum_{\substack{\mathbf{z} \in \Zn \\ \mathbf{z} \not \equiv \mu \mathbf{x},\ \mu=0,1,2}} \prob \{H_f\mathbf{x}^T \equiv \zero^T \bmod p, H_f\mathbf{z}^T \equiv \zero^T \bmod p, \mathbf{z} \in \B\}  \\
    & \stackrel{(b)}{=} \sum_{\mathbf{x} \in S} \left( \frac{1}{p} \right)^{n(R_f-R)} \\
    & \ \ \ \ \ \ \ \ \ \ \ \ \sum_{\substack{\mathbf{z} \in \Zn \\ \mathbf{z} \not \equiv \mu \mathbf{x},\ \mu=0,1,2}} \prob \{H_f\mathbf{x}^T \equiv \zero^T \bmod p, H_f\mathbf{z}^T \equiv \zero^T \bmod p\} \prob\{ \mathbf{z} \in \B\};  
  \end{align*}
  $(a)$ holds true because the random entries of $H$ are all i.i.d.\ and the events converning $H'$ and $H_f$ are independent; $(b)$ is justified by the fact that the events related to the random choice of $H_f$ and the event related to the random noise are independent.

  Recall that $\B$ is a random object, that depends on $\mathbf{x}$ and $\mathbf{w}$. We have already observed that $\mathbf{x}$ lies inside it with very high probability. Given this, $\mathbf{z}$ cannot be simultaneously inside the ball and further than twice the radius of $\B$ from $\mathbf{x}$. For this reason we restrict our sum to the $\mathbf{z}$ inside the sphere $\B' = B_{\mathbf{x},n}(2\rho_{\dec})$. We will show that
  \begin{equation}
    \label{eq:same_limit}
    \begin{aligned}
      \lim_{n \to \infty} \sum_{\mathbf{x} \in S} & \left( \frac{1}{p} \right)^{n(R_f-R)} \\
      & \sum_{\substack{\mathbf{z} \in \B'\cap \Zn \\ \mathbf{z} \not \equiv \mu \mathbf{x},\ \mu=0,1,2}}\prob \{H_f\mathbf{x}^T \equiv \zero^T \bmod p, H_f\mathbf{z}^T \equiv \zero^T \bmod p\} \prob\{ \mathbf{z} \in \B\} = 0.
    \end{aligned}
  \end{equation}
  
  There are now two possible situations. If $\mathbf{z} \not \equiv \mu \mathbf{x} \bmod p$ for every $\mu \in \Fp$, %the events $\{H_f\mathbf{x}^T \equiv \zero^T \bmod p\}$ and $\{H_f\mathbf{z}^T \equiv \zero^T \bmod p\}$ are independent, 
  then
  \begin{align*}
    & \prob \{H_f\mathbf{x}^T \equiv \zero^T \bmod p, H_f\mathbf{z}^T \equiv \zero^T \bmod p\} \\
    & = \prob \{H_f\mathbf{x}^T \equiv \zero^T \bmod p\} \prob\{H_f\mathbf{z}^T \equiv \zero^T \bmod p\} \\
    & = \left( \frac{1}{p} \right)^{2n(1-R_f)}.
  \end{align*}
  If instead %On the other hand, it is easy to see that the two events are dependent if and only if 
  $\mathbf{z} \equiv \mu \mathbf{x} \bmod p$ for some $\mu \in \Fp$%. In that case
  , the fact that $\mathbf{x}$ belongs to $\Lambda_f$ automatically implies that $\mathbf{z}$ belongs to $\Lambda_f$, too. Hence,
  \begin{align*}
    & \prob \{H_f\mathbf{x}^T \equiv \zero^T \bmod p, H_f\mathbf{z}^T \equiv \zero^T \bmod p\} \\
    & = \prob \{H_f\mathbf{x}^T \equiv \zero^T \bmod p\} \\
    & = \left( \frac{1}{p} \right)^{n(1-R_f)}.
  \end{align*}
  Now, let $S'$ be the subset of $S$ %$(\B_{\eff}\cap \Zn) \smallsetminus p\Zn$ 
  of all the points $\mathbf{x}$ for which there exists at least one $\mathbf{z} \in \B'$ such that $\mathbf{z} \equiv \mu \mathbf{x} \bmod p$ (for some $\mu \neq 0,1,2$ by definition of $S$). %Lemma \ref{lem:good_points} states that, when $\mu \neq 1$,
  %\begin{equation}
  %  \label{eq:cardinality_of_S}
  %  |S| \lesssim C^n n^{n \lambda( \eta (1-R_f) + (1-\eta)(1-R))} + 2 \sqrt{2} \eta n^{\frac{3}{2}} 2^{(2\eta + \frac{1}{2} - \gamma)n} |\B_{\eff} \cap \Zn|,
  %\end{equation}
  %for any two fixed constants $0 < \eta < \gamma < 1$ and for some constant $C$.
  Summarizing what we have elaborated till now, we are left to show that
  \begin{equation}
    \label{eq:first_addend}
    \lim_{n\to \infty} \sum_{\mathbf{x} \in S \smallsetminus S'} \left( \frac{1}{p} \right)^{n(R_f-R)} \sum_{\substack{\mathbf{z} \in \B'\cap \Zn \\ \mathbf{z} \not \equiv \mu \mathbf{x} }} \left( \frac{1}{p} \right)^{2n(1-R_f)} \prob\{ \mathbf{z} \in \B \} = 0
  \end{equation}
  and 
  \begin{equation}
    \label{eq:second_addend}
    \lim_{n\to \infty} \sum_{\mathbf{x} \in S'} \left( \frac{1}{p} \right)^{n(R_f-R)} \sum_{\substack{\mathbf{z} \in \B'\cap \Zn \\ \mathbf{z} \equiv \mu \mathbf{x},\ \mu\neq 0,1,2}} \left( \frac{1}{p} \right)^{n(1-R_f)} \prob\{ \mathbf{z} \in \B \} = 0.
  \end{equation}

  {\bf Proof of \eqref{eq:first_addend}.} Recall that $\B = B_{\alpha \mathbf{y},n}(\rho_{\dec})$ and $\mathbf{y} = \mathbf{x} + \mathbf{w}$; therefore,
  \begin{equation}
    \label{eq:sum_balls}
    \sum_{\substack{\mathbf{z} \in \B'\cap \Zn \\ \mathbf{z} \not \equiv \mu \mathbf{x} }} \prob\{ \mathbf{z} \in \B \} = \sum_{\substack{\mathbf{z} \in \B'\cap \Zn \\ \mathbf{z} \not \equiv \mu \mathbf{x} }} \prob\{ \alpha\mathbf{y} \in B_{\mathbf{z},n}(\rho_{\dec}) \} = \sum_{\substack{\mathbf{z} \in \B'\cap \Zn \\ \mathbf{z} \not \equiv \mu \mathbf{x} }} \prob \left \{ \alpha \mathbf{w} \in B_{\mathbf{z} - \alpha \mathbf{x},n} \left(\rho_{\dec} \right) \right \}.
  \end{equation}
  If we call
  \begin{equation*}
    \mathbf{z}' = \arg \max_{\substack{\mathbf{z} \in \B'\cap \Zn \\ \mathbf{z} \not \equiv \mu \mathbf{x} }} \prob \left \{ \alpha \mathbf{w} \in B_{\mathbf{z} - \alpha \mathbf{x},n} \left(\rho_{\dec} \right) \right \} \text{\ \ \ \ \ and\ \ \ \ \ } B = B_{\mathbf{z}' - \alpha \mathbf{x},n} \left(\rho_{\dec} \right)
  \end{equation*}
  and if $p(w)$ is the (Gaussian) probability density function of $\alpha \mathbf{w}$, then the previous sum is bounded as follows:
  \begin{align}
    \eqref{eq:sum_balls} & \leq \sum_{\substack{\mathbf{z} \in \B'\cap \Zn \\ \mathbf{z} \not \equiv \mu \mathbf{x} }} \prob \left \{ \alpha \mathbf{w} \in B \right \} = \sum_{\substack{\mathbf{z} \in \B'\cap \Zn \\ \mathbf{z} \not \equiv \mu \mathbf{x} }} \int_{B}p(w)\mathrm{d}w = \int_{B} \sum_{\substack{\mathbf{z} \in \B'\cap \Zn \\ \mathbf{z} \not \equiv \mu \mathbf{x} }} p(w)\mathrm{d} w \nonumber \\
    \label{eq:inequality_balls}
    & = \int_{B} \sum_{\substack{\mathbf{z} \in B \cap \B'\cap \Zn \\ \mathbf{z} \not \equiv \mu \mathbf{x} }} p(w)\mathrm{d} w \leq |B \cap \Zn| \int_{B} p(w)\mathrm{d} w \\
    & \leq \vol(B_{\zero,n}(\rho_{\dec}+\sqrt{n}/2)), \nonumber
  \end{align}
  where, the latter inequality comes from Lemma \ref{lem:integer_sphere_points}.

  Going back to \eqref{eq:first_addend} and using what we have just deduced, we have
  \begin{align}
    & \sum_{\mathbf{x} \in S \smallsetminus S'} \left( \frac{1}{p} \right)^{n(R_f-R)} \sum_{\substack{\mathbf{z} \in \B'\cap \Zn \\ \mathbf{z} \not \equiv \mu \mathbf{x} }} \left( \frac{1}{p} \right)^{2n(1-R_f)} \prob\{ \mathbf{z} \in \B\} \nonumber \\
    \label{eq:appreciable_situation}
    & \leq \left( |\Zn \cap \B_{\eff}| \left( \frac{1}{p} \right)^{n(1-R)} \right) \left( \vol(B_{\zero,n}(\rho_{\dec}+\sqrt{n}/2)) \left( \frac{1}{p} \right)^{n(1-R_f)} \right).
  \end{align}
  The left factor is very similar to \eqref{eq:standard_volume_power_pf_p_ratio} (it differs only by a modification of a sign in the radius) and can be shown to go to infinity subexponentially in $n$. On the other hand, the right term exponentially decreases to $0$, just like \eqref{eq:same_term_bis} does. As a result, the dominating term is the latter and the whole product vanishes when $n$ tends to infinity.

  {\bf Proof of \eqref{eq:second_addend}.}
  %Observe that the distance between the $\mathbf{x}$ and the $\mathbf{z}$ that we consider is bounded by twice the radius of $\B$. Hence,
  We have
  \begin{align}
    \label{eq:only_multiple_z}
    & \sum_{\mathbf{x} \in S'} \left( \frac{1}{p} \right)^{n(R_f-R)} \sum_{\substack{\mathbf{z} \in \B'\cap \Zn \\ \mathbf{z} \equiv \mu \mathbf{x},\ \mu \neq 0,1,2}} \left( \frac{1}{p} \right)^{n(1-R_f)} \prob\{ \mathbf{z} \in \B \}  \\
    & \leq \sum_{\mathbf{x} \in S'} \left( \frac{1}{p} \right)^{n(R_f-R)} \sum_{\substack{\mathbf{z} \in \B'\cap \Zn \\ \mathbf{z} \equiv \mu \mathbf{x},\ \mu \neq 0,1,2}} \left( \frac{1}{p} \right)^{n(1-R_f)} \nonumber \\
    \label{eq:intermediate}
    & \leq \sum_{\mathbf{x} \in S'}\left( \frac{1}{p} \right)^{n(1-R)}|\{\mathbf{z} \in \B' : \mathbf{z} \equiv \mu \mathbf{x} \bmod p, \exists \mu \in \Fp \smallsetminus \{0,1,2\}\}|.
  \end{align}
  Lemma \ref{lem:equivalent_points} provides the following upper bound of every fixed $\mu$:
  \begin{equation*}
    |\{\mathbf{z} \in \B' : \mathbf{z} \equiv \mu \mathbf{x} \bmod p\}| \leq 1 + \frac{16\rho_{\dec}^2}{p^2}\left( \frac{32n\rho_{\dec}^2}{p^2} \right)^{16\rho_{\dec}^2/p^2},
  \end{equation*}
  hence
  \begin{align*}
    |\{\mathbf{z} \in \B' : \mathbf{z} \equiv \mu \mathbf{x} \bmod p, \exists \mu \in \Fp \smallsetminus \{0,1,2\}\}| & \leq p + \frac{16\rho_{\dec}^2}{p}\left( \frac{32n\rho_{\dec}^2}{p^2} \right)^{16\rho_{\dec}^2/p^2} \\
    %& = O\left(n^{(1-2\lambda R_f)n^{(1-2\lambda R_f)}}\right).
    & = O\left(n^{En^{(1-2\lambda R_f)}}\right),
  \end{align*}
  for some constant $E$. Let us call $t(n)$ this last term, which does not grow more than subexponentially fast in $n$. Going on from \eqref{eq:intermediate}, we get
  \begin{equation}
    \label{eq:cardinality_of_S'}
    \sum_{\mathbf{x} \in S'}\left( \frac{1}{p} \right)^{n(1-R)}|\{\mathbf{z} \in \B' : \mathbf{z} \equiv \mu \mathbf{x} \bmod p, \exists \mu \in \Fp \smallsetminus \{0,1,2\}\}| \leq \frac{|S'|t(n)}{p^{n(1-R)}},
    %\lesssim & \frac{\left(C^n n^{n \lambda( \eta (1-R_f) + (1-\eta)(1-R))} \right)pT_2(n)}{p^{n(1-R)}} \\
    %& + \frac{|\B_{\eff} \cap \Zn|}{p^{n(1-R)}} \left(2 \sqrt{2} \eta n^{\frac{3}{2}} 2^{(2\eta + \frac{1}{2} - \gamma)n} \right) pT_2(n) 
  \end{equation}
  which vanishes asymptotically in $n$ because of Lemma \ref{lem:good_points}, since by definition $|S'|$ is equal to $N$ defined in \eqref{eq:big_N}.

  Putting together the estimations of \eqref{eq:P1} and \eqref{eq:P2}, we can derive that
  \begin{equation*}
    \lim_{n\to \infty} P_e(\mathbf{s}) = 0,
  \end{equation*}
  \emph{quod erat demonstrandum}.
\end{IEEEproof}

\section{Interlude: expansion properties of bipartite graphs}
\label{sec:graph_theoretical_tools}

We have achieved our main result on random Construction-A Voronoi constellations. Before moving to the low-density construction, we need to treat in this self-contained section a graph-theoretical problem that will have relevant applications in the sequel. Let $\mathcal{G}=(V_L,V_R,E)$ be an undirected bipartite graph; $V_L \cup V_R$ is its set of (left and right) vertices and $E$ its set of edges. Let $|V_L|=n$ and $|V_R|=fn$, for some constant fraction $f\in\mathbb{Q} \smallsetminus \{0\}$ (that can be bigger than $1$). Parallel edges are accepted: there might be two or more edges connecting the same two vertices. 

\begin{definition}[Neighborhood]
  If $S$ is a subset of vertices of a graph $\mathcal{G}$, its \emph{neighborhood} $N(S)$ is defined as the set of vertices of the graph that are incident to a vertex of $S$.
\end{definition}

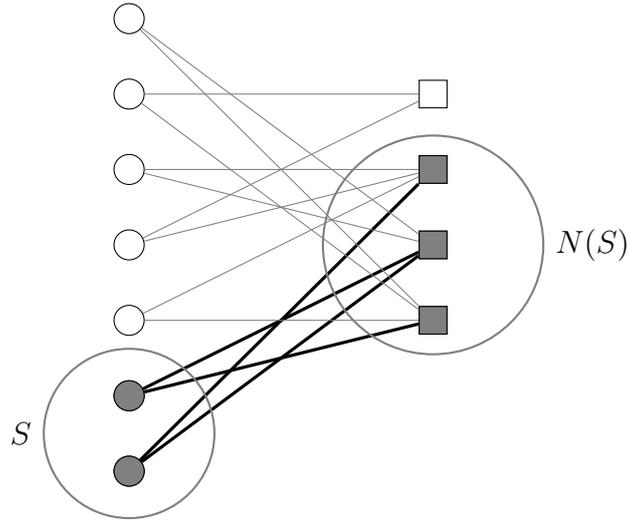
\begin{figure}[!t]
  \begin{center}
    \begin{tikzpicture}
      [every fit/.style={circle,draw,inner sep=-2pt,text width=2cm}]
      \foreach \xitem in {1,2,...,7}
               { \node[style={circle,draw}] at (0,\xitem) (a\xitem) {};   
               }%
               \foreach \xitem in {3,4,...,6}
                        {%
                          \node[style={draw},scale=1.3] at (4,\xitem) (b\xitem) {};   
                        }%
                        
                        \node[style={fill=gray,circle,draw}] at (0,2) (a2) {};
                        \node[style={fill=gray,draw},scale=1.3] at (4,3) (b3) {};
                        \node[style={fill=gray,draw},scale=1.3] at (4,4) (b4) {};
                        \node[style={fill=gray,circle,draw}] at (0,1) (a1) {};
                        \node[style={fill=gray,draw},scale=1.3] at (4,5) (b5) {};
                        
                        % connections
                        \draw(a1)[very thick] -- (b4);
                        \draw(a1)[very thick] -- (b5);
                        \draw(a2)[very thick] -- (b3);
                        \draw(a2)[very thick] -- (b4);
                        \draw(a3)[gray] -- (b3);
                        \draw(a3)[gray] -- (b5);
                        \draw(a4)[gray] -- (b5);
                        \draw(a4)[gray] -- (b6);
                        \draw(a5)[gray] -- (b4);
                        \draw(a5)[gray] -- (b5);
                        \draw(a6)[gray] -- (b3);
                        \draw[gray](a6) -- (b6);
                        \draw(a7)[gray] -- (b3);
                        \draw(a7)[gray] -- (b4);
                        
                        \node[thick,gray,fit=(a1) (a2)](cerchio1) {};
                        \node[thick,gray,fit=(b3) (b4) (b5)](cerchio2) {};
                        \node[left] at (cerchio1.west) {$S$};
                        \node[right] at (cerchio2.east) {$N(S)$};
    \end{tikzpicture}  
  \end{center}
  \caption[A bipartite graph with an example of neighborhood of a subset of vertices.]{\label{fig:bipartite_graph} A bipartite graph with an example of neighborhood of a subset of vertices. $V_L$ is the set of round vertices, $V_R$ is the set of square vertices. Observe that $S \subseteq N(N(S))$ and the inclusion is generally strict.
  }
\end{figure}

In a bipartite graph $\mathcal{G}=(V_L,V_R,E)$, it is clear that $N(S) \subseteq V_R$ for every $S \subseteq V_L$ and vice versa $N(T) \subseteq V_L$ for every $T \subseteq V_R$. See Fig.~\ref{fig:bipartite_graph} for a simple example.

From now on, we will consider only graphs with the following variation of the \emph{biregularity} property:
the number of edges incident to any single vertex of $V_R$ (resp. $V_L$) has constant
cardinality $\Delta$ (resp. $f\Delta$). Consequently, the neighborhood of any single
vertex of $V_R$ (resp. $V_L$) has cardinality at most $\Delta$ (resp. $f\Delta$). If the graph
has no parallel edges, these cardinalities are exactly $\Delta$ and $f\Delta$ and
the graph is biregular, according to the standard definition. Denote by
$\mathcal{F}(n, f, \Delta)$ the family of graphs just defined.

We are interested in some particular \emph{expansion} properties of this kind of graph. In other words, we are interested in studying what graphs are such that any ``small'' set of vertices has a ``big enough'' neighborhood. Thus we give the following definition:
\begin{definition}[$D$-good graphs]
  \label{def:good_graphs}
  Let $D > 0$ be a constant. 
  We say that a bipartite graph of $\mathcal{F}(n, f, \Delta)$ is \emph{$D$-good from left to right} if 
  \begin{equation}
    \label{eq:condition_1}
    \forall S \subseteq V_L\ \text{such that } |S| \leq \frac{n}{D+1},\ \text{then } |N(S)| \geq fD|S|. 
  \end{equation}
  Analogously, it is \emph{$D$-good from right to left} if
  \begin{equation}
    \label{eq:condition_2}
    \forall T \subseteq V_R\ \text{such that } |T| \leq \frac{fn}{D+1},\ \text{then } |N(T)| \geq \frac{D|T|}{f}. 
  \end{equation}
  We say that a graph of $\mathcal{F}(n, f, \Delta)$ is \emph{$D$-good} if it is both $D$-good from left to right and from right to left.
\end{definition}
Important remark: notice that the two conditions above imply that every subset of nodes at least as big as a fraction of $1/(D+1)$ of the total number of nodes on its side of the graph, has a neighborhood at least as big as a fraction of $D/(D+1)$ of the number of nodes on the other side. 

\begin{lemma}
  \label{lem:good_graphs}
  Let $\mathcal{G}$ be a graph in $\mathcal{F}(n, f, \Delta)$, chosen uniformly at random in the family. If $D\geq 1$ and
  \begin{equation}
    \label{eq:Delta_expansion}
    \Delta > \max \left \{ \left(1+\frac{1}{f}\right) \left(1 - \frac{Dh\left(\frac{1}{D}\right)}{(D+1)h\left(\frac{1}{D+1}\right)}\right)^{-1}, D^2+\frac{1}{f}\right\},
  \end{equation}
  then
  \begin{equation*}
    \lim_{n \to \infty}\prob\{\text{\emph{$\mathcal{G}$ is not} $D$-\emph{good from left to right}}\} = 0.
  \end{equation*}
\end{lemma}
The proof of the previous lemma can be found in Appendix \ref{sec:proof_good_graphs} and uses the same main ideas that Bassalygo applies in \cite{Bassalygo1981}. Nevertheless, our statement is slightly different and some elements of the proof are modified with respect to Bassalygo's one. The reader may also be interested in comparing this lemma with Theorem 8.7 of \cite[p. 431]{Richardson2008} and reading therein about the construction of \emph{expander codes}. 
%because we add a condition that Bassalygo claimed to be trivially satisfied, but which is actually not (in the proof of its lemma, with his notation take for example $\alpha=1/3, \beta=16, p=8, q=1$ and $\psi$ turns out to be negative).

\begin{corollary}
  \label{cor:good_graphs}
  Let $\mathcal{G}$ be a graph in $\mathcal{F}(n, f, \Delta)$, chosen uniformly at random in the family. If $D\geq 1$ and
  \begin{equation*}
    \Delta > \max \left \{ \left(1+\frac{1}{f}\right) \left(1 - \frac{Dh\left(\frac{1}{D}\right)}{(D+1)h\left(\frac{1}{D+1}\right)}\right)^{-1}, D^2+\frac{1}{f}, \frac{D^2}{f} + 1\right\},
  \end{equation*}
  then
  \begin{equation*}
    \lim_{n \to \infty}\prob\{\text{\emph{$\mathcal{G}$ is not} $D$-\emph{good}}\} = 0.
  \end{equation*}
\end{corollary}
\begin{IEEEproof}
  Lemma \ref{lem:good_graphs} states that $\mathcal{G}$ is $D$-good from left to right (asymptotically, with probability tending to $1$). We only need to prove that it is also $D$-good from right to left. But this is simply the application of Lemma \ref{lem:good_graphs} to the family $\mathcal{F}(m,f',\Delta')$ with $m=fn$, $f'=f^{-1}$, and $\Delta' = f \Delta$, which represents $\mathcal{F}(n, f, \Delta)$ with the nodes and their degree distributions switched from left to right and vice versa.
\end{IEEEproof}

\section{Achieving capacity with LDA lattices}%------------------ SECTION - FINITE CONSTELLATIONS LDA ---------------------
\label{sec:capacity_for_LDA}

From now on, we will adapt the results of the previous sections to the family of \emph{LDA lattices}:
\begin{definition}[LDA lattice]
  \label{def:LDA}
  A lattice $\Lambda \subseteq \Rn$ is called a \emph{Low-Density Construction-A} (or briefly \emph {LDA}) lattice if it is built with Construction A from an LDPC code.
\end{definition}
We recall that \emph{Low-Density Parity-Check codes} are linear codes whose parity-check matrix is sparse, i.e., whose great majority of the entries is equal to zero \cite{Gallager1963}. 

\begin{figure}[!t]
  \centering
    \includegraphics[width=10cm, angle=-90]{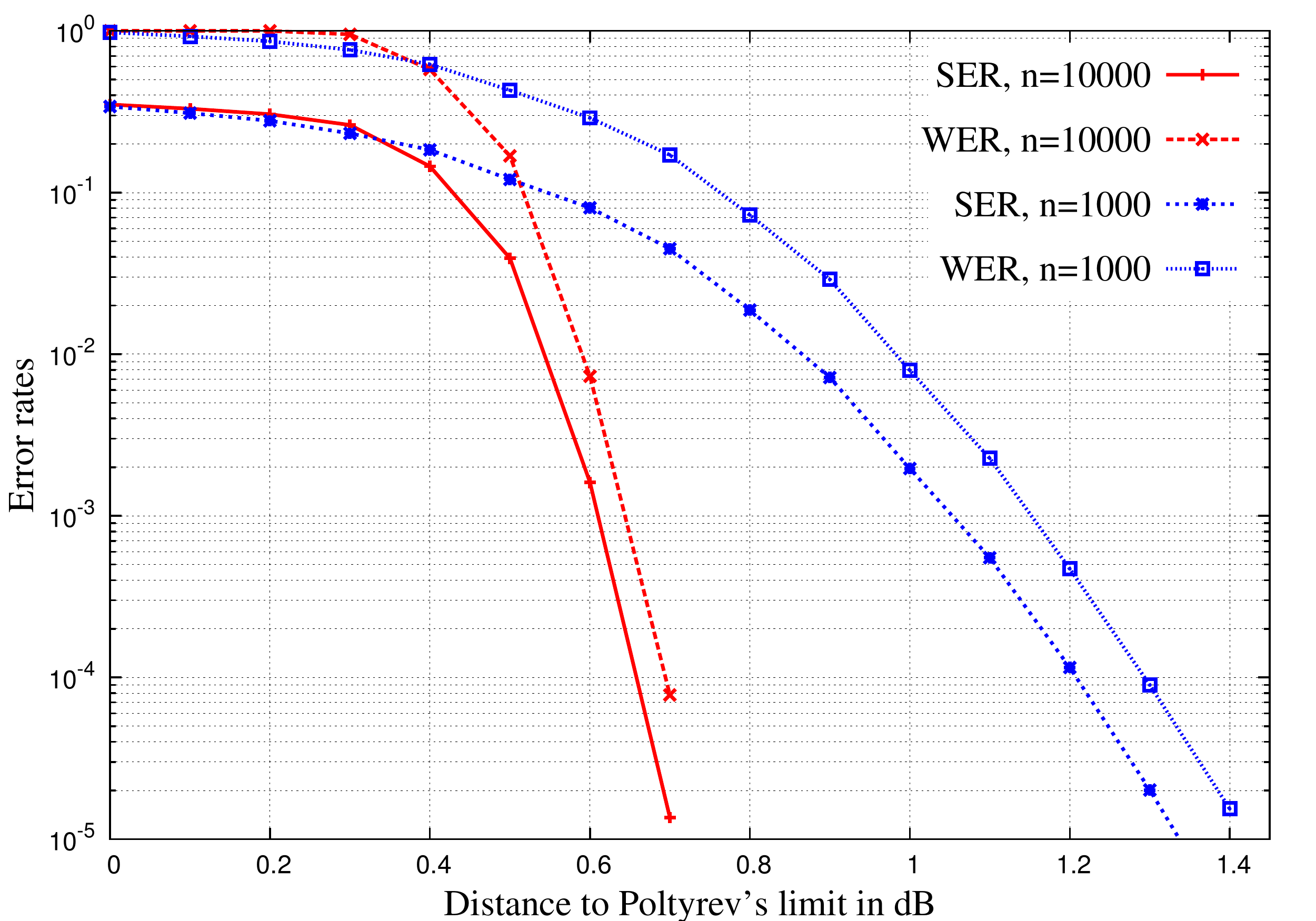}
    \caption{Symbol Error Rate (SER) and Word Error Rate (WER) of two infinite LDA constellations in dimension $n=1000$ and $n=10000$. The underlying LDPC codes are $(2,5)$-regular over $\F_{11}$.}
    \label{fig:LDA_perf}
\end{figure}

As we have anticipated in Section \ref{sec:intro}, infinite constellations of LDA lattices have already been shown to be very well-performing under iterative decoding \cite{diPietro2012}. An example of their performance, obtained with the decoding algorithm presented in \cite{diPietro2012}, can be found in Fig.~\ref{fig:LDA_perf}. 
The possibility of achieving Poltyrev limit with LDA lattices was shown in \cite{diPietro2013} and \cite{diPietro2013bis}. Our main goal here is to prove that they can achieve capacity of the AWGN channel under MMSE lattice decoding with similar hypotheses to the ones of Theorem \ref{thm:awgn_capacity}. The geometrical approach to demonstrate our result, as well as the encoding and decoding scheme, will be the very same that we have used for the more general Construction-A ensemble in the previous sections. Therefore, we will go once again along the same steps that have led to the proof of Theorem \ref{thm:awgn_capacity}. Nevertheless, some of these will need to be modified and adapted to the low-density structure of the parity-check matrices of the LDA lattices. In particular, we will extensively employ the \emph{expansion properties} of the random Tanner graphs \cite{Richardson2008} associated with them. We strongly emphasize this point: the $D$-goodness hypothesis (cf. Definition \ref{def:good_graphs}) of our Tanner graphs has to be considered one of the most novel tool of this entire work. It is used here in a clearer, more complete, and more elegant way than in the preliminary versions \cite{diPietro2013bis,diPietro2014}.

Finally, we point out that for this finite-constellation result the degree of the parity-check nodes of the Tanner graphs associated with our LDA lattices is constant. As said in Section \ref{sec:intro}, this is not a negligible detail, since the complexity of the iterative decoding algorithm is proportional to the parity-check degree and it is important to keep it bounded. This also contrasts sharply, and somewhat surprisingly, with the behavior of binary LDPC codes that need growing row weights to achieve capacity.

\section{The random LDA ensemble}%------------------------- SECTION - RANDOM LDA ENSEMBLE ---------------------
\label{sec:random_LDA_ensemble}

Once again, our lattice codes are given by Voronoi constellations of nested Construction-A lattices. However, this time we restrict our construction to LDA lattices. The random ensemble of \emph{fine lattices} (cf. Definition \ref{def:voronoi_constellation}) is built as follows:
\begin{enumerate}
\item Fix some constant $0 < R_f < 1$.
\item Consider a bipartite graph with $n$ left nodes (variable nodes) and $n(1-R_f)$ right nodes (check nodes).
\item The check nodes have degree $\Delta_P$, the variable nodes have degree $\Delta_P(1-R_f)$.
\item The edges are fixed once for all by taking a permutation of $\{1,2,\ldots,n(1-R_f)\Delta_P\}$ at random and connecting the left $n(1-R_f)\Delta_P$ sockets to the right $n(1-R_f)\Delta_P$ sockets according to the permutation.
\item Contingent parallel edges are unified.
\item Consider the binary parity-check matrix that has this graph as its Tanner graph.
\item Substitute each $1$ in the binary matrix with a random variable with uniform distribution over $\{0,1,\ldots,p-1\}$; notice that this is equivalent to assigning to every edge of the Tanner graph a random label chosen in $\{0,1,\ldots,p-1\}$.
\item Our random LDA fine lattice $\Lambda_f$ will be the lattice obtained with Construction A from the $p$-ary LDPC code defined by the previous random $p$-ary parity-check matrix and associated with the previous Tanner graph.
\end{enumerate}
We emphasize that the positions of the random entries in the parity-check matrix is deterministically fixed by the permutation. The randomness in the matrix is only given by its random non-zero entries.

Now, let us build the random ensemble of LDA \emph{shaping lattices}:
\begin{enumerate}
\item Fix some constant $R$ such that $0 < R < R_f$.
\item Use the same procedure as before to build a graph with $n(R_f-R)$ check nodes of degree $\Delta_P$ and $n$ variable nodes of degree $\Delta_P(R_f-R)$.
\item \label{item:previous_step} Put some random $p$-ary labels on the deterministically fixed edges of the graph and associate with it a random parity-check matrix of dimension $n(R_f-R) \times n$.
\item Our random LDA shaping lattice $\Lambda$ will be the lattice obtained with Construction A from the LDPC code whose random $p$-ary parity-check matrix of dimension $n(1-R) \times n$ is the superposition of the matrix built at step \ref{item:previous_step} and the previously created fine-lattice-generating matrix.
\end{enumerate}

The deterministic part of the construction is represented by the following binary matrix of dimension $n(1-R) \times n$:
\begin{equation*}
  H = \left(
  \begin{array}{c}
    H' \\ \hline
    H_f
  \end{array}
  \right);
\end{equation*}
$H_f$ is the lower submatrix formed by the last $n(1-R_f)$ rows and corresponding to the (unlabeled) Tanner graph of the fine lattice. $H$ is fixed once for all, according to the choice of the permutations that create the corresponding graphs. It has $\Delta_P$ non-zero entries in each row and $\Delta_V = \Delta_P(1-R)$ non-zero entries in each column (explaining why the associated $p$-ary codes are LDPC). Substituting to each $1$ in $H$ a random variable which takes equiprobable values in $\{0,1,\ldots,p-1\}$, we obtain the random matrix
\begin{equation*}
  \HH = \left(
  \begin{array}{c}
    \HH' \\ \hline
    \HH_f
  \end{array}
  \right).
\end{equation*}
\begin{definition}[Skeleton matrix]
  \label{def:skeleton}
  In this context, we call the binary matrix $H$ (resp. $H_f$) the {\em skeleton} of the random matrix $\HH$ (resp. $\HH_f$). 
\end{definition}
The random fine lattice of our ensemble is $\Lambda_f$, generated by $\HH_f$, while the random shaping lattice is $\Lambda$, generated by $\HH$. They are nested lattices ($\Lambda \subseteq \Lambda_f$) and the Voronoi constellations that we will deal with are given by $\Lambda_f/\Lambda$. Observe the numerous similarities with respect to the construction of the random ensemble of Section \ref{sec:the_random_ensemble}. 

As we have already anticipated, the proof of Theorem \ref{thm:awgn_capacity_LDA} is strongly based on the fact that the graph that underlies our random ensemble of lattices has some particular expansion properties: Corollary \ref{cor:good_graphs} of Section \ref{sec:graph_theoretical_tools} guarantees that (for $n$ tending to infinity and with probability tending to $1$) the Tanner graph associated with $\Lambda_f$ is \emph{$D$-good} for every $D \geq 1$ such that:
%\begin{align}
%  \Delta_P & > \max \left \{\frac{2-R_f}{1-R_f}\left(1-\frac{Dh\left(\frac{1}{D}\right)}{(D+1)h\left(\frac{1}{D+1}\right)}\right)^{-1} , D^%2+\frac{1}{1-R_f}, \frac{D^2}{1-R_f} + 1 \right\} \\
%  \label{eq:bound_Delta}
%  & = \max \left \{ \frac{2-R_f}{1-R_f}\left(1-\frac{Dh\left(\frac{1}{D}\right)}{(D+1)h\left(\frac{1}{D+1}\right)}\right)^{-1}, \frac{D^2}{1-R_f} + 1\right\}.
%\end{align}
\begin{equation}
  \label{eq:bound_Delta}
  \Delta_P > \max \left \{ \frac{2-R_f}{1-R_f}\left(1-\frac{Dh\left(\frac{1}{D}\right)}{(D+1)h\left(\frac{1}{D+1}\right)}\right)^{-1}, \frac{D^2}{1-R_f} + 1\right\}.
\end{equation}
Notice that since $0 < R < R_f < 1$,
\begin{equation*}
  \frac{2-R_f}{1-R_f} > \frac{2-R}{1-R} \text{\ \ \ and \ \ \ } \frac{D^2}{1-R_f} > \frac{D^2}{1-R};
\end{equation*}
this implies that \eqref{eq:bound_Delta} is a sufficient condition for the asymptotic $D$-goodness of the family $\mathcal{F}(n,1-R,\Delta_P)$ (cf. the notation of Section \ref{sec:graph_theoretical_tools}). A simple application of Stirling's formula shows that the number of graphs in $\mathcal{F}(n,1-R,\Delta_P)$ and the number of possible Tanner graphs associated with $H$ are asymptotically the same:
\begin{equation*}
  (n(1-R)\Delta_P)! \sim (n(1-R_f)\Delta_P)!\ (n(R_f-R)\Delta_P)!.
\end{equation*}
For this reason, \eqref{eq:bound_Delta} is also sufficient to claim that the graph associated with $H$ is $D$-good with probability tending to $1$ when $n$ tends to infinity.

Remark: from now on, we will always assume that the Tanner graphs associated with the skeleton matrices $H$ and $H_f$ are $D$-good, neglecting the probabilistic aspect of this assertion. When $n$ is big enough, this will be almost always the case and we can assume that good permutations in the graph construction are chosen. Moreover, this is not a loss in generality for the construction of the lattice ensemble, since its randomness comes from the random entries of the matrices and not from the position of these entries in the matrix, which are fixed once for all.

A consequence of the $D$-goodness of the Tanner graphs is that we can find a lower bound of the minimum Hamming distance of the LDPC codes underlying our LDA construction:
\begin{lemma}[Asymptotic goodness of non-binary LDPC codes]
  \label{lem:min_H_dist}
  Let $\Lambda_f$ be our random $n$-dimensional LDA fine lattice ($p = n^\lambda$) with
  \begin{equation}
    \label{eq:condition_D_lambda}
    D > \frac{1}{1-R_f} \text{\ \ \ and\ \ \ } \lambda > \frac{1}{D(1-R_f)-1}.
  \end{equation}
  Suppose also that \eqref{eq:bound_Delta} holds true:
  \begin{equation*}
    \Delta_P > \max \left \{ \frac{2-R_f}{1-R_f}\left(1-\frac{Dh\left(\frac{1}{D}\right)}{(D+1)h\left(\frac{1}{D+1}\right)}\right)^{-1}, \frac{D^2}{1-R_f} + 1\right\}.
  \end{equation*}
  Moreover, for every $\mathbf{x} \in \Lambda_f$, let $w(\mathbf{x}) = |\{i:x_i \neq 0\}|$. Then, for every $\delta < D(1-R_f)/(D+1)$,
  \begin{equation*}
    \lim_{n \to \infty} \prob \left\{\mathbf{x} \in p\Zn \big| w(\mathbf{x}) \leq \delta n \right\} = 1.
  \end{equation*}
  In other words, the minimum Hamming distance of the LDPC code underlying our construction is typically lower bounded by $D(1-R_f)n/(D+1) - o(1)$.
\end{lemma}
Remark: we invite the reader to pay particular attention to the proof of this lemma. The argument used here is a prototype of the application of expansion properties to the more general techniques utilized in Lemma \ref{lem:typical_norm_LDA} and Theorem \ref{thm:awgn_capacity_LDA}. In what follows, it is easy to understand how the probability of an integer point to belong to an LDA lattice (or an LDPC code) is estimated thanks to the $D$-goodness of the associated graphs, in spite of the difficulties arising from the low density.
\begin{IEEEproof}
  Let $\Lambda_f = C_f + p\Zn$, where $C_f$ is the random LDPC code associated with the random parity-check matrix $\HH_f$. 
  Let $X$ be the random variable that counts the number of points of $C_f$ of Hamming weight $1 \leq w(\mathbf{x}) \leq \delta n$. For any $\mathbf{x} \in \Fp^n$, consider the random variable
  \begin{equation*}
    X_{\mathbf{x}} =
    \begin{cases}
      1, & \text{if } \mathbf{x} \in C_f \\ 
      0, & \text{otherwise} \\
    \end{cases}.
  \end{equation*}
  Consequently,
  \begin{equation*}
    X = \sum_{\substack{\mathbf{x} \in \Fpn \\ 1 \leq w(\mathbf{x}) \leq \delta n}} X_{\mathbf{x}}.
  \end{equation*}
  Notice that we only need to prove that
  \begin{equation*}
    \lim_{n \to \infty} \prob \left\{ X > 0 \right\} = 0
  \end{equation*}
  and, to do it, it is sufficient to show that
  \begin{equation*}
    \lim_{n \to \infty} \E[X] = \lim_{n \to \infty} \sum_{\substack{\mathbf{x} \in \Fpn \\ 1 \leq w(\mathbf{x}) \leq \delta n}} \prob \left\{ \mathbf{x} \in C_f \right\} = 0.
  \end{equation*}
  We will split the previous sum into two smaller sums and show that both of them converge to $0$.
  
  {\bf Case 1: $w(\mathbf{x}) \leq n/(D+1)$.} By definition of parity-check matrix,
  \begin{equation*}
    \prob \left\{ \mathbf{x} \in C_f \right\} = \prob \left\{ \HH_f \mathbf{x}^T \equiv \zero^T \bmod p \right\}. 
  \end{equation*}
  Let us call $\supp(\mathbf{x})=\{j : x_j \neq 0\}$ and, if $\mathbf{h}_i=(h_1,\ldots,h_n)$ is a row of $\HH_f$, let us define $\supp(\mathbf{h}_i)=\{j : h_j \text { is a random variable}\}$. Then, since $h_j = 0$ for every $j \not \in \supp(\mathbf{h_i})$, we deduce that
  \begin{equation*}
    \prob \left\{ \mathbf{h}_i \mathbf{x}^T \equiv 0 \bmod p \right\} =
    \begin{cases}
      1, & \text{if } \supp(\mathbf{x}) \cap \supp(\mathbf{h}_i) = \emptyset\\
      \frac{1}{p}, & \text{otherwise}
    \end{cases}.
  \end{equation*}
  Now, the rows of $\HH_f$ such that $\supp(\mathbf{x}) \cap \supp(\mathbf{h}_i) \neq \emptyset$ are exactly $|N(\supp(\mathbf{x}))|$ and the events $\left\{ \mathbf{h}_i \mathbf{x}^T \equiv 0 \bmod p \right\}_{i=1,\ldots,n(1-R_f)}$ are pairwise independent, therefore
  %and $N(\supp(\mathbf{x}))$ is its neighborhood in the Tanner graph associated with $\HH_f$, notice that
  \begin{equation*}
    \prob \left\{ \HH_f \mathbf{x}^T \equiv \zero^T \bmod p \right\} = \left( \frac{1}{p} \right)^{|N(\supp(\mathbf{x}))|} \leq \left( \frac{1}{p} \right)^{D(1-R_f)|\supp(\mathbf{x})|};
  \end{equation*}
  %where $(a)$ comes from the fact that all the parity-check equations connected to only-$0$ (modulo $p$) variables are satisfied with probability $1$, whereas the others are such that, calling $\mathbf{h}$ the corresponding row of $\HH_f$,
  %\begin{equation*}
  %  \prob \left\{ \mathbf{h} \mathbf{x}^T \equiv 0 \bmod p \right\} = \frac{1}{p}.
  %\end{equation*}
  %Moreover, 
  the inequality is a consequence of the $D$-goodness of the Tanner graph: simply apply \eqref{eq:condition_1} to $S = \supp(\mathbf{x})$ with $f = 1-R_f$. Therefore,
  \begin{align*}
    \sum_{\substack{\mathbf{x} \in \Fpn \\ 1 \leq w(\mathbf{x}) \leq n/(D+1)}} \prob \left\{ \mathbf{x} \in C_f \right\} & \leq \sum_{w=1}^{\lfloor n/(D+1) \rfloor} \sum_{\substack{\mathbf{x} \in \Fpn \\ w(\mathbf{x})=w}} \left( \frac{1}{p} \right)^{D(1-R_f)w} \\ 
    & = \sum_{w=1}^{\lfloor n/(D+1) \rfloor} \binom{n}{w}(p-1)^w \left( \frac{1}{p} \right)^{D(1-R_f)w} \\
    & \leq \sum_{w=1}^{\lfloor n/(D+1) \rfloor} \left(n^{1 - \lambda(D(1-R_f)-1)} \right)^w \rightarrow 0,
  \end{align*}
  because of \eqref{eq:condition_D_lambda}. 

  {\bf Case 2: $ n/(D+1) < w(\mathbf{x}) \leq \delta n$.} In this case, applying \eqref{eq:condition_1} to any $S \subseteq \supp(\mathbf{x})$ of size $n/(D+1)$, the only property guaranteed by $D$-goodness of the Tanner graph is:
  \begin{equation*}
    |N(\supp(\mathbf{x}))| \geq |N(S)| \geq \frac{D(1-R_f)n}{D+1},
  \end{equation*}
  Therefore,
  \begin{align*}
    \sum_{\substack{\mathbf{x} \in \Fpn \\ n/(D+1) < w(\mathbf{x}) \leq \delta n}} \prob \left\{ \mathbf{x} \in C_f \right\} & \leq \sum_{w=\lfloor n/(D+1) \rfloor +1}^{\lfloor \delta n \rfloor} \sum_{\substack{\mathbf{x} \in \Fpn \\ w(\mathbf{x})=w}} \left( \frac{1}{p} \right)^{\frac{D(1-R_f)n}{(D+1)}} \\
    & = \sum_{w=\lfloor n/(D+1) \rfloor +1}^{\lfloor \delta n \rfloor} \binom{n}{w}(p-1)^w \left( \frac{1}{p} \right)^{\frac{D(1-R_f)n}{(D+1)}} \\
    & \leq n 2^n p^{n \left( \delta - \frac{D(1-R_f)}{(D+1)} \right)} \rightarrow 0,
  \end{align*}
  because $\delta < D(1-R_f)/(D+1)$ by hypothesis.
\end{IEEEproof}

\begin{corollary}[Fundamental gain of LDA lattices]
  \label{cor:fundamental_gain_LDA}
  Let $\Lambda_f$ be our random $n$-dimensional LDA fine lattice ($p = n^\lambda$) for some
  \begin{equation}
    \label{eq:lambda_D_Hermite}
    D > \frac{1}{1-R_f} + 2 \text{\ \ \ and\ \ \ } \frac{1}{D(1-R_f)-1} < \lambda < \frac{1}{2(1-R_f)}.
  \end{equation}
  Moreover, let us impose \eqref{eq:bound_Delta}:
  \begin{equation*}
    \Delta_P > \max \left \{ \frac{2-R_f}{1-R_f}\left(1-\frac{Dh\left(\frac{1}{D}\right)}{(D+1)h\left(\frac{1}{D+1}\right)}\right)^{-1}, \frac{D^2}{1-R_f} + 1\right\}.
  \end{equation*}
  Then, the fundamental gain \eqref{eq:fundamental_gain} of $\Lambda_f$ is such that
  \begin{equation*}
    \lim_{n \to \infty} \prob \left\{\gamma(\Lambda_f) \rightarrow +\infty \right\} = 1.
  \end{equation*}
\end{corollary}
\begin{IEEEproof}
  Let $\Lambda_f = C_f + p\Zn$. If we call $\dHmin(C_f)$ the minimum Hamming distance of $C_f$, it is clear that the minimum Euclidean distance of $\Lambda_f$ satisfies
  \begin{equation*}
    \dEmin (\Lambda_f) \geq \min \left\{ p,\sqrt{\dHmin(C_f)} \right\}.
  \end{equation*}
  Lemma \ref{lem:min_H_dist} states that with probability tending to $1$, the minimum Hamming distance of $C_f$ satisfies
  \begin{equation*}
    \dHmin(C_f) \gtrsim \frac{D(1-R_f)n}{D+1}. 
  \end{equation*}
  The volume of $\Lambda_f$ is known to be $p^{n(1-R_f)}$, therefore if $\lambda < 1/2$, then $p = o(\sqrt{n})$ and we have almost surely that $\dEmin (\Lambda_f) = p$; thus,
  \begin{equation*}
    \gamma (\Lambda_f) = \frac{\dEmin (\Lambda_f)^2}{\vol(\Lambda_f)^{\frac{2}{n}}} = p^{2R_f} \rightarrow +\infty.
  \end{equation*}
  Otherwise, when $\lambda \geq 1/2$, thanks to \eqref{eq:lambda_D_Hermite} with probability tending to $1$ we have
  \begin{equation*}
    \gamma (\Lambda_f) \gtrsim \frac{D(1-R_f)n}{(D+1)n^{2\lambda(1-R_f)}} \rightarrow +\infty.
  \end{equation*}
\end{IEEEproof}

Remark: the previous lemma and corollary hold true also for the shaping lattice $\Lambda$, if we substitute $R_f$ with $R$ in the formulae.

\section{LDA lattices achieve capacity - Detailed proof}%------------------------- SECTION - LDA ACHIEVE CAPACITY ---------------------
\label{sec:LDA_capacity_detailed_proof}

\subsection{The encoding and decoding scheme}%------------------------- SECTION - ENCODING SCHEME ---------------------
\label{sec:new_encoding_scheme}

The encoding and decoding scheme that we apply to LDA Voronoi constellations is the same that we have described in Section \ref{sec:encoding_and_decoding} and summarized in Fig.~\ref{fig:enc_dec_scheme} of Section \ref{sec:capacity_for_random_construction_A} for the case of more general Construction-A lattices. Nothing changes at all and the fact that the lattices that we deal with now are LDA does not affect the information transmission scheme.

\subsection{A useful lemma}%------------------------- SUBSECTION - USEFUL LEMMA ---------------------
\label{subsec:a_useful_lemma}

In the sequel we will often need to compare the volumes of two spheres with the same radius, but different dimensions. This lemma contains once for all the computation that leads to this comparison and its simple proof is in Appendix \ref{sec:proof_volume_ratio}.

\begin{lemma}
  \label{lem:volume_ratio}
  Consider the two balls $B_{\mathbf{c},n}(\rho)$ and $B_{\mathbf{c}',n-m}(\rho)$, with the same given radius $\rho$, but with different dimensions $n$ and $n-m$. Suppose also that $0 \leq m \leq n/2$. Then, if $\rho > \sqrt{n}/2$,
  \begin{equation*}
    \frac{|\Z^{n-m} \cap B_{\mathbf{c}',n-m}(\rho)|}{|\Zn \cap B_{\mathbf{c},n}(\rho)|} \lesssim \frac{(\sqrt{n})^{n+1}}{(\sqrt{n-m})^{n-m+1}} \left(\sqrt{2\pi e}\right)^{-m} \rho^{-m} \left(1 + \frac{2\sqrt{n}}{2\rho-\sqrt{n}} \right)^n.
  \end{equation*}
\end{lemma}

\subsection{The typical norm of a constellation point}%------------------------- SUBSECTION - TYPICAL POINT NORM ---------------------
The next lemma states that our Voronoi LDA constellation points have the same typical norm of the more general Construction-A constellation points of Section \ref{sec:detailed_proof}. The proof of the lemma follows that of Lemma \ref{lem:typical_norm}, but needs to be adapted to the LDA setting in which we work. This requires some tricky combinatorial analysis of the structure of the Tanner graphs associated with the random lattices. The most interesting argument is probably the variance estimation that starts from \eqref{eq:variance_starting_point} and goes on till the end of the proof. Similar reasonings will be used in the proof of Theorem \ref{thm:awgn_capacity_LDA}.

Like in Section \ref{sec:detailed_proof}, let $\rho_{\eff}$ denote the asymptotic effective radius of the shaping lattice associated with the parity-check matrix $\HH$:
\begin{equation*}
  \rho_{\eff} = \frac{\sqrt{n}p^{(1-R)}}{\sqrt{2\pi e}}.
\end{equation*}

\begin{lemma}[Typical norm of an LDA-constellation point]
  \label{lem:typical_norm_LDA}
  In the setting fixed in Section \ref{sec:random_LDA_ensemble} and \ref{sec:LDA_capacity_detailed_proof}, consider a non-zero syndrome $\mathbf{s} = (s_1, s_2, \ldots, s_{n(R_f-R)},0,\ldots,0) \neq \zero$ associated with a message and a constellation point. Suppose that $p=n^{\lambda}$ for some $\lambda > 0$ and let $0 < \omega < 1$. Fix the constant $D$ to be
  \begin{equation}
    \label{eq:further_condition_D}
    D > \max \left\{ \frac{1}{1-R_f}, 2 \right\}
  \end{equation}
  and suppose that \eqref{eq:bound_Delta} is true:
  \begin{equation*}
    \Delta_P > \max \left \{ \frac{2-R_f}{1-R_f}\left(1-\frac{Dh\left(\frac{1}{D}\right)}{(D+1)h\left(\frac{1}{D+1}\right)}\right)^{-1}, \frac{D^2}{1-R_f} + 1\right\}.
  \end{equation*}
  If $\mathbf{x}$ is the random LDA constellation point whose syndrome is $\mathbf{s}$ (cf. \eqref{eq:phi}) and if $\lambda$ satisfies
  \begin{equation}
    \label{eq:lambda_typical_norm_LDA}
    \lambda > \max \left\{ \frac{1}{D(1-R_f)-1}, \frac{1}{2R}, \frac{1}{1-R}, \frac{1}{D-2}, \left(1-\frac{1}{D^2-1}-\frac{1}{D(1-R)}\right)^{-1} \right\},
  \end{equation} 
  then
  \begin{equation}
    \label{eq:typical_norm_x}
    \lim_{n\to \infty} \prob\left\{\rho_{\eff}\left(1-\frac{1}{n^\omega}\right) \leq \Vert\mathbf{x}\Vert \leq \rho_{\eff}\left(1+\frac{1}{n^\omega}\right)\right\} = 1.
  \end{equation}
\end{lemma}

Remark: the hypotheses of the lemma imply that the Tanner graphs associated with both the fine and the shaping (random) lattices can be assumed to be $D$-good. Moreover, the hypotheses of Lemma \ref{lem:min_H_dist} are met. Finally, if we compare this statement to Lemma \ref{lem:typical_norm}, notice that \eqref{eq:omega_lambda} reduces to $\omega < 1$ because of \eqref{eq:lambda_typical_norm_LDA}.

\begin{IEEEproof}
  First of all, let us consider the Tanner graph associated with $\HH$ and see what properties derive from its $D$-goodness. If $V$ is its set of variable nodes (of cardinality $n$) and $P$ its set of check nodes (of cardinality $n(1-R)$), \eqref{eq:condition_1} and \eqref{eq:condition_2} with $f= (1-R)$ imply:
  \begin{itemize}
  \item $\forall S \subseteq V\ \text{such that } |S| \leq \frac{n}{D+1},\ \text{then } |N(S)| \geq D(1-R)|S|$;
  \item $\forall S \subseteq V\ \text{such that } |S| \geq \frac{n}{D+1},\ \text{then } |N(S)| \geq \frac{Dn(1-R)}{D+1}$;
  \item $\forall T \subseteq P\ \text{such that } |T| \leq \frac{n(1-R)}{D+1},\ \text{then } |N(T)| \geq \frac{D|T|}{1-R}$;
  \item $\forall T \subseteq P\ \text{such that } |T| \geq \frac{n(1-R)}{D+1},\ \text{then } |N(T)| \geq \frac{Dn}{D+1}$. 
  \end{itemize}
  We will extensively use these expansion properties in this proof.
  
  Now, let $X_\rho$ be the random variable that counts the number of points with syndrome $\mathbf{s}$ in the $n$-dimensional ball $B_{\zero, n}(\rho)$. For any $\rho \geq 0$ and for any $\mathbf{x} \in \Zn \cap B_{\zero,n}(\rho)$, consider the random variable
  \begin{equation*}
    X_{\mathbf{x}} =
    \begin{cases}
      1, & \text{if } \HH \mathbf{x}^T \equiv \mathbf{s}^T \bmod p \\ 
      0, & \text{otherwise} \\
    \end{cases}.
  \end{equation*}
  Consequently,
  \begin{equation}
    \label{eq:X_rho_sum}
    X_{\rho} = \sum_{\mathbf{x} \in \Zn \cap B_{\zero,n}(\rho)}X_{\mathbf{x}} = \sum_{\mathbf{x} \in \Zn \cap B_{\zero,n}(\rho) \smallsetminus p\Zn}X_{\mathbf{x}},
  \end{equation}
  because the probability that the points of $p\Zn$ have syndrome $\mathbf{s} \neq \zero$ is $0$.
  Let us also define the \emph{support} of $\mathbf{x}$:
  \begin{equation}
    \label{eq:support}
    \supp(\mathbf{x})=\{j : x_j \not\equiv 0 \bmod p\}
  \end{equation}
  and, if $\mathbf{h}$ is a row of $\HH$,
  \begin{equation*}
    \supp(\mathbf{\mathbf{h}}) = \{j : h_j \text{ is a random variable}\}.
  \end{equation*}
  If we call $\mathbf{h}_i$ the $i$-th row of $\HH$ and $s_i$ is the $i$-th coordinate of $\mathbf{s}$, supposing that $\mathbf{x}$ is a given point of $ \Zn \cap B_{\zero,n}(\rho) \smallsetminus p\Zn$, we can deduce that:
  \begin{itemize}
  \item If $\supp(\mathbf{x}) \cap \supp(\mathbf{h}_i) \neq \emptyset$, then $\prob\{\mathbf{h}_i\mathbf{x}^T \equiv s_i \bmod p\}=1/p$.
  \item If $\supp(\mathbf{x}) \cap \supp(\mathbf{h}_i) = \emptyset$ and $s_i = 0$, then $\prob\{\mathbf{h}_i\mathbf{x}^T \equiv s_i \bmod p\}=1$.
  \item If $\supp(\mathbf{x}) \cap \supp(\mathbf{h}_i) = \emptyset$ and $s_i \neq 0$, then $\prob\{\mathbf{h}_i\mathbf{x}^T \equiv s_i \bmod p\}=0$.
  \end{itemize}
  In order to quantify $\prob \{X_\mathbf{x} =1 \}$, it is then important to know the size of the set %Let $\ell$ be the number of rows of $\HH$ whose support intersects the support of $\mathbf{x}$ and let
  \begin{equation*}
    T_{\mathbf{x}} = \{i\in\{1,2,\ldots,n(1-R)\} : \supp(\mathbf{h}_i)\cap\supp(\mathbf{x}) \neq\emptyset \}.
  \end{equation*}
  $T_{\mathbf{x}}$ is identified with the set of the parity-check equation nodes of the Tanner graph associated with $\HH$ whose support intersects the support of $\mathbf{x}$, then $T_{\mathbf{x}} \subseteq P$. Let us suppose for a moment that
  \begin{equation}
    \label{eq:T_small}
    |T_{\mathbf{x}}| \leq n(1-R)\left(\frac{D^2+D(1-R_f)-1}{D(D+1)}\right) = A(n)
  \end{equation}
  or, equivalently, that
  \begin{equation}
    \label{eq:P_T_big}
    |P \smallsetminus T_{\mathbf{x}}| \geq n(1-R)\left( \frac{DR_f+1}{D(D+1)} \right) = n(1-R) - A(n).
  \end{equation}
  Since $(DR_f+1)/(D(D+1)) < 1/(D+1)$ because of \eqref{eq:further_condition_D}, the $D$-goodness of the Tanner graph associated with $\HH$ implies that
  \begin{equation*}
    |N(P \smallsetminus T_{\mathbf{x}})| \geq \frac{D}{1-R} n(1-R)\left( \frac{DR_f+1}{D(D+1)} \right) = \frac{n(DR_f+1)}{D+1} = n\left(1-\frac{D(1-R_f)}{D+1} \right).
  \end{equation*}
  Now notice that for any fixed $\mathbf{x}$, all its coordinates that belong to $N(P \smallsetminus T_{\mathbf{x}}) \subseteq V$ have to be equal to $0$ (modulo $p$) by definition of $T_{\mathbf{x}}$, because all the non-zero coordinates of $\mathbf{x}$ are connected via an edge in the Tanner graph to an equation of $T_{\mathbf{x}}$. Therefore, %if $V$ is the set of variable nodes in the Tanner graph,
  \begin{equation*}
    |\supp(\mathbf{x})| \leq |V \smallsetminus N(P \smallsetminus T_{\mathbf{x}})| \leq n - n\left(1-\frac{D(1-R_f)}{D+1} \right) = \frac{nD(1-R_f)}{D+1}.
  \end{equation*}
  By Lemma \ref{lem:min_H_dist}, we can assume without loss of generality that there is no point of the fine lattice $\Lambda_f$ (except for some points of $p\Zn$) with such a small support. Hence, for every $\mathbf{x} \in \Zn \cap B_{\zero,n}(\rho)\smallsetminus p\Zn$ satisfying \eqref{eq:T_small},
  \begin{align*}
    \prob\{X_{\mathbf{x}} = 1\} & = \prob\{\HH\mathbf{x}^T \equiv \mathbf{s}^T \bmod p\} \leq \prob\{\HH_f\mathbf{x}^T \equiv \zero^T \bmod p\} = \prob\{\mathbf{x} \in \Lambda_f\} = 0.
  \end{align*}
  For this reason and because the events $\{\mathbf{h}_i\mathbf{x}^T \equiv s_i \bmod p\}_{i=1,\ldots,n(1-R)}$ are independent, we can write
  \begin{equation}
    \label{eq:prob_Xx_1}
    \prob\{X_{\mathbf{x}} = 1\} =
    \begin{cases}
      0, & \text{if } \exists i : \supp(\mathbf{x}) \cap \supp(\mathbf{h}_i) = \emptyset \text{ and } s_i \neq 0\\ 
      0, & \text{if } |T_{\mathbf{x}}| \leq A(n)\\
      \left(\frac{1}{p}\right)^{|T_{\mathbf{x}}|}, & \text{otherwise} \\
    \end{cases}.
  \end{equation}
  %Moreover, $\prob\{X_{\mathbf{x}} = 1\} = 0$ for every $\mathbf{x} \in p\Zn \cap B_{\zero,n}(\rho)$, since its syndrome is $\zero \neq \mathbf{s}$. This is in particular the case for $\ell=0$, which coincides with $\mathbf{x} \in p\Zn$.
  
  Like for \eqref{eq:first_part} and \eqref{eq:second_part} in Lemma \ref{lem:typical_norm}, we will split the proof into two parts. First of all, we deduce that
  \begin{equation}
    \label{eq:first_part_LDA}
    \lim_{n\to \infty} \prob \left\{X_{\rho_{\eff}\left(1-\frac{1}{n^\omega}\right)} > 0 \right\} = 0.
  \end{equation}
  Later, that
  \begin{equation}
    \label{eq:second_part_LDA}
    \lim_{n\to \infty} \prob \left\{X_{\rho_{\eff}\left(1+\frac{1}{n^\omega}\right)} = 0 \right\} = 0.
  \end{equation}
  These two conditions together imply \eqref{eq:typical_norm_x}.
  
  {\bf Proof of \eqref{eq:first_part_LDA}.} Now $\rho = \rho_{\eff}\left(1-1/n^\omega\right)$. Using \eqref{eq:X_rho_sum} and \eqref{eq:prob_Xx_1}, we deduce that  
  \begin{align}
    \E[X_{\rho}]  & = \sum_{\mathbf{x} \in \Zn \cap B_{\zero,n}(\rho)\smallsetminus p\Zn} \prob\{X_{\mathbf{x}}=1\} \nonumber \\
    & = \sum_{\substack{\mathbf{x} \in \Zn \cap B_{\zero,n}(\rho)\smallsetminus p\Zn \\ |T_{\mathbf{x}}| \geq A(n)}} \prob\{X_{\mathbf{x}}=1\} \nonumber \\
    & \leq \sum_{\ell=\lceil A(n) \rceil}^{n(1-R)} \sum_{\substack{\mathbf{x} \in \Zn \cap B_{\zero,n}(\rho)\smallsetminus p\Zn\\|T_{\mathbf{x}}|=\ell}} \left(\frac{1}{p}\right)^\ell \nonumber \\
    \label{eq:u_instead_of_j}
    & = \sum_{u = 0}^{n(1-R) - \lceil A(n) \rceil} \sum_{\substack{\mathbf{x} \in \Zn \cap B_{\zero,n}(\rho)\smallsetminus p\Zn\\|T_{\mathbf{x}}|=n(1-R)-u}} \left(\frac{1}{p}\right)^{n(1-R)-u}.
  \end{align}
  Notice that $u = |P \smallsetminus T_{\mathbf{x}}|$ and the fact that
  \begin{equation*}
    u \leq n(1-R) - A(n) \leq n(1-R)/(D+1) 
  \end{equation*}
  implies by the expansion properties that
  \begin{equation*}
    |N(P \smallsetminus T_{\mathbf{x}})| \geq \frac{D|P \smallsetminus T_{\mathbf{x}}|}{1-R}.
  \end{equation*}
  This means that once $P \smallsetminus T_{\mathbf{x}}$ is fixed, at least $D|P \smallsetminus T_{\mathbf{x}}|/(1-R)$ coordinates of $\mathbf{x}$ are equal to $0$ (modulo $p$). Hence
  \begin{align*}
    &|\{\mathbf{x} \in \Zn \cap B_{\zero,n}(\rho) \smallsetminus p\Zn: |T_{\mathbf{x}}|=n(1-R)-u\}| \\
    & \ \ \ \ \ = |\{\mathbf{x} \in \Zn \cap B_{\zero,n}(\rho) \smallsetminus p\Zn: |P \smallsetminus T_{\mathbf{x}}|=u\}| \\
    & \ \ \ \ \ \leq \binom{n(1-R)}{u} |\Z^{n-Du/(1-R)} \cap B_{\zero,n-Du/(1-R)}(\rho)| \\
    & \ \ \ \ \ \leq n^{u}|\Z^{n-Du/(1-R)} \cap B_{\zero,n-Du/(1-R)}(\rho)|.
  \end{align*}
  Applying Lemma \ref{lem:volume_ratio} and substituting the real value of $\rho$ to obtain \eqref{eq:volume_ratio_1}, we deduce that
  \begin{align}
    \eqref{eq:u_instead_of_j} & \leq \sum_{u = 0}^{n(1-R) - \lceil A(n) \rceil} n^{u}|\Z^{n-Du/(1-R)} \cap B_{\zero,n-Du/(1-R)}(\rho)| \left(\frac{1}{p}\right)^{n(1-R)-u} \nonumber \\
    & = \sum_{u = 0}^{n(1-R) - \lceil A(n) \rceil} n^{u}p^u\frac{|\Z^{n-Du/(1-R)} \cap B_{\zero,n-Du/(1-R)}(\rho)|}{|\Z^n \cap B_{\zero,n}(\rho)|} |\Z^n \cap B_{\zero,n}(\rho)|\left(\frac{1}{p}\right)^{n(1-R)} \nonumber \\
    & \lesssim |\Z^n \cap B_{\zero,n}(\rho)|\left(\frac{1}{p}\right)^{n(1-R)} \nonumber \\
    \label{eq:volume_ratio_1}
    & \ \ \ \ \ \ \ \ \ \ \ \ \ \ \ \ \ \ \ \cdot \sum_{u = 0}^{n(1-R) - \lceil A(n) \rceil} \left(1-\frac{1}{n^\omega}\right)^{-\frac{Du}{1-R}}\left(\sqrt{\frac{n}{n-Du/(1-R)}}\right)^{n-\frac{Du}{1-R}+1} \frac{n^{u}p^u}{p^{Du}} \\
    & = |\Z^n \cap B_{\zero,n}(\rho)|\left(\frac{1}{p}\right)^{n(1-R)} \nonumber \\
    \label{eq:sum_bounded_by_n}
    & \ \ \ \cdot \sum_{u = 0}^{n(1-R) - \lceil A(n) \rceil} \left(1-\frac{1}{n^\omega}\right)^{-\frac{Du}{1-R}}\left(1+\frac{Du/(1-R)}{n-Du/(1-R)} \right)^{(n-\frac{Du}{1-R}+1)/2} n^{u(1 -\lambda(D-1))}.  
  \end{align}
  Now, it is easy to show (and we leave the details to the reader) that 
  \begin{equation*}
    \left(1-\frac{1}{n^\omega}\right)^{-\frac{Du}{1-R}} \left(1+\frac{Du/(1-R)}{n-Du/(1-R)} \right)^{(n-\frac{Du}{1-R}+1)/2} n^{u(1 -\lambda(D-1))} \lesssim 1
  \end{equation*}
  and, in particular, it is $o(1)$ whenever $u>0$, provided that $1 -\lambda(D-1) < 0$. This is guaranteed by \eqref{eq:lambda_typical_norm_LDA} and by the fact that $D-1 > 0$. Thus, we can crudely state that \eqref{eq:sum_bounded_by_n} is less than $n$, whereas we already know that
  \begin{equation*}
    |\Z^n \cap B_{\zero,n}(\rho)|\left(\frac{1}{p}\right)^{n(1-R)} \rightarrow 0
  \end{equation*}
  subexponentially fast in $n$: this computation was already been carried out in the proof of Lemma \ref{lem:typical_norm}, from \eqref{eq:standard_volume_power_pf_p_ratio} to \eqref{eq:effective_radius_volume}. Consequently, the whole sum tends to $0$ when $n$ tends to infinity. 

  Summarizing, we have shown that $\E[X_\rho]$ is asymptotically vanishing and, considering that $\prob\{X_{\rho}>0\} \leq \E[X_{\rho}]$, we finally have
  \begin{equation*}
    \lim_{n\to \infty} \prob \left\{X_{\rho_{\eff}\left(1-\frac{1}{n^\omega}\right)} > 0 \right\} = 0.
  \end{equation*}

  {\bf Proof of \eqref{eq:second_part_LDA}.} Now, let $\rho = \rho_{\eff}\left(1+1/n^\omega\right)$. %Using the fact that $|\Zn \cap B_{\zero,n}(\rho) \smallsetminus p\Zn| \sim |\Zn \cap B_{\zero,n}(\rho)|$,
  We have
  \begin{align*}
    \E[X_{\rho}]  & = \sum_{\mathbf{x} \in \Zn \cap B_{\zero,n}(\rho)\smallsetminus p\Zn} \prob\{X_{\mathbf{x}}=1\} \\
    & \geq \sum_{\substack{\mathbf{x} \in \Zn \cap B_{\zero,n}(\rho) \smallsetminus p\Zn \\ |T_{\mathbf{x}}|=n(1-R)}} \left(\frac{1}{p}\right)^{n(1-R)} \\
    & \geq \sum_{\substack{\mathbf{x} \in \Zn \cap B_{\zero,n}(\rho) \smallsetminus p\Zn \\ \forall i, x_i \neq 0}} \left(\frac{1}{p}\right)^{n(1-R)} \\
    & = |\{\mathbf{x} \in \Zn \cap B_{\zero,n}(\rho) \smallsetminus p\Zn :  x_i \neq 0, \forall i=1,2,\ldots,n\}|\left(\frac{1}{p}\right)^{n(1-R)} \\
    & = \big( |\Zn \cap B_{\zero,n}(\rho) \smallsetminus p\Zn| - |\{\mathbf{x} \in \Zn \cap B_{\zero,n}(\rho) \smallsetminus p\Zn :  x_i = 0, \exists i\}|\big) \left(\frac{1}{p}\right)^{n(1-R)} \\
    & \geq \left(|\Zn \cap B_{\zero,n}(\rho) \smallsetminus p\Zn| - \sum_{i=1}^n|\{\mathbf{x} \in \Zn \cap B_{\zero,n}(\rho) \smallsetminus p\Zn :  x_i = 0\}|\right) \left(\frac{1}{p}\right)^{n(1-R)} \\
    & = |\Zn \cap B_{\zero,n}(\rho) \smallsetminus p\Zn|\left(1 - n\frac{|\Z^{n-1} \cap B_{\zero,n-1}(\rho) \smallsetminus p\Z^{n-1}|}{|\Zn \cap B_{\zero,n}(\rho) \smallsetminus p\Zn|} \right)\left(\frac{1}{p}\right)^{n(1-R)}.
  \end{align*}
  Now, we have already computed from \eqref{eq:general_term_average} to \eqref{eq:lower_bound_mean} that
  \begin{equation*}
    \lim_{n \to \infty} |\Zn \cap B_{\zero,n}(\rho) \smallsetminus p\Zn| \left(\frac{1}{p}\right)^{n(1-R)} = +\infty.
  \end{equation*}
  What about
  \begin{equation*}
    n\frac{|\Z^{n-1} \cap B_{\zero,n-1}(\rho) \smallsetminus p\Z^{n-1}|}{|\Zn \cap B_{\zero,n}(\rho) \smallsetminus p\Zn|}\ ?
  \end{equation*}
  By Lemma \ref{lem:volume_ratio}, introducing the actual value of $\rho$ and recalling that $\lambda > (1-R)^{-1}$ by \eqref{eq:lambda_typical_norm_LDA}, we can deduce that
  \begin{align*}
    n\frac{|\Z^{n-1} \cap B_{\zero,n-1}(\rho) \smallsetminus p\Z^{n-1}|}{|\Zn \cap B_{\zero,n}(\rho) \smallsetminus p\Zn|} \lesssim \frac{n}{p^{(1-R)}}\left(\sqrt{\frac{n}{n-1}}\right)^n \sim \frac{n\sqrt{e}}{p^{1-R}} \to 0.
  \end{align*}
  This allows us to conclude that in this case
  \begin{equation*}
    \lim_{n \to \infty}\E[X_{\rho}] = +\infty.
  \end{equation*}
  
  After that, we need to carry out a detailed estimation of $\var(X_{\rho})$, like we did in the proof of Lemma \ref{lem:typical_norm} for the more general Construction-A constellations. We have
  \begin{align}
    \label{eq:variance_starting_point}
    \var(X_{\rho}) & = \var\left(\sum_{\mathbf{x} \in \Zn \cap B_{\zero,n}(\rho)\smallsetminus p\Zn} X_{\mathbf{x}}\right) \\
    & = \sum_{\mathbf{x},\mathbf{z} \in \Zn \cap B_{\zero,n}(\rho)\smallsetminus p\Zn} \cov(X_{\mathbf{x}},X_{\mathbf{z}}) \nonumber \\
    & \leq \sum_{\mathbf{x},\mathbf{z} \in \Zn \cap B_{\zero,n}(\rho)\smallsetminus p\Zn} \E[X_{\mathbf{x}}X_{\mathbf{z}}] \nonumber \\
    & = \sum_{\mathbf{x},\mathbf{z} \in \Zn \cap B_{\zero,n}(\rho)\smallsetminus p\Zn} \prob\{X_{\mathbf{x}}X_{\mathbf{z}}=1\} \nonumber \\
    & = \sum_{\mathbf{x},\mathbf{z} \in \Zn \cap B_{\zero,n}(\rho)\smallsetminus p\Zn} \prob\{X_{\mathbf{x}}=1,X_{\mathbf{z}}=1\} \nonumber \\
    & = \sum_{\mathbf{x},\mathbf{z} \in \Zn \cap B_{\zero,n}(\rho)\smallsetminus p\Zn} \prob\{\HH\mathbf{x}^T \equiv \mathbf{s}^T \bmod p, \HH\mathbf{z}^T \equiv \mathbf{s}^T \bmod p\}. \nonumber
  \end{align}
  Now, let $\mathbf{h}$ be a generic row of $\HH$; it represents a parity-check equation and we also write $\mathbf{h} \in P$. %, where $P$ is the set of vertices of the Tanner graph associated with $\HH$ corresponding to the parity-check equations (recall that the notation is the same of Section \ref{sec:graph_theoretical_tools}). 
  %This is a little abuse in notation, but it will help us to transfer some arguments onto the Tanner graph and make clearer our demonstration. 
  For a given $\mathbf{x} \in \Zn \cap B_{\zero,n}(\rho)\smallsetminus p\Zn$, let $\mathbf{x}_{\mathbf{h}}$ be the subvector of $\mathbf{x}$ made only of the coordinates of $\mathbf{x}$ that belong to the neighborhood $N(\mathbf{h})$ of $\mathbf{h}$ in the graph. In other words, these are the coordinates of $\mathbf{x}$ that correspond to ones in the row of the skeleton matrix $H$ of $\HH$ corresponding to $\mathbf{h}$.

  Let us fix $\mathbf{x}, \mathbf{z} \in \Zn \cap B_{\zero,n}(\rho)\smallsetminus p\Zn$ and a row $\mathbf{h}$ of $\HH$ and consider %let $\langle \mathbf{x}_{\mathbf{h}},\mathbf{z}_{\mathbf{h}} \rangle$ denote
  the vector space generated by $\mathbf{x}_{\mathbf{h}}$ and $\mathbf{z}_{\mathbf{h}}$, which can have dimension $0$, $1$, or $2$ over $\R$. We call the latter $\dim(\mathbf{x},\mathbf{z}|\mathbf{h})$. Hence, denoting $s$ the syndrome coordinate corresponding to $\mathbf{h}$, we have:
  \begin{itemize}
  \item if $\dim(\mathbf{x},\mathbf{z}|\mathbf{h}) = 0$ and $s=0$, then $\prob\{\mathbf{h}\mathbf{x}^T \equiv s \bmod p, \mathbf{h}\mathbf{z}^T \equiv s \bmod p\}=1$;
  \item if $\dim(\mathbf{x},\mathbf{z}|\mathbf{h}) = 0$ and $s \neq 0$, then $\prob\{\mathbf{h}\mathbf{x}^T \equiv s \bmod p, \mathbf{h}\mathbf{z}^T \equiv s \bmod p\}=0$;
  \item if $\dim(\mathbf{x},\mathbf{z}|\mathbf{h}) = 1$ and $s=0$, then $\prob\{\mathbf{h}\mathbf{x}^T \equiv s \bmod p, \mathbf{h}\mathbf{z}^T \equiv s \bmod p\}= 1/p$;
  \item if $\dim(\mathbf{x},\mathbf{z}|\mathbf{h}) = 1$ and $s \neq 0$, then $\prob\{\mathbf{h}\mathbf{x}^T \equiv s \bmod p, \mathbf{h}\mathbf{z}^T \equiv s \bmod p\}= 1/p$ if $\mathbf{z}_{\mathbf{h}} \equiv \mathbf{x}_{\mathbf{h}} \bmod p$, otherwise it is $0$;
  \item if $\dim(\mathbf{x},\mathbf{z}|\mathbf{h}) = 2$, then $\prob\{\mathbf{h}\mathbf{x}^T \equiv s \bmod p, \mathbf{h}\mathbf{z}^T \equiv s \bmod p\}= 1/p^2$.
  \end{itemize}
  Summarizing, given $\mathbf{x}, \mathbf{z} \in \Zn \cap B_{\zero,n}(\rho)\smallsetminus p\Zn$, we can consider the partition of the set of parity-check equations $P$ given by the following three sets:
  \begin{align}
    \label{eq:J_x_z}
    J_{\mathbf{x},\mathbf{z}} & = \{ \mathbf{h} \in P : \dim(\mathbf{x},\mathbf{z}|\mathbf{h}) = 0 \}, \\
    \label{eq:I_x_z}
    I_{\mathbf{x},\mathbf{z}} & = \{ \mathbf{h} \in P : \dim(\mathbf{x},\mathbf{z}|\mathbf{h}) = 1 \}, \\
    \label{eq:T_x_z}
    T_{\mathbf{x},\mathbf{z}} & = \{ \mathbf{h} \in P : \dim(\mathbf{x},\mathbf{z}|\mathbf{h}) = 2 \};
  \end{align}
  Notice that the coordinates of $\mathbf{x}$ and $\mathbf{z}$ that belong to $N(J_{\mathbf{x},\mathbf{z}})$ have to be equal to $0$ (modulo $p$). Hence, if we suppose that
  \begin{equation}
    \label{eq:J_big}
    |J_{\mathbf{x},\mathbf{z}}| \geq n(1-R) - A(n),
  \end{equation}
  we can use the very same argument used before in the study of $|P \smallsetminus T_{\mathbf{x}}|$ (from \eqref{eq:P_T_big} on), to prove that
  \begin{align}
    \prob\{\HH\mathbf{x}^T \equiv \mathbf{s}^T \bmod p, \HH\mathbf{z}^T \equiv \mathbf{s}^T \bmod p\} & \leq \prob\{\HH_f\mathbf{x}^T \equiv \zero^T \bmod p, \HH_f\mathbf{z}^T \equiv \zero^T \bmod p\} \nonumber \\
    \label{eq:prob_x_z_lambda_f_0}
    & = \prob\{\mathbf{x} \in \Lambda_f, \mathbf{z} \in \Lambda_f\} = 0.
  \end{align}
  Instead, for the $\mathbf{x}$ and $\mathbf{z}$ that satisfy the opposite of \eqref{eq:J_big}, recalling that $\HH$ has $n(1-R)=|P|$ rows,
  \begin{equation}
    \label{eq:prob_I_J_T}
    \prob\{\HH\mathbf{x}^T \equiv \mathbf{s}^T \bmod p, \HH\mathbf{z}^T \equiv \mathbf{s}^T \bmod p\} \leq \left(\frac{1}{p}\right)^{2n(1-R)-|I_{\mathbf{x},\mathbf{z}}|-2|J_{\mathbf{x},\mathbf{z}}|} = \left(\frac{1}{p}\right)^{2|T_{\mathbf{x},\mathbf{z}}| + |I_{\mathbf{x},\mathbf{z}}|}.
  \end{equation}
  More precisely, if the equality above does not hold, then the probability is $0$. Thanks to this information, we can write
  \begin{align}
    \var(X_{\rho}) & \leq \sum_{\mathbf{x},\mathbf{z} \in \Zn \cap B_{\zero,n}(\rho) \smallsetminus p\Zn} \prob\{\HH\mathbf{x}^T \equiv \mathbf{s}^T \bmod p, \HH\mathbf{z}^T \equiv \mathbf{s}^T \bmod p\} \nonumber \\
    & = \sum_{\substack{\mathbf{x},\mathbf{z} \in \Zn \cap B_{\zero,n}(\rho) \smallsetminus p\Zn \\ |J_{\mathbf{x},\mathbf{z}}| \leq n(1-R) - A(n)}} \prob\{\HH\mathbf{x}^T \equiv \mathbf{s}^T \bmod p, \HH\mathbf{z}^T \equiv \mathbf{s}^T \bmod p\} \nonumber \\
    \label{eq:variance_sum}
    & \leq \sum_{\substack{\mathbf{x},\mathbf{z} \in \Zn \cap B_{\zero,n}(\rho) \smallsetminus p\Zn \\ |J_{\mathbf{x},\mathbf{z}}| \leq n(1-R) - A(n)}} \left(\frac{1}{p}\right)^{2n(1-R)-|I_{\mathbf{x},\mathbf{z}}|-2|J_{\mathbf{x},\mathbf{z}}|}.
    %& = \sum_{\substack{\mathbf{x},\mathbf{z} \in \Zn \cap B_{\zero,n}(\rho) \smallsetminus p\Zn \\ |J_{\mathbf{x},\mathbf{z}}| \leq n(1-R) - A(n)}} \left(\frac{1}{p}\right)^{2|T_{\mathbf{x},\mathbf{z}}|+|I_{\mathbf{x},\mathbf{z}}|}.
  \end{align}

  Before estimating the sum, we will need to investigate the structure of $I_{\mathbf{x},\mathbf{z}}$ and its neighborhood. For this purpose, consider the graph $\mathcal{G}'_{\mathbf{x},\mathbf{z}}$ that consists of the bipartite subgraph of the whole Tanner graph (called $\mathcal{G}$) given by the parity-check equation nodes of $I_{\mathbf{x},\mathbf{z}}$, the variable nodes of $N(I_{\mathbf{x},\mathbf{z}})$, and the edges connecting them. \emph{A priori}, $\mathcal{G}'_{\mathbf{x},\mathbf{z}}$ can be made of many different (bipartite) connected components, depending for example on the size of $I_{\mathbf{x},\mathbf{z}}$ (even if $\mathcal{G}$ is connected with very high probability, tending to $1$ when $n$ tends to infinity). The set of vertices of each one of these components is made of a subset of $N(I_{\mathbf{x},\mathbf{z}})$ (variable nodes) and a subset of $I_{\mathbf{x},\mathbf{z}}$ (parity-check equation nodes). The connected components can be (trivially) partitioned into two kinds: the ones whose set of parity-check equations has size bigger than $n(1-R)/(D+1)$ and the ones for which this does not hold. So, if $\mathcal{C}$ is the generic connected component of $\mathcal{G}'_{\mathbf{x},\mathbf{z}}$ and $P_{\mathcal{C}} \subseteq P$ is its set of parity-check equation nodes, let us define:
  \begin{align}
    & \mathcal{K}_{\mathbf{x},\mathbf{z}} = \left\{\mathcal{C} \subseteq \mathcal{G}'_{\mathbf{x},\mathbf{z}} : |P_\mathcal{C}| \leq \frac{n(1-R)}{D+1} \right\} \text{ and} \nonumber \\
    \label{eq:definition_of_M}
    & \mathcal{M}_{\mathbf{x},\mathbf{z}} = \left\{\mathcal{C} \subseteq \mathcal{G}'_{\mathbf{x},\mathbf{z}} : |P_\mathcal{C}| > \frac{n(1-R)}{D+1} \right\}.
  \end{align}
  Of course, $\mathcal{G}'_{\mathbf{x},\mathbf{z}} = \mathcal{K}_{\mathbf{x},\mathbf{z}} \cup \mathcal{M}_{\mathbf{x},\mathbf{z}}$ and the union is disjoint. If we define
  \begin{align*}
    & K_{\mathbf{x},\mathbf{z}} = \bigcup \{P_\mathcal{C} : \mathcal{C} \in \mathcal{K}_{\mathbf{x},\mathbf{z}}\} \subseteq P \text{ and} \\
    & M_{\mathbf{x},\mathbf{z}} = \bigcup \{P_\mathcal{C} : \mathcal{C} \in \mathcal{M}_{\mathbf{x},\mathbf{z}}\} \subseteq P,
  \end{align*}
  then we can also write $I_{\mathbf{x},\mathbf{z}} = K_{\mathbf{x},\mathbf{z}} \cup M_{\mathbf{x},\mathbf{z}}$ and again the union is disjoint.

  Now, by definition and by the expansion properties, every %$P_{\mathcal{C}} \subseteq \mathcal{C} \in \mathcal{K}_{\mathbf{x},\mathbf{z}}$ 
  $P_{\mathcal{C}} \subseteq K_{\mathbf{x},\mathbf{z}}$ is such that $|N(P_\mathcal{C})| \geq D|P_\mathcal{C}|/(1-R)$, so this holds for the whole $K_{\mathbf{x},\mathbf{z}}$, too (in $\mathcal{G}'_{\mathbf{x},\mathbf{z}}$ and \emph{a fortiori} in $\mathcal{G}$): 
  \begin{equation}
    \label{eq:marker}
    |N(K_{\mathbf{x},\mathbf{z}})| \geq \frac{D|K_{\mathbf{x},\mathbf{z}}|}{1-R}.
  \end{equation}
  Another useful observation is that $|\mathcal{M}_{\mathbf{x},\mathbf{z}}| \leq 1$; in other words, there cannot be more than one connected component whose parity-check equation set is ``big''. Indeed, each one of these sets is such that its neighborhood has size at least $Dn/(D+1)$. If there were two (or more) connected components in $\mathcal{M}_{\mathbf{x},\mathbf{z}}$, the union of these neighborhoods would exceed the size of the whole set of variable nodes of the Tanner graph itself, which is impossible.

We will consider separately the two cases $|\mathcal{M}_{\mathbf{x},\mathbf{z}}| = 0$ and $|\mathcal{M}_{\mathbf{x},\mathbf{z}}| = 1$ and split the summation into two parts:
\begin{align}
  \eqref{eq:variance_sum} & = \sum_{\substack{\mathbf{x},\mathbf{z} \in \Zn \cap B_{\zero,n}(\rho) \smallsetminus p\Zn \\ |J_{\mathbf{x},\mathbf{z}}| \leq n(1-R) - A(n)}} \left(\frac{1}{p}\right)^{2n(1-R)-|I_{\mathbf{x},\mathbf{z}}|-2|J_{\mathbf{x},\mathbf{z}}|} \nonumber \\
  \label{eq:M_0}
  & = \sum_{\substack{\mathbf{x},\mathbf{z} \in \Zn \cap B_{\zero,n}(\rho) \smallsetminus p\Zn \\ |J_{\mathbf{x},\mathbf{z}}| \leq n(1-R) - A(n) \\ |\mathcal{M}_{\mathbf{x},\mathbf{z}}| = 0 }} \left(\frac{1}{p}\right)^{2n(1-R)-|I_{\mathbf{x},\mathbf{z}}|-2|J_{\mathbf{x},\mathbf{z}}|} \\
  \label{eq:M_1}
  & \ \ \ + \sum_{\substack{\mathbf{x},\mathbf{z} \in \Zn \cap B_{\zero,n}(\rho) \smallsetminus p\Zn \\ |J_{\mathbf{x},\mathbf{z}}| \leq n(1-R) - A(n) \\ |\mathcal{M}_{\mathbf{x},\mathbf{z}}| = 1 }} \left(\frac{1}{p}\right)^{2n(1-R)-|I_{\mathbf{x},\mathbf{z}}|-2|J_{\mathbf{x},\mathbf{z}}|}.
\end{align}
A small remark before proceeding with the estimation of \eqref{eq:M_0} and \eqref{eq:M_1}: \emph{a priori}, we are summing also over the $\mathbf{x}$ and $\mathbf{z}$ such that $|I_{\mathbf{x},\mathbf{z}}|=0=|J_{\mathbf{x},\mathbf{z}}|$. This implies that $|T_{\mathbf{x},\mathbf{z}}|=n(1-R)$ and that
\begin{equation*}
  \prob\{\HH\mathbf{x}^T \equiv \mathbf{s}^T \bmod p, \HH\mathbf{z}^T \equiv \mathbf{s}^T \bmod p\} = \prob\{\HH\mathbf{x}^T \equiv \mathbf{s}^T \bmod p\} \prob\{\HH\mathbf{z}^T \equiv \mathbf{s}^T \bmod p\}.
\end{equation*}
The consequence is that in this particular case $\cov(X_{\mathbf{x}},X_{\mathbf{x}}) = 0$ and the actual contribution to the variance of these couples of $\mathbf{x}$ and $\mathbf{z}$ is null. Consequently, when needed and without loss of generality, we will restrict the sum to the case $|I_{\mathbf{x},\mathbf{z}}|+|J_{\mathbf{x},\mathbf{z}}| \neq 0$. We will recall this observation in the sequel.

\begin{enumerate}
\item If $|\mathcal{M}_{\mathbf{x},\mathbf{z}}| = 0$, then $I_{\mathbf{x},\mathbf{z}} = K_{\mathbf{x},\mathbf{z}}$ and $|N(I_{\mathbf{x},\mathbf{z}})| = |N(K_{\mathbf{x},\mathbf{z}})| \geq D|I_{\mathbf{x},\mathbf{z}}|/(1-R)$. Let us estimate in this context the number of $\mathbf{x}$ and $\mathbf{z}$ for a given value of $|J_{\mathbf{x},\mathbf{z}}| \leq n(1-R) - A(n) \leq n(1-R)/(D+1)$ in this case. $|J_{\mathbf{x},\mathbf{z}}|$ is ``small'' and the expansion properties imply that 
  \begin{equation*}
    |N(J_{\mathbf{x},\mathbf{z}})| \geq \frac{D|J_{\mathbf{x},\mathbf{z}}|}{1-R}.
  \end{equation*}
  By definition of $J_{\mathbf{x},\mathbf{z}}$, this implies that at least $D|J_{\mathbf{x},\mathbf{z}}|/(1-R)$ coordinates of $\mathbf{x}$ and $\mathbf{z}$ are fixed to $0$ (modulo $p$). Fixing these coordinates is equivalent to fixing the parity-check equations of $J_{\mathbf{x},\mathbf{z}}$ inside $P$. 

    On the other hand, what can we say about $\mathbf{z}$? Observe that, by definition, $\mathbf{x}_{\mathbf{h}}$ and $\mathbf{z}_{\mathbf{h}}$ are multiple modulo $p$ for every parity-check equation $\mathbf{h}$ that corresponds to a vertex of $I_{\mathbf{x},\mathbf{z}}$. Moreover, the condition $\lambda > 2R$, contained in \eqref{eq:lambda_typical_norm_LDA}, implies that $2\rho = o(p)$, which in turn implies that are no couples of integer points of $B_{\zero,n}(\rho)$ that are equivalent modulo $p$ (a shift of a simple coordinate modulo $p$ from a value to a different value in the same equivalence class is a shift of more than the diameter of the ball and brings the point out of it). Hence, for a fixed $\mathbf{x}$, the $\mathbf{z}$ that we take into account cannot take more than $p$ different values with respect to $\mathbf{x}$ in the coordinates that correspond to $N(I_{\mathbf{x},\mathbf{z}})$ (and we know that these coordinates are at least $D|I_{\mathbf{x},\mathbf{z}}|/(1-R)$). Fixing them is the same as fixing the parity-check equations of $I_{\mathbf{x},\mathbf{z}}$ inside $P$.

Putting together all of these observations%and noticing that $|I_{\mathbf{x},\mathbf{z}}| \leq n(1-R)-|J_{\mathbf{x},\mathbf{z}}|$
, we obtain that when $|\mathcal{M}_{\mathbf{x},\mathbf{z}}| = 0$,
    \begin{align}
      & |\{\mathbf{x},\mathbf{z} \in \Zn \cap B_{\zero,n}(\rho) \smallsetminus p\Zn : |I_{\mathbf{x},\mathbf{z}}| = i, |J_{\mathbf{x},\mathbf{z}}| = j\}| \nonumber \\
      & \leq \binom{n(1-R)}{j}|\Z^{n-Dj/(1-R)} \cap B_{\zero,n-Dj/(1-R)}(\rho)| \cdot \nonumber \\
      & \ \ \ \ \ \cdot \binom{n(1-R)}{i}p^i |\Z^{n-D(i+j)/(1-R)} \cap B_{\zero,n-D(i+j)/(1-R)}(\rho)| \nonumber \\
      \label{eq:bound_I_J}
      & \leq n^{(j+i)} p^i |\Z^{n-Dj/(1-R)} \cap B_{\zero,n-Dj/(1-R)}(\rho)||\Z^{n-D(i+j)/(1-R)} \cap B_{\zero,n-D(i+j)/(1-R)}(\rho)|.
     \end{align}
    
    Let us define the quantity
    \begin{equation}
      \label{eq:E_rho}
      \mathcal{E}(\rho) = \sum_{\mathbf{x},\mathbf{z} \in \Zn \cap B_{\zero,n}(\rho)} \left(\frac{1}{p}\right)^{2n(1-R)} = |\Zn \cap B_{\zero,n}(\rho)|^2 \left(\frac{1}{p}\right)^{2n(1-R)} \lesssim \E[X_\rho]^2.
    \end{equation}
    We will use it in the estimation of \eqref{eq:M_0}: 
    \begin{align}
      \eqref{eq:M_0} & = \sum_{\substack{\mathbf{x},\mathbf{z} \in \Zn \cap B_{\zero,n}(\rho) \smallsetminus p\Zn \\ |J_{\mathbf{x},\mathbf{z}}| \leq n(1-R) - A(n) \\ |\mathcal{M}_{\mathbf{x},\mathbf{z}}| = 0 }} \left(\frac{1}{p}\right)^{2n(1-R)-|I_{\mathbf{x},\mathbf{z}}|-2|J_{\mathbf{x},\mathbf{z}}|} \nonumber \\
      & = \sum_{\substack{\mathbf{x},\mathbf{z} \in \Zn \cap B_{\zero,n}(\rho) \smallsetminus p\Zn \\ |J_{\mathbf{x},\mathbf{z}}| \leq n(1-R) - A(n) \\ |\mathcal{M}_{\mathbf{x},\mathbf{z}}| = 0 }} \frac{\mathcal{E}(\rho)}{|\Zn \cap B_{\zero,n}(\rho)|^2} p^{|I_{\mathbf{x},\mathbf{z}}|+2|J_{\mathbf{x},\mathbf{z}}|} \nonumber \\
      & \leq \sum_{j=0}^{\lfloor n(1-R) - A(n) \rfloor}\sum_{i=0}^{n(1-R)-j} \sum_{\substack{\mathbf{x},\mathbf{z} \in \Zn \cap B_{\zero,n}(\rho) \smallsetminus p\Zn \\ |J_{\mathbf{x},\mathbf{z}}| = j, |I_{\mathbf{x},\mathbf{z}}| = i \\ |\mathcal{M}_{\mathbf{x},\mathbf{z}}| = 0}} \frac{\mathcal{E}(\rho)}{|\Zn \cap B_{\zero,n}(\rho)|^2} p^{i+2j} \nonumber \\
      \label{eq:half_sum_1}
      & \leq \sum_{j=0}^{\lfloor n(1-R)-A(n) \rfloor}\sum_{i=0}^{n(1-R)-j} \mathcal{E}(\rho) \frac{|\Z^{n-Dj/(1-R)} \cap B_{\zero,n-Dj/(1-R)}(\rho)|}{|\Zn \cap B_{\zero,n}(\rho)|} \cdot  \\
      \label{eq:half_sum_2}
      & \ \ \ \ \ \ \ \ \ \ \ \ \ \ \ \ \ \ \ \ \ \cdot \frac{|\Z^{n-D(i+j)/(1-R)} \cap B_{\zero,n-D(i+j)/(1-R)}(\rho)|}{|\Zn \cap B_{\zero,n}(\rho)|} n^{(j+i)} p^i p^{i+2j} \\
      \label{eq:fn_p_E}
      & \lesssim  \mathcal{E}(\rho) \sum_{j=0}^{\lfloor n(1-R) - A(n) \rfloor}\sum_{i=0}^{n(1-R)-j} f(n) p^{-D(2j+i)} n^{(j+i)} p^{2(i+j)},
    \end{align}
    where the last asymptotic inequality comes from Lemma \ref{lem:volume_ratio} and
    \begin{align*}
      f(n) & = \frac{(\sqrt{n})^{2(n+1)}}{(\sqrt{n-Dj/(1-R)})^{n-\frac{Dj}{1-R}+1}(\sqrt{n-D(i+j)/(1-R)})^{n-\frac{D(i+j)}{1-R}+1}}\cdot \\
      & \ \ \ \ \cdot \left(\sqrt{2\pi e}\right)^{-\frac{D(2j+i)}{1-R}} \left(1 + \frac{2\sqrt{n}}{2\rho-\sqrt{n}} \right)^{2n} \frac{\rho^{-\frac{D(2j+i)}{1-R}}}{p^{-D(2j+i)}} \\
      & \leq \frac{(\sqrt{n})^{2(n+1)-\frac{D(2j+i)}{1-R}}}{(\sqrt{n-Dj/(1-R)})^{n-\frac{Dj}{1-R}+1}(\sqrt{n-D(i+j)/(1-R)})^{n-\frac{D(i+j)}{1-R}+1}} \cdot \\
      & \ \ \ \ \cdot \left(1 + \frac{2\sqrt{n}}{2\rho-\sqrt{n}} \right)^{2n},
    \end{align*}
    recalling that
    \begin{equation*}
      \rho = \frac{\sqrt{n}p^{(1-R)}}{\sqrt{2 \pi e}} \left(1+\frac{1}{n^\omega}\right).
    \end{equation*}

    Let us go back to \eqref{eq:fn_p_E}: besides $f(n)$ and $\mathcal{E}(\rho)$, in the sum we have
    \begin{align*}
       p^{-Dj}p^{-D(j+i)} n^{(j+i)} p^{2(i+j)} = n^{-j \lambda D} n^{(j+i)(1-\lambda(D-2))}
    \end{align*}
    and the exponent is \emph{strictly} negative because \eqref{eq:further_condition_D} and \eqref{eq:lambda_typical_norm_LDA} impose that
    \begin{equation}
      \label{eq:condition_B_lambda}
      D > 2 \text{\ \ \ \ \ and\ \ \ \ \ } \lambda > \frac{1}{D-2}
    \end{equation}
    (recall also that, as previously explained, we do not take into consideration the case $j+i = 0$).
    %, because the contribution to the variance of couples of $\mathbf{x}$ and $\mathbf{z}$ corresponding to $i=j=0$ is actually $0$).

    What can we say about $f(n)$? First of all that
    \begin{align*}
      \left(1 + \frac{2\sqrt{n}}{2\rho-\sqrt{n}} \right)^{2n} \leq \left(1+\frac{2\sqrt{2\pi e}}{2p^{(1-R)}-\sqrt{2 \pi e}}\right)^{2n} \to 1,
    \end{align*}
    because we have imposed that $\lambda > (1-R)^{-1}$, always by \eqref{eq:lambda_typical_norm_LDA}. 
    %Moreover,
    %\begin{equation}
    %  \left(1+\frac{1}{n^\omega}\right)^{-B(2j+i)} \leq 1.
    %\end{equation}
    Now, consider the term
    \begin{equation*}
      f_j(n) = \left(\sqrt{\frac{n}{n-Dj/(1-R)}}\right)^{n-\frac{Dj}{1-R}+1} = \left(1 + \frac{Dj/(1-R)}{n-Dj/(1-R)}\right)^{\frac{n}{2}-\frac{Dj}{2(1-R)}+\frac{1}{2}};
    \end{equation*}
    it is easy to show that if $j \neq 0$
    \begin{equation*}
      f_j(n)n^{-j \lambda D} = o(1),
    \end{equation*}
    otherwise it is $1$. Similarly, defining %if $i \neq 0$,
    \begin{align*}
      f_{i+j}(n) & = \left(\sqrt{\frac{n}{n-D(i+j)/(1-R)}}\right)^{n-\frac{D(i+j)}{1-R}+1} \\
      & = \left(1 + \frac{D(i+j)/(1-R)}{n-D(i+j)/(1-R)}\right)^{\frac{n}{2}-\frac{D(i+j)}{2(1-R)}+\frac{1}{2}},
    \end{align*}
    we have
    \begin{equation*}
      f_{i+j}(n)n^{(i+j)(1-\lambda(D-2))} = o(1),
    \end{equation*}
    never equal to $1$ under our assumption that $i+j \neq 0$. As a consequence, 
    \begin{equation*}
      f(n) p^{-D(2j+i)} n^{(j+i)} p^{2(i+j)} = f_j(n)f_{i+j}(n) n^{-j \lambda D} n^{(j+i)(1-\lambda(D-2))} = o(1).
    \end{equation*}
    Furthermore, we will not perform it here in all details, but a more precise analysis of the series in \eqref{eq:fn_p_E} shows that \eqref{eq:condition_B_lambda} is actually sufficient to conclude that
    \begin{equation}
      \label{eq:second_inequality_E}
      \mathcal{E}(\rho) \sum_{j=0}^{\lfloor n(1-R)-A(n) \rfloor}\sum_{\substack{i=0 \\ (i,j)\neq(0,0)}}^{n(1-R)-j} f(n) p^{-D(2j+i)} n^{(j+i)} p^{2(i+j)} \lesssim o(1) \mathcal{E}(\rho).
    \end{equation}
    %COMMENT TO BE ERASED: to solve the series above, see it as a double geometric series like
    %\begin{equation*}
    %  \sum_j o(n^{-C})^j \sum_i o(n^{-Cj})^i.
    %\end{equation*}
    We will need this inequality later, after the estimation of the variance for the case $|\mathcal{M}_{\mathbf{x},\mathbf{z}}| = 1$.
  
  \item If $|\mathcal{M}_{\mathbf{x},\mathbf{z}}| = 1$, then the graph $\mathcal{G}'_{\mathbf{x},\mathbf{z}}$ contains a ``big'' connected component and $|M_{\mathbf{x},\mathbf{z}}| > n(1-R)/(D+1)$, which implies by the expansion properties that
    \begin{equation}
      \label{eq:neighbor_M}
      |N(I_{\mathbf{x},\mathbf{z}} \cup J_{\mathbf{x},\mathbf{z}})| \geq |N(I_{\mathbf{x},\mathbf{z}})| \geq |N(M_{\mathbf{x},\mathbf{z}})| \geq \frac{Dn}{D+1}.
    \end{equation}
    If we call $R_{\mathbf{x},\mathbf{z}} = V \smallsetminus N(I_{\mathbf{x},\mathbf{z}} \cup J_{\mathbf{x},\mathbf{z}})$, we have that $|R_{\mathbf{x},\mathbf{z}}| \leq n/(D+1)$. Moreover, $N(R_{\mathbf{x},\mathbf{z}}) \subseteq T_{\mathbf{x},\mathbf{z}}$ and the expansion properties of the graph guarantee that $|N(R_{\mathbf{x},\mathbf{z}})| \geq D(1-R)|R_{\mathbf{x},\mathbf{z}}|$, from which we deduce that 
    \begin{equation*}
      |T_{\mathbf{x},\mathbf{z}}| + |J_{\mathbf{x},\mathbf{z}}| \geq |T_{\mathbf{x},\mathbf{z}}| \geq |N(R_{\mathbf{x},\mathbf{z}})| \geq D(1-R)|R_{\mathbf{x},\mathbf{z}}|.
    \end{equation*}
    These considerations will help us in counting the number of $\mathbf{x}$ and $\mathbf{z}$ such that $|J_{\mathbf{x},\mathbf{z}}| \leq n(1-R) - A(n) \leq n(1-R)/(D+1)$ and $|\mathcal{M}_{\mathbf{x},\mathbf{z}}| = 1$. First of all, the same argument of the case $|\mathcal{M}_{\mathbf{x},\mathbf{z}}| = 0$ holds: at least $D|J_{\mathbf{x},\mathbf{z}}|/(1-R)$ of the coordinates of $\mathbf{x}$ and $\mathbf{z}$ are fixed to be $0$ (modulo $p$) and these coordinates are identified by the parity-check equations in $J_{\mathbf{x},\mathbf{z}}$. Concerning $\mathbf{z}$, given a fixed $\mathbf{x}$, its coordinates are fixed to $0$ in the neighborhood of $J_{\mathbf{x},\mathbf{z}}$ and can take up to $p$ different values in the neighborhood of $I_{\mathbf{x},\mathbf{z}}$ (these values are the multiples modulo $p$ of the coordinates of $\mathbf{x}$). This allows us to conclude that
    \begin{align}
      & |\{\mathbf{x},\mathbf{z} \in \Zn \cap B_{\zero,n}(\rho) \smallsetminus p\Zn : |J_{\mathbf{x},\mathbf{z}}|=j, |T_{\mathbf{x},\mathbf{z}}|=t, |R_{\mathbf{x},\mathbf{z}}|=r, |K_{\mathbf{x},\mathbf{z}}|=k\}| \nonumber \\
      & \leq \binom{n(1-R)}{j}|\Z^{n-\frac{Dj}{1-R}} \cap B_{\zero,n-Dj/(1-R)}(\rho)| \binom{n(1-R)}{t}p^{k+1} |\Z^{r} \cap B_{\zero,r}(\rho)| \nonumber \\
      \label{eq:erre_kappa}
      & \leq n^{j+t} p^{k+1+r} |\Z^{n-\frac{Dj}{1-R}} \cap B_{\zero,n-Dj/(1-R)}(\rho)|.
    \end{align}
    We are always implicitly using the fact that $2\rho = o(p)$ and that a fixed coordinate of an integer point inside a ball of radius $\rho$ cannot take more than $p$ different values (from which we get, for example, the crude estimation: $|\Z^{r} \cap B_{\zero,r}(\rho)| \leq p^r$).

    Now, we would like to estimate $k = |K_{\mathbf{x},\mathbf{z}}|$. By definition of $K_{\mathbf{x},\mathbf{z}}$ and $M_{\mathbf{x},\mathbf{z}}$, we have that $N(K_{\mathbf{x},\mathbf{z}}) \subseteq V \smallsetminus N(M_{\mathbf{x},\mathbf{z}})$; moreover, \eqref{eq:neighbor_M} tells us that $|N(M_{\mathbf{x},\mathbf{z}})| \geq Dn/(D+1)$. This implies that $|N(K_{\mathbf{x},\mathbf{z}})| \leq n-|N(M_{\mathbf{x},\mathbf{z}})| \leq n/(D+1)$. Then, by the expansion properties, $|N(N(K_{\mathbf{x},\mathbf{z}}))| \geq D(1-R) |N(K_{\mathbf{x},\mathbf{z}})|$. Notice that $K_{\mathbf{x},\mathbf{z}}$ is ``small'' by definition and, thanks to the expansion properties, we have that $|N(K_{\mathbf{x},\mathbf{z}})| \geq D|K_{\mathbf{x},\mathbf{z}}|/(1-R)$. Since $N(N(K_{\mathbf{x},\mathbf{z}})) \subseteq P \smallsetminus M_{\mathbf{x},\mathbf{z}} = J_{\mathbf{x},\mathbf{z}} \cup K_{\mathbf{x},\mathbf{z}} \cup T_{\mathbf{x},\mathbf{z}}$, we deduce that
    \begin{equation*}
      |J_{\mathbf{x},\mathbf{z}}|+|K_{\mathbf{x},\mathbf{z}}|+|T_{\mathbf{x},\mathbf{z}}| \geq |N(N(K_{\mathbf{x},\mathbf{z}}))| \geq D(1-R) |N(K_{\mathbf{x},\mathbf{z}})| \geq D^2|K_{\mathbf{x},\mathbf{z}}|,
    \end{equation*}
    or, equivalently,
    \begin{equation*}
      |K_{\mathbf{x},\mathbf{z}}| \leq \frac{|J_{\mathbf{x},\mathbf{z}}|+|T_{\mathbf{x},\mathbf{z}}|}{D^2-1}.
    \end{equation*}
    If we apply this estimation to \eqref{eq:erre_kappa}, also recalling that $|J_{\mathbf{x},\mathbf{z}}| + |T_{\mathbf{x},\mathbf{z}}| \geq D(1-R)|R_{\mathbf{x},\mathbf{z}}|$, we obtain:
    \begin{align}
      \label{eq:case_M_1_j_small_a}
      &|\{\mathbf{x},\mathbf{z} \in \Zn \cap B_{\zero,n}(\rho) \smallsetminus p\Zn : |J_{\mathbf{x},\mathbf{z}}|=j, |T_{\mathbf{x},\mathbf{z}}|=t\}|\\
      \label{eq:case_M_1_j_small_b}
      & \ \ \ \ \leq n^{t+j} p^{(t+j)/(D^2-1)+1} p^{(t+j)/(D(1-R))}|\Z^{n-\frac{Dj}{1-R}} \cap B_{\zero,n-Dj/(1-R)}(\rho)|.
    \end{align}
    
    We can now go back to the main estimation and, again, introduce the quantity $\mathcal{E}(\rho)$:
    \begin{align}
      \eqref{eq:M_1} & = \sum_{\substack{\mathbf{x},\mathbf{z} \in \Zn \cap B_{\zero,n}(\rho) \smallsetminus p\Zn \\ |J_{\mathbf{x},\mathbf{z}}| \leq n(1-R) - A(n) \\ |\mathcal{M}_{\mathbf{x},\mathbf{z}}| = 1 }} \left(\frac{1}{p}\right)^{2n(1-R)-|I_{\mathbf{x},\mathbf{z}}|-2|J_{\mathbf{x},\mathbf{z}}|} \nonumber \\
      & = \sum_{\substack{\mathbf{x},\mathbf{z} \in \Zn \cap B_{\zero,n}(\rho) \smallsetminus p\Zn \\ |J_{\mathbf{x},\mathbf{z}}| \leq n(1-R)-A(n) \\ |\mathcal{M}_{\mathbf{x},\mathbf{z}}| = 1 }} \left(\frac{1}{p}\right)^{n(1-R)-2|J_{\mathbf{x},\mathbf{z}}|+|T_{\mathbf{x},\mathbf{z}}|+|J_{\mathbf{x},\mathbf{z}}|} \nonumber \\
      & = \sum_{j=0}^{\lfloor n(1-R)-A(n) \rfloor}\sum_{t=0}^{n(1-R)-j} \sum_{\substack{\mathbf{x},\mathbf{z} \in \Zn \cap B_{\zero,n}(\rho) \smallsetminus p\Zn \\ |J_{\mathbf{x},\mathbf{z}}| = j, |T_{\mathbf{x},\mathbf{z}}| = t \\ |\mathcal{M}_{\mathbf{x},\mathbf{z}}| = 1}} \left(\frac{1}{p}\right)^{n(1-R)-2j+t+j} \nonumber \\
      & \leq \sum_{j=0}^{\lfloor n(1-R)-A(n) \rfloor}\sum_{t=0}^{n(1-R)-j} n^{t+j} p^{(t+j)/(D^2-1)+1} p^{(t+j)/(D(1-R))} \cdot \nonumber \\
      & \ \ \ \ \ \ \ \ \ \ \ \ \ \ \ \ \ \cdot \frac{|\Z^{n-\frac{Dj}{1-R}} \cap B_{\zero,n-Dj/(1-R)}(\rho)|}{|\Z^n \cap B_{\zero,n}(\rho)|} p^{2j-t-j} \sqrt{\mathcal{E}(\rho)} \nonumber \\
      \label{eq:case_M_1_j_small}
      & \lesssim \sum_{j=0}^{\lfloor n(1-R)-A(n) \rfloor}\sum_{t=0}^{n(1-R)-j} \frac{g(n) p}{\sqrt{\mathcal{E}(\rho)}} \left(\frac{p^{2}}{p^D}\right)^j \left(\frac{n p^{1/(D^2-1)} p^{1/(D(1-R))}}{p}\right)^{t+j} \mathcal{E}(\rho),
    \end{align}
    where we have applied Lemma \ref{lem:volume_ratio} to obtain the latter asymptotic estimation and $g(n)$ is the analogue of $f(n)$:
    \begin{equation*}
      g(n) = \left(\sqrt{\frac{n}{n-Dj/(1-R)}}\right)^{n-\frac{Dj}{1-R}+1} \left(1 + \frac{2\sqrt{n}}{2\rho-\sqrt{n}} \right)^n \left(1+\frac{1}{n^\omega}\right)^{-\frac{Dj}{1-R}}.
    \end{equation*}
    Now, very similarly to what happens in the case $|\mathcal{M}_{\mathbf{x},\mathbf{z}}| = 0$ (we omit the details), conditions
    \begin{equation*}
      D > 2 \text{\ \ \ \ \ and\ \ \ \ \ } \lambda > \left(1-\frac{1}{D^2-1}-\frac{1}{D(1-R)}\right)^{-1},
    \end{equation*}
    implied by \eqref{eq:further_condition_D} and \eqref{eq:lambda_typical_norm_LDA}, allow us to deduce that
    \begin{equation}
      \label{eq:third_inequality_E}
      \eqref{eq:case_M_1_j_small} \lesssim o(1)\mathcal{E}(\rho).
    \end{equation}
    Notice that from \eqref{eq:E_rho} the quantity $\mathcal{E}(\rho)$ is known to tend at least subexponentially to infinity when $n$ grows and so does its square root.
\end{enumerate}

We are finally very close to the end of the proof. Starting from \eqref{eq:variance_sum}, putting together \eqref{eq:second_inequality_E} and \eqref{eq:third_inequality_E}, we obtain that
\begin{equation*}
  \var(X_{\rho}) \lesssim o(1)\mathcal{E}(\rho).
\end{equation*}
By the means of the Chebyshev's inequality %(cf. Lemma \ref{lem:chebyshev_inequality}) 
and since $\mathcal{E}(\rho) \leq \E[X_\rho]^2$, we can conclude:
\begin{align*}
  \prob\{X_{\rho} = 0\} & \leq \prob\{|X_{\rho} - \E[X_{\rho}]| \geq \E[X_{\rho}]\} \\
  & \leq \frac{\var(X_{\rho})}{\E[X_{\rho}]^2} \\
  & \lesssim \frac{o(1)\mathcal{E}(\rho)}{\E[X_{\rho}]^2} \\
  & \lesssim \frac{o(1)\E[X_{\rho}]^2}{\E[X_{\rho}]^2} \longrightarrow 0,
\end{align*}
that is,
\begin{equation*}
  \lim_{n\to \infty} \prob \left\{X_{\rho_{\eff}\left(1+\frac{1}{n^\omega}\right)} = 0 \right\} = 0.
\end{equation*}
\end{IEEEproof}

\subsection{The proof that capacity is achieved with LDA lattices}%------------------------- SUBSUBSECTION - MAIN RESULT ------------------
\label{sec:main_theorem}
Now that we have proved that in the case of LDA Voronoi constellations the sent point has the same typical norm of the constellation points of the more general Construction A, we are ready to prove the result that LDA lattices can achieve the capacity of the AWGN channel under MMSE lattice decoding. We repeat that the transmission scheme is the same of Section \ref{sec:encoding_and_decoding} and the proof of the theorem is then very similar to the one of Theorem \ref{thm:awgn_capacity}. Nevertheless, we will have to adapt it to the LDPC structure that gives rise to LDA lattices, just like we had to adapt the proof of the previous lemma.

\begin{theorem}
  \label{thm:awgn_capacity_LDA}
  Fix $1 > R_f > R > 1/2$ and a constant $D$ such that
  \begin{equation*}
    D > \frac{1}{1-R_f}.
  \end{equation*}
  Choose a degree $\Delta_P$ that satisfies \eqref{eq:bound_Delta}: 
  \begin{equation*}
    \Delta_P > \max \left \{ \frac{2-R_f}{1-R_f}\left(1-\frac{Dh\left(\frac{1}{D}\right)}{(D+1)h\left(\frac{1}{D+1}\right)}\right)^{-1}, \frac{D^2}{1-R_f} + 1\right\}.
  \end{equation*}
  If $p = n^{\lambda}$, with
  \begin{equation}
    \label{eq:lambda_Rf_LDA}
    \lambda > \max \left\{ \frac{1}{D(1-R_f)-1}, \frac{1}{1-R_f}, \left(1-\frac{1}{D^2-1}-\frac{1}{D(1-R_f)}\right)^{-1} \right\},
  \end{equation} 
  then the random ensemble of nested LDA lattices presented in Section \ref{sec:random_LDA_ensemble} achieves capacity of the AWGN channel under MMSE lattice decoding, when $\snr > 1$.
\end{theorem}

Remark: the proof of this theorem strongly relies on the techniques that we have already applied in the proofs of Theorem \ref{thm:awgn_capacity} and Lemma \ref{lem:typical_norm_LDA}. For this reason, we will skip some details and some technical computations that would have the disadvantage of making it much longer and less readable. Everything which is not completely developed is a straightforward modification of some well-referenced computations that were previously carried out. We strongly recommend to get familiar with the arguments used in the demonstrations of Theorem \ref{thm:awgn_capacity} and Lemma \ref{lem:typical_norm_LDA} before reading the sequel in depth.

\begin{IEEEproof}
  The geometric and probabilistic strategy to prove this theorem is the same that we have applied to prove Theorem \ref{thm:awgn_capacity}. Namely, the beginnings of the two proofs are identical and almost everything coincides; the small differences can be easily solved by a slight adaptation of what is done in the proof of Theorem \ref{thm:awgn_capacity}. For this reason, we claim that the only thing that we need to prove is that
  \begin{equation}
    \label{eq:target_2}
    \begin{aligned}
      \lim_{n \to \infty} \Bigg( & \sum_{\mathbf{x} \in S} \prob \{\HH'\mathbf{x}^T \equiv \mathbf{m}^T \bmod p\} \cdot \\
      & \ \cdot\sum_{\substack{\mathbf{z} \in \B' \cap \Zn \\\mathbf{z} \not \equiv \mu \mathbf{x},\ \mu =0,1,2}} \prob \{\HH_f\mathbf{x}^T \equiv \zero^T \bmod p, \HH_f\mathbf{z}^T \equiv \zero^T \bmod p\} \prob\{ \mathbf{z} \in \B\} \Bigg) = 0.
    \end{aligned}
  \end{equation}
  This formula is the LDA-equivalent of \eqref{eq:same_limit}. For the notation, we recall that:
  \begin{itemize}
  \item $\B_{\eff}$ is the $n$-dimensional ball centered at $\zero$ with radius
    \begin{equation*}
      %\label{eq:rho}
      \rho = \rho_{\eff}\left(1+\frac{1}{n^\omega}\right) = \frac{\sqrt{n}p^{(1-R)}}{\sqrt{2 \pi e}}\left(1+\frac{1}{n^\omega}\right),
    \end{equation*}
    where $\omega$ is the same constant of Lemma \ref{lem:typical_norm_LDA}.
  \item $\mathbf{m}$ is the non-zero part of the syndrome $\mathbf{s} = (\mathbf{m}\ |\ \zero)$ (see also Fig.~\ref{fig:enc_dec_scheme} at the beginning of Section \ref{sec:capacity_for_random_construction_A}).
  \item $\B'$ is the $n$-dimensional ball centered at $\mathbf{x}$, with radius $2\rho_{\dec}$ equal to twice the radius of the decoding sphere $\B = B_{\alpha \mathbf{y},n}(\rho_{\dec})$:
    \begin{equation*}
      2 \rho_{\dec} = 2\sqrt{n}p^{(1-R_f)}(1-\delta)(1+\varepsilon)/\sqrt{2 \pi e},
    \end{equation*}
    where $\delta$ is the constant that ``represents'' the distance between the constellation rate and capacity and $\varepsilon$ is a positive constant that can be taken as small as wanted (compare with \eqref{eq:alpha_sigma_max} and what follows).
    %\item $\sigma = \sigma_{\max}(1-\delta)$, for some $0<\delta<1$ and $\sigma_{\max}$ is defined like in \eqref{eq:new_sigma_max}.
    %\item $\B$ is the \emph{decoding sphere}, centered at $\alpha \mathbf{y}$ (the MMSE-scaled channel output) with radius $\rho_{\dec}$.
  \item $S$ is defined as in \eqref{eq:S}:
    \begin{equation*}
      S = \{\mathbf{x} \in (\B_{\eff}\cap \Zn) \smallsetminus p\Zn : \mathbf{z} \equiv \mu\mathbf{x} \bmod p \text{ produces no error}, \forall \mu \in \{0,1,2\}\}. 
    \end{equation*}
  \end{itemize}

  First of all, let us deduce something about the non-zero subsyndrome $\mathbf{m}$: how many are the $\mathbf{m} \in \Fp^{n(R_f-R)}$ such that $m_i \neq 0$ for every $i$? We have:
  \begin{align*}
    |\{ \mathbf{m} \in \Fp^{n(R_f-R)} : m_i \neq 0, \forall i\}| & = (p-1)^{n(R_f-R)} \\
    & = \left(1 - \frac{1}{p}\right)^{n(R_f-R)} p^{n(R_f-R)} \\
    & = \left(1 - \frac{1}{n^\lambda}\right)^{n(R_f-R)} p^{n(R_f-R)} \\
    & \sim p^{n(R_f-R)},
  \end{align*}
  because $\lambda > 1$ as a consequence of \eqref{eq:lambda_Rf_LDA}. This means that the proportion of $\mathbf{m}$ that contain some zero coordinates is vanishing with respect to the total number of subsyndromes. For this reason, the contribution to the average error probability of this messages is vanishing and we only need to show \eqref{eq:target_2} for the $\mathbf{m}$ such that $m_i \neq 0$ for every $i$. From now on, we make this hypothesis, which implies that
  \begin{equation*}
    \prob \{\HH'\mathbf{x}^T \equiv \mathbf{m}^T \bmod p\} = \left( \frac{1}{p} \right)^{n(R_f-R)},
  \end{equation*}
  since the intersection of the supports of $\mathbf{x}$ and any row of $\HH'$ is never empty. %Note that, if the inequality is strict, then the probability is $0$.
    
  Now, we would like to express the probabilities of \eqref{eq:target_2} that $\mathbf{x}$ and $\mathbf{z}$ have a certain subsyndrome in the same form as in the proof of Lemma \ref{lem:typical_norm_LDA}. For this purpose, given a fixed $\mathbf{x}$ and a fixed $\mathbf{z}$, let
  \begin{align*}
    & J_{\mathbf{x},\mathbf{z}}^f = \{ \mathbf{h} \text{ row of } \HH_f : \dim(\mathbf{x},\mathbf{z}|\mathbf{h}) = 0 \}, \\
    & I_{\mathbf{x},\mathbf{z}}^f = \{ \mathbf{h} \text{ row of } \HH_f : \dim(\mathbf{x},\mathbf{z}|\mathbf{h}) = 1 \}, \\
    & T_{\mathbf{x},\mathbf{z}}^f = \{ \mathbf{h} \text{ row of } \HH_f : \dim(\mathbf{x},\mathbf{z}|\mathbf{h}) = 2 \},
  \end{align*}
  where the definition of $\dim(\mathbf{x},\mathbf{z}|\mathbf{h})$ is the same that we have given in the proof of Lemma \ref{lem:typical_norm_LDA} (see also \eqref{eq:J_x_z}, \eqref{eq:I_x_z}, and \eqref{eq:T_x_z}). We will employ the very same expansion arguments used in the proof of Lemma \ref{lem:typical_norm_LDA}, but this time applied to the $D$-good Tanner graph associated with $\HH_f$, instead of $\HH$. From now on, we will call $V$ its set of variable nodes and $P$ its set of check nodes. Furthermore, notice that \eqref{eq:bound_Delta} is assumed in order to guarantee that both of them are $D$-good, as anticipated in Section \ref{sec:random_LDA_ensemble}. First of all, we can argue like we did from \eqref{eq:J_big} to \eqref{eq:prob_x_z_lambda_f_0} to claim that
  \begin{equation*}
    \prob\{\HH_f\mathbf{x}^T \equiv \zero^T \bmod p, \HH_f\mathbf{z}^T \equiv \zero^T \bmod p\} = 0
  \end{equation*}
  for every couple of $\mathbf{x}$ and $\mathbf{z}$ such that
  \begin{equation*}
    |J_{\mathbf{x},\mathbf{z}}| \geq n(1-R) - B(n), \text{\ \ \ with\ \ \ } B(n) = n(1-R_f)\left(\frac{D^2+D(1-R_f)-1}{D(D+1)}\right).
  \end{equation*}
  Thus, if we define for a fixed $\mathbf{x}$ the set
  \begin{equation*}
    Z = \{ \mathbf{z} \in \B'\cap\Zn : \mathbf{z} \not \equiv \mu \mathbf{x} \bmod p, \forall \mu\in \{0,1,2\},\ \text{and\ } |J_{\mathbf{x},\mathbf{z}}| \leq n(1-R) - B(n)\},
  \end{equation*}
  we can compute $\prob \{\HH_f\mathbf{x}^T \equiv \zero^T \bmod p, \HH_f\mathbf{z}^T \equiv \zero^T \bmod p\}$ analogously to \eqref{eq:prob_I_J_T} and obtain that the sum in \eqref{eq:target_2} is equal to
  \begin{align}
    & \sum_{\mathbf{x} \in S} \prob \{\HH'\mathbf{x}^T \equiv \mathbf{m}^T \bmod p\} \sum_{\mathbf{z} \in Z} \prob \{\HH_f\mathbf{x}^T \equiv \zero^T \bmod p, \HH_f\mathbf{z}^T \equiv \zero^T \bmod p\} \prob\{ \mathbf{z} \in \B\} \nonumber \\
    \label{eq:reference_for_small_T}
    & \ \ \ \ \ = \sum_{\mathbf{x} \in S} \left(\frac{1}{p}\right)^{n(R_f-R)} \sum_{\mathbf{z} \in Z} \prob \{\HH_f\mathbf{x}^T \equiv \zero^T \bmod p, \HH_f\mathbf{z}^T \equiv \zero^T \bmod p\} \prob\{ \mathbf{z} \in \B\}
    %& \leq \sum_{\substack{i,j,t \\ i + j + t = n(1-R_f)}} \sum_{\substack{\mathbf{x} \in S \\ \mathbf{z} \in Z \\ |I_{\mathbf{x},\mathbf{z}}^f|=i, |J_{\mathbf{x},\mathbf{z}}^f|=j,|T_{\mathbf{x},\mathbf{z}}^f|=t}} \left(\frac{1}{p}\right)^{n(R_f-R)} \left(\frac{1}{p}\right)^{2n(1-R_f)-i-2j} \prob\{ \mathbf{z} \in \B \smallsetminus \{\mathbf{x}\}\}.
  \end{align}

  From now on, we take inspiration from the proof of Lemma \ref{lem:typical_norm_LDA} %. There, the ``second case'' was split into two more cases. We will do the same here 
  and bound \eqref{eq:reference_for_small_T} in two different ways, depending on the fact that $\mathcal{M}_{\mathbf{x},\mathbf{z}}^f$ is equal to $0$ or $1$. The definition of $\mathcal{M}_{\mathbf{x},\mathbf{z}}^f$ corresponds to the definition of $\mathcal{M}_{\mathbf{x},\mathbf{z}}$ in the proof of Lemma \ref{lem:typical_norm_LDA} (cf. \eqref{eq:definition_of_M}); the only difference is that all the graph-theoretical arguments are based on the Tanner graph associated with $\HH_f$ instead of $\HH$. Nonetheless, all definitions can be straight transposed to the present setting and do not need to be repeated.
\begin{enumerate}
\item Let us suppose that $|\mathcal{M}_{\mathbf{x},\mathbf{z}}^f|=0$. Notice that the terms of \eqref{eq:reference_for_small_T} corresponding to this case are upper bounded as follows:
  \begin{align}
    \label{eq:even_before_cases}
    & \sum_{\mathbf{x} \in S} \left(\frac{1}{p}\right)^{n(R_f-R)} \sum_{\substack{\mathbf{z} \in Z \\ |\mathcal{M}_{\mathbf{x},\mathbf{z}}^f|=0}} \prob \{\HH_f\mathbf{x}^T \equiv \zero^T \bmod p, \HH_f\mathbf{z}^T \equiv \zero^T \bmod p\} \prob\{ \mathbf{z} \in \B\} \\
    & \ \ \ \ \leq \sum_{\substack{\mathbf{x} \in S,\ \mathbf{z} \in Z \\ |\mathcal{M}_{\mathbf{x},\mathbf{z}}^f|=0}} \left(\frac{1}{p}\right)^{n(R_f-R)} \left(\frac{1}{p}\right)^{2n(1-R_f)-|I_{\mathbf{x},\mathbf{z}}^f| - 2|J_{\mathbf{x},\mathbf{z}}^f|} \prob\{ \mathbf{z} \in \B \} \nonumber \\
    \label{eq:before_cases}
    & \ \ \ \ \leq \sum_{j=0}^{\lfloor n(1-R_f) - B(n)\rfloor} \sum_{i = 0}^{n(1-R_f)-j} \left(\frac{1}{p}\right)^{n(R_f-R)} \left(\frac{1}{p}\right)^{2n(1-R_f)-i - 2j} \sum_{\substack{\mathbf{x}\in S,\ \mathbf{z}\in Z \\ |\mathcal{M}_{\mathbf{x},\mathbf{z}}^f|=0 \\ |J_{\mathbf{x},\mathbf{z}}^f| = j,\ |I_{\mathbf{x},\mathbf{z}}^f| = i}} \prob\{ \mathbf{z} \in \B \}.
  \end{align}
  
  Now, using the very same notation of the computation that led from \eqref{eq:sum_balls} to \eqref{eq:inequality_balls}, we can write:
  \begin{align*}
    \sum_{\substack{\mathbf{x} \in S,\ \mathbf{z} \in Z \\ |\mathcal{M}_{\mathbf{x},\mathbf{z}}^f|=0 \\ |J_{\mathbf{x},\mathbf{z}}^f| = j,\ |I_{\mathbf{x},\mathbf{z}}^f| = i}} \prob\{ \mathbf{z} \in \B\} & = \sum_{\substack{\mathbf{x} \in S,\ \mathbf{z} \in Z \\ |\mathcal{M}_{\mathbf{x},\mathbf{z}}^f|=0 \\ |J_{\mathbf{x},\mathbf{z}}^f| = j,\ |I_{\mathbf{x},\mathbf{z}}^f| = i}} \prob\{ \alpha\mathbf{w} \in B_{\mathbf{z}-\alpha \mathbf{x},n}(\rho_{\dec}) \} %\sum_{\substack{B \in \mathcal{S} \\ \mathbf{z}\in B}}\prob\{ \B = B\} 
    \nonumber \\
    %\label{eq:intermediate_PB}
    & \leq \int_{B} \sum_{\substack{\mathbf{x} \in S,\ \mathbf{z} \in B \cap Z \\ |\mathcal{M}_{\mathbf{x},\mathbf{z}}^f|=0 \\ |J_{\mathbf{x},\mathbf{z}}^f| = j,\ |I_{\mathbf{x},\mathbf{z}}^f| = i}} p(w) \mathrm{d}w \\
    & \leq Z_{ij},
    %& = \sum_{B\in \mathcal{S}} \ \ \  \sum_{\substack{\mathbf{x} \in S,\ \mathbf{z} \in B \cap Z \\ |\mathcal{M}_{\mathbf{x},\mathbf{z}}^f|=0 \\ |J_{\mathbf{x},\mathbf{z}}^f| = j,\ |I_{\mathbf{x},\mathbf{z}}^f| = i}} \prob\{ \B = B\}.
  \end{align*}
  where we define %Let us define the quantity
  \begin{equation*}
    %Z_{ij} = \sup_{B \in \mathcal{S}}|\{(\mathbf{x},\mathbf{z}) \in S \times (B \cap Z) : |\mathcal{M}_{\mathbf{x},\mathbf{z}}^f|=0, |I_{\mathbf{x},\mathbf{z}}^f|=i, |J_{\mathbf{x},\mathbf{z}}^f|=j\}|;
    Z_{ij} = |\{(\mathbf{x},\mathbf{z}) \in S \times (B \cap Z) : |\mathcal{M}_{\mathbf{x},\mathbf{z}}^f|=0, |I_{\mathbf{x},\mathbf{z}}^f|=i, |J_{\mathbf{x},\mathbf{z}}^f|=j\}|.
  \end{equation*}
  %it is clear that
  %\begin{equation*}
  %  \eqref{eq:intermediate_PB} \leq \sum_{B \in \mathcal{S}} Z_{ij} \prob\{ \B = B\} = Z_{ij}
  %\end{equation*}
  %and
  Hence,
  \begin{equation}
    \label{eq:sum_Z_ij}
    \eqref{eq:before_cases} \leq \sum_{j=0}^{\lfloor n(1-R_f) - B(n)\rfloor} \sum_{i = 0}^{n(1-R_f)-j} Z_{ij} \left(\frac{1}{p}\right)^{n(R_f-R)} \left(\frac{1}{p}\right)^{2n(1-R_f)-i - 2j}.
  \end{equation}
  A straightforward adaptation of the arguments used in the proof of Lemma \ref{lem:typical_norm_LDA} for the estimation of $|N(J_{\mathbf{x},\mathbf{z}})|$ and $|N(I_{\mathbf{x},\mathbf{z}})|$ %for the corresponding case $|\mathcal{M}_{\mathbf{x},\mathbf{z}}|=0$ 
  says that
  \begin{equation*}
    |N(J_{\mathbf{x},\mathbf{z}}^f)| \geq \frac{D|J_{\mathbf{x},\mathbf{z}}^f|}{1-R_f} \text{\ \ \ and\ \ \ } |N(I_{\mathbf{x},\mathbf{z}}^f)| \geq \frac{D|I_{\mathbf{x},\mathbf{z}}^f|}{1-R_f}.
  \end{equation*}
  %\begin{align}
  %  \label{eq:case_M_0}
  %  & \sum_{\substack{\mathbf{x} \in S,\ \mathbf{z} \in Z \\ |J_{\mathbf{x},\mathbf{z}}^f| \leq n(1-R_f)/(D+1) \\ |\mathcal{M}_{\mathbf{x},\mathbf{z}}^f|=0}} \left(\frac{1}{p}\right)^{n(R_f-R)} \left(\frac{1}{p}\right)^{2n(1-R_f)-|I_{\mathbf{x},\mathbf{z}}^f| - 2|J_{\mathbf{x},\mathbf{z}}^f|} \prob\{ \mathbf{z} \in \B \} \\
  %  & \ \ \ \ \ \ \ \ \ \ \ \ \leq \sum_{j=0}^{\lfloor n(1-R_f)/(D+1) \rfloor} \sum_{i=0}^{n(1-R_f)-j} Z_{ij} \left(\frac{1}{p}\right)^{n(R_f-R)} \left(\frac{1}{p}\right)^{2n(1-R_f)-i - 2j} \\
  %  \label{eq:intermediate_i_j}
  %  & \ \ \ \ \ \ \ \ \ \ \ \ = \sum_{j=0}^{\lfloor n(1-R_f)/(D+1) \rfloor} \sum_{i=0}^{n(1-R_f)-j} Z_{ij} \left(\frac{1}{p}\right)^{n(1-R)} \left(\frac{1}{p}\right)^{n(1-R_f)}p^{i+2j}.
  %\end{align}
  Now, the same arguments used to deduce \eqref{eq:bound_I_J} also imply that
  \begin{align*}
     Z_{ij} & \leq n^{(j+i)} p^i |\Z^{n-Dj/(1-R_f)} \cap B_{\zero,n-Dj/(1-R_f)}(\rho)| \cdot \\
     & \ \ \ \ \ \ \ \cdot |\Z^{n-D(i+j)/(1-R_f)} \cap B_{\zero,n-D(i+j)/(1-R_f)}(\rho_{\dec})|.
  \end{align*}
  Let us define the analogue of $\mathcal{E}(\rho)$ in \eqref{eq:E_rho}:
  \begin{equation}
    \label{eq:Q_rho}
    \mathcal{Q}(\rho_{\eff},\rho_{\dec}) = |\Zn \cap \B_{\eff}| \left(\frac{1}{p}\right)^{n(1-R)} |\Zn \cap \B| \left(\frac{1}{p}\right)^{n(1-R_f)}.
  \end{equation}
  We can write
  \begin{align*}
    \eqref{eq:sum_Z_ij} & \leq \sum_{j=0}^{\lfloor n(1-R_f)-B(n) \rfloor} \sum_{i=0}^{n(1-R_f)-j} \frac{|\Z^{n-Dj/(1-R_f)} \cap B_{\zero,n-Dj/(1-R_f)}(\rho)|}{|\Zn \cap \B_{\eff}|} \cdot  \\
      & \ \ \ \ \ \ \ \ \ \ \ \cdot \frac{|\Z^{n-D(i+j)/(1-R_f)} \cap B_{\zero,n-D(i+j)/(1-R_f)}(\rho_{\dec})|}{|\Zn \cap \B|} n^{(j+i)} p^{2(i+j)} \mathcal{Q}(\rho_{\eff},\rho_{\dec}).
  \end{align*}
  The previous sum can be studied in the same way as \eqref{eq:half_sum_1} and \eqref{eq:half_sum_2}, i.e., since
  \begin{equation*}
    D > \frac{1}{1-R_f} > 2 \text {\ \ \ and\ \ \ } \lambda > \frac{1}{D(1-R_f)-1} > \frac{1}{D-2},
  \end{equation*}
  we have that, when $|\mathcal{M}_{\mathbf{x},\mathbf{z}}^f|=0$,
  \begin{equation*}
    \eqref{eq:even_before_cases} \lesssim o(1) \mathcal{Q}(\rho_{\eff},\rho_{\dec}).
  \end{equation*}
  Now, notice that we have already shown in the proof of Theorem \ref{thm:awgn_capacity} that
  \begin{equation*}
    \lim_{n \to \infty} \mathcal{Q}(\rho_{\eff},\rho_{\dec}) = 0;
  \end{equation*}
  indeed, it is bounded from above by \eqref{eq:appreciable_situation}, which was shown to be vanishing when $n$ tends to infinity. %This concludes the analysis of the case $|\mathcal{M}_{\mathbf{x},\mathbf{z}}^f|=0$.

\item Let $|\mathcal{M}_{\mathbf{x},\mathbf{z}}^f|=1$ and suppose for now that $|T_{\mathbf{x},\mathbf{z}}^f|<n^\nu$ for some $\nu < 1$. Consider the set of check nodes of the Tanner graph associated with $\HH_f$ given by $I_{\mathbf{x},\mathbf{z}}^f \cup J_{\mathbf{x},\mathbf{z}}^f$ and the bipartite subgraph $\mathcal{H}_{\mathbf{x},\mathbf{z}}$ that it induces, whose set of check nodes is $I_{\mathbf{x},\mathbf{z}}^f \cup J_{\mathbf{x},\mathbf{z}}^f$, whose set of variable nodes is $N(I_{\mathbf{x},\mathbf{z}}^f \cup J_{\mathbf{x},\mathbf{z}}^f)$ and whose edges are all the edges of the original Tanner graph beween these two sets. \emph{A priori} this graph may be not connected; if we denote $\mathcal{C}$ one of its connected components and $P_{\mathcal{C}}$ its set of check nodes, we can partition $\mathcal{H}_{\mathbf{x},\mathbf{z}}$ into the disjoint union of the two following graphs:
  \begin{align*}
    & \mathcal{L}_{\mathbf{x},\mathbf{z}}^f = \left\{\mathcal{C} \subseteq \mathcal{H}_{\mathbf{x},\mathbf{z}} : |P_\mathcal{C}| \leq \frac{n(1-R_f)}{D+1}\right\} \text{ and} \\
    & \mathcal{D}_{\mathbf{x},\mathbf{z}}^f = \left\{\mathcal{C} \subseteq \mathcal{H}_{\mathbf{x},\mathbf{z}} : |P_\mathcal{C}| > \frac{n(1-R_f)}{D+1}\right\}.
  \end{align*}
  As a consequence, $I_{\mathbf{x},\mathbf{z}}^f \cup J_{\mathbf{x},\mathbf{z}}^f$ is the disjoint union of
  \begin{align*}
    L_{\mathbf{x},\mathbf{z}}^f = \bigcup \{P_\mathcal{C} : \mathcal{C} \in \mathcal{L}_{\mathbf{x},\mathbf{z}}^f\} \text{\ \ \ \ \  and\ \ \ \ \ } D_{\mathbf{x},\mathbf{z}}^f = \bigcup \{P_\mathcal{C} : \mathcal{C} \in \mathcal{D}_{\mathbf{x},\mathbf{z}}^f\}.
  \end{align*}
  The first observation that we can make is that since $|J_{\mathbf{x},\mathbf{z}}^f| \leq n(1-R_f) - B(n) \leq n(1-R_f)/(D+1)$ and $|T_{\mathbf{x},\mathbf{z}}^f|<n^\nu$, then $|\mathcal{D}_{\mathbf{x},\mathbf{z}}^f|=1$. Indeed, $|\mathcal{D}_{\mathbf{x},\mathbf{z}}^f|\leq 1$ because the expansion properties imply that $|N(P_\mathcal{C})| \geq Dn/(D+1)$ for every $\mathcal{C}\in \mathcal{D}_{\mathbf{x},\mathbf{z}}^f$; hence, if there were two ore more, the union of their $N(P_\mathcal{C})$ would exceed the size of the set of variable nodes in $\mathcal{H}_{\mathbf{x},\mathbf{z}}$, which is obviously impossible (compare to what follows \eqref{eq:marker} in the proof of Lemma \ref{lem:typical_norm_LDA}). Moreover, $\mathcal{D}_{\mathbf{x},\mathbf{z}}^f \neq \emptyset$ because otherwise $L_{\mathbf{x},\mathbf{z}}^f = I_{\mathbf{x},\mathbf{z}}^f \cup J_{\mathbf{x},\mathbf{z}}^f$ and these two conditions would hold (at least asymptotically): 
  \begin{enumerate}
    \item $L_{\mathbf{x},\mathbf{z}}^f$ has size $n(1-R_f)-|T_{\mathbf{x},\mathbf{z}}^f| \geq n(1-R_f)-n^\nu$.
    \item $n \geq |N(L_{\mathbf{x},\mathbf{z}}^f)| \geq D|L_{\mathbf{x},\mathbf{z}}^f|/(1-R_f) \geq Dn(1-D/(n^{1-\nu}(1-R_f)) \sim Dn > n$.
  \end{enumerate}
  The second one is clearly a nonsense and proves that $|\mathcal{D}_{\mathbf{x},\mathbf{z}}^f|=1$.
  We go on with this analysis and we claim that $L_{\mathbf{x},\mathbf{z}}^f$ is actually quite small. By definition of $L_{\mathbf{x},\mathbf{z}}^f$ and $D_{\mathbf{x},\mathbf{z}}^f$, we have that $N(L_{\mathbf{x},\mathbf{z}}^f) \subseteq V \smallsetminus N(D_{\mathbf{x},\mathbf{z}}^f)$ (recall that $V$ is the set of variable nodes of the Tanner graph associated with $\HH_f$ and $P$ its set of check nodes); moreover, $|N(D_{\mathbf{x},\mathbf{z}}^f)| \geq Dn/(D+1)$. This implies that $|N(L_{\mathbf{x},\mathbf{z}}^f)| \leq n-|N(D_{\mathbf{x},\mathbf{z}}^f)| \leq n/(D+1)$. Then, by the expansion properties, $|N(N(L_{\mathbf{x},\mathbf{z}}^f))| \geq D(1-R_f) |N(L_{\mathbf{x},\mathbf{z}}^f)|$. At the same time, we have that $|N(L_{\mathbf{x},\mathbf{z}}^f)| \geq D|L_{\mathbf{x},\mathbf{z}}^f|/(1-R_f)$. Since $N(N(L_{\mathbf{x},\mathbf{z}}^f)) \subseteq P \smallsetminus D_{\mathbf{x},\mathbf{z}}^f = T_{\mathbf{x},\mathbf{z}}^f \cup L_{\mathbf{x},\mathbf{z}}^f$, we deduce that
  \begin{equation*}
    |T_{\mathbf{x},\mathbf{z}}^f|+|L_{\mathbf{x},\mathbf{z}}^f| \geq |N(N(L_{\mathbf{x},\mathbf{z}}^f))| \geq D(1-R_f) |N(L_{\mathbf{x},\mathbf{z}}^f)| \geq D^2|L_{\mathbf{x},\mathbf{z}}^f|,
  \end{equation*}
  or, equivalently,
  \begin{equation*}
    |L_{\mathbf{x},\mathbf{z}}^f| \leq \frac{|T_{\mathbf{x},\mathbf{z}}^f|}{D^2-1} < \frac{n^{\nu}}{D^2-1}.
  \end{equation*}
  Substantially, we have just proved that when $|J_{\mathbf{x},\mathbf{z}}^f| \leq n(1-R_f)/(D+1)$ and $|T_{\mathbf{x},\mathbf{z}}^f|<n^\nu$, then the parity-check equations associated with the ``big'' connected component $\mathcal{D}_{\mathbf{x},\mathbf{z}}$ of $\mathcal{H}_{\mathbf{x},\mathbf{z}}$ are almost all the equations of the matrix $\HH_f$; the size of what is left (the set $|T_{\mathbf{x},\mathbf{z}}^f|$ plus the equations of the ``small'' connected components of $I_{\mathbf{x},\mathbf{z}}^f \cup J_{\mathbf{x},\mathbf{z}}^f$) is $O(n^\nu)$.
  %The next thing we would like to estimate is the number of $\mathbf{x} \in S$ such that there exists some $\mathbf{z} \in Z$ satisfying the hypotheses on $|T_{\mathbf{x},\mathbf{z}}^f|$ and $|J_{\mathbf{x},\mathbf{z}}^f|$. Hence, let
  %\begin{equation}
  %  \mathcal{N}_j = \Bigg|\left\{\mathbf{x} \in S : \exists \mathbf{z}\in Z \text{ such that } |T_{\mathbf{x},\mathbf{z}}^f| < n^\nu \text{ and } |J_{\mathbf{x},\mathbf{z}}^f| =j \leq \frac{n(1-R_f)}{D+1} \right\}\Bigg|.
  %\end{equation}
  Moreover, $\mathbf{x}$ and $\mathbf{z}$ have to be multiple modulo $p$ on all the coordinates of $N(D_{\mathbf{x},\mathbf{z}}^f)$. Indeed, this holds by definition of $I_{\mathbf{x},\mathbf{z}}^f$ on the coordinates of $N(I_{\mathbf{x},\mathbf{z}}^f \cap D_{\mathbf{x},\mathbf{z}}^f)$ and by the fact that they are fixed to $0$ modulo $p$ on the coordinates of $N(J_{\mathbf{x},\mathbf{z}}^f \cap D_{\mathbf{x},\mathbf{z}}^f)$. In other terms, there exists $\mu \in \{3,4,\ldots,p-1\}$ - recall that the values $0,1$ and $2$ are excluded by the definition of $S$ and $Z$ - such that
  \begin{align*}
    |\{l\in\{1,2,\ldots,n\} : x_l \equiv \mu z_l \bmod p\}| & \geq n - |N(L_{\mathbf{x},\mathbf{z}}^f \cup T_{\mathbf{x},\mathbf{z}}^f)| \\
    & \geq n - \frac{D}{1-R_f} (|L_{\mathbf{x},\mathbf{z}}^f| + |T_{\mathbf{x},\mathbf{z}}^f|)\\
    & \geq n - |T_{\mathbf{x},\mathbf{z}}^f|\left(1+\frac{1}{D^2 -1} \right) \\
    & > n - 2n^\nu.
  \end{align*}
  This also implies that $|\supp(\mathbf{x}-\mu \mathbf{z})| \leq 2n^\nu$ (recall the definition of support: \eqref{eq:support}); but the LDPC code underlying the construction of $\Lambda_f$ can be supposed to be \emph{asymptotically good} by Lemma \ref{lem:min_H_dist}. In other words, all the points of $\Lambda_f \smallsetminus p\Zn$ have a support of size linear in $n$. This means that 
  \begin{equation*}
    \prob \{\HH_f\mathbf{x}^T \equiv \zero^T \bmod p, \HH_f\mathbf{z}^T \equiv \zero^T \bmod p\} = 0
  \end{equation*}
  for every couple of points $\mathbf{x}$ and $\mathbf{z}$ such that $0 < |\supp(\mathbf{x}-\mu \mathbf{z})| \leq 2n^\nu$, because $\mathbf{x}-\mu \mathbf{z}$ has to belong to $\Lambda_f$ if $\mathbf{x}$ and $\mathbf{z}$ do. Therefore, when we suppose $|\mathcal{M}_{\mathbf{x},\mathbf{z}}^f|=1$ and $|T_{\mathbf{x},\mathbf{z}}^f| < n^\nu$, the only $\mathbf{z}$ that contribute to \eqref{eq:reference_for_small_T} with a non-zero term are the ones that belong to
  \begin{equation*}
    Z' = \{\mathbf{z}\in Z : \mathbf{z} \equiv \mu \mathbf{x} \bmod p,\ \exists \mu \in \{3,4,\ldots,p-1\} \}.
  \end{equation*}
  So, concerning the corresponding terms in \eqref{eq:reference_for_small_T}, we can conclude that
  \begin{align}
    \label{eq:vanishing_sum_T_small}
    & \sum_{\mathbf{x} \in S} \left(\frac{1}{p}\right)^{n(R_f-R)} \sum_{\substack{\mathbf{z} \in Z \\ |\mathcal{M}_{\mathbf{x},\mathbf{z}}^f|=1 \\ |T_{\mathbf{x},\mathbf{z}}^f| < n^\nu}} \prob \{\HH_f\mathbf{x}^T \equiv \zero^T \bmod p, \HH_f\mathbf{z}^T \equiv \zero^T \bmod p\} \prob\{ \mathbf{z} \in \B\}\\
    %& \sum_{j=0}^{\lfloor n(1-R_f) - B(n)\rfloor} \sum_{i = 0}^{n(1-R_f)-j} \left(\frac{1}{p}\right)^{n(R_f-R)} \left(\frac{1}{p}\right)^{2n(1-R_f)-i - 2j} \sum_{\substack{\mathbf{x}\in S,\mathbf{z}\in Z \\ |J_{\mathbf{x},\mathbf{z}}^f| = j, |I_{\mathbf{x},\mathbf{z}}^f| = i, |T_{\mathbf{x},\mathbf{z}}^f| < n^\nu}} \prob\{ \mathbf{z} \in \B \}\\
    %& \sum_{\substack{\mathbf{x} \in S,\ \mathbf{z} \in Z \\ |\mathcal{M}_{\mathbf{x},\mathbf{z}}^f|=1 \\ |T_{\mathbf{x},\mathbf{z}}^f| < n^\nu}} \left(\frac{1}{p}\right)^{n(R_f-R)} \left(\frac{1}{p}\right)^{2n(1-R_f)-|I_{\mathbf{x},\mathbf{z}}^f| - 2|J_{\mathbf{x},\mathbf{z}}^f|} \prob\{ \mathbf{z} \in \B \} \\
    & \ \ \ \ \ \ \ \ = \sum_{\substack{\mathbf{x} \in S,\ \mathbf{z} \in Z' \\ |\mathcal{M}_{\mathbf{x},\mathbf{z}}^f|=1 \\ |J_{\mathbf{x},\mathbf{z}}^f| = |T_{\mathbf{x},\mathbf{z}}^f| = 0}} \left(\frac{1}{p}\right)^{n(R_f-R)} \left(\frac{1}{p}\right)^{n(1-R_f)} \prob\{ \mathbf{z} \in \B \}. \nonumber 
  \end{align}
  This sum is vanishing, because it is upper bounded by \eqref{eq:only_multiple_z}, which was vanishing, too.

  We are left to study the terms of \eqref{eq:reference_for_small_T} corresponding to $|T_{\mathbf{x},\mathbf{z}}^f| \geq n^\nu$:
  \begin{align}
    \label{eq:last_case_T_big}
    & \sum_{\mathbf{x} \in S} \left(\frac{1}{p}\right)^{n(R_f-R)} \sum_{\substack{\mathbf{z} \in Z \\ |\mathcal{M}_{\mathbf{x},\mathbf{z}}^f|=1 \\ |T_{\mathbf{x},\mathbf{z}}^f| \geq n^\nu}} \prob \{\HH_f\mathbf{x}^T \equiv \zero^T \bmod p, \HH_f\mathbf{z}^T \equiv \zero^T \bmod p\} \prob\{ \mathbf{z} \in \B\}\\
    \label{eq:big_t}
    & \ \ \ \ \ \leq \sum_{\substack{\mathbf{x} \in S,\ \mathbf{z} \in Z \\ |\mathcal{M}_{\mathbf{x},\mathbf{z}}^f|=1 \\ |T_{\mathbf{x},\mathbf{z}}^f| \geq n^\nu}} \left(\frac{1}{p}\right)^{n(R_f-R)} \left(\frac{1}{p}\right)^{2n(1-R_f)-|I_{\mathbf{x},\mathbf{z}}^f| - 2|J_{\mathbf{x},\mathbf{z}}^f|} \prob\{ \mathbf{z} \in \B \}.
  \end{align} 
  For this estimation, we rely once again on the similar computations already done in the proof of Lemma \ref{lem:typical_norm_LDA}: we do not show explicitly all the details, but at this point it should be clear how to use \eqref{eq:case_M_1_j_small_a} and \eqref{eq:case_M_1_j_small_b}, together with \eqref{eq:Q_rho} and some of the strategies used in this proof to show that \eqref{eq:big_t} converges to $0$ when $n$ tends to infinity.
\end{enumerate}

{\bf Conclusion.} The sum of \eqref{eq:even_before_cases}, \eqref{eq:vanishing_sum_T_small}, and \eqref{eq:last_case_T_big} is a vanishing upper bound of \eqref{eq:reference_for_small_T}. Therefore, the limit in \eqref{eq:target_2} holds true and the theorem is proved. 
\end{IEEEproof}

\section{Conclusions}
\label{sec:conclusion}
In the first part of this paper, we have given a new proof that random Construction-A Voronoi constellations achieve the capacity of the AWGN channel with lattice encoding and decoding. We have obtained this result without employing the dithering technique, thus simplifying the information transmission scheme with respect to other solutions proposed in the literature. Also, we have explicitly shown how the prime number $p$ that underlies Construction A has to grow as a function of the lattice dimension $n$: it is of the same order as $n^\lambda$, for a positive constant $\lambda$ whose lower bound varies between $1/2$ and $2/3$ depending on the rate of the linear code involved in the construction. The proof of this capacity result is also based on a lemma which states the sphericity of our Voronoi random constellation: its points typically lie very close to the surface of a ball, called the shaping sphere, whose radius is the effective radius of the shaping lattice.

The second part of the paper is dedicated to LDA lattices. We have adapted the arguments used in the case of random Construction A to show the novel result that there exists a family of LDA Voronoi constellations which is capacity-achieving under lattice encoding and decoding. Again, we have expressed in formulae the dependence of $p$ on $n$, finding lower bounds of $\lambda$ that are still constant, but larger than the ones of the random-Construction-A case. Furthermore, we have shown the sphericity of LDA Voronoi constellations, too. One important feature of our LDA  ensemble is that the row and column Hamming weights of the associated parity-check matrices are well-determined constants and do not need to grow with $n$.

The probabilistic arguments used for dealing with the technical difficulties that arise from the low density of the LDA parity-check matrices are based on what we have called the $D$-goodness of the associated Tanner graphs. This property is a crucial tool in our analysis and has also made possible an estimation of the minimum Hamming distance and the fundamental gain of our LDA-lattice family.

The analysis of lattice decoding of capacity-achieving LDA lattices that we carried out relies on ML/MAP block-wise decoding of the embedded LDPC ensemble. Modern coding theory offers low-complexity iterative decoding methods for LDPC codes which make LDA decoding practically feasible. For this reason, it could be interesting in the future to investigate theoretically the performance of LDA lattices under iterative decoding. Nevertheless, at the present moment and to the best of our knowledge, no theoretical tools are available to prove that LDA ensembles are capacity-achieving with iterative message-passing decoding. The latter attains the MAP performance of an LDPC ensemble when applied to the associated spatially-coupled ensemble for binary codes and binary symmetric 
memoryless channels \cite{Kudekar2013}. Hence, in practical applications non-binary spatial coupling is a potential 
way to enhance the performance of LDA lattices. However, any future theoretical 
breakthrough on iterative LDA decoding is conditioned on finding an exact solution of 
density evolution for non-binary codes on graphs.

%----------------------------------------------------------------------------------------------------
%----------------------------------------------------------------------------------------------------

% if have a single appendix:
%\appendix[Proof of the Zonklar Equations]
% or
%\appendix  % for no appendix heading
% do not use \section anymore after \appendix, only \section*
% is possibly needed

% use appendices with more than one appendix
% then use \section to start each appendix
% you must declare a \section before using any
% \subsection or using \label (\appendices by itself
% starts a section numbered zero.)
%

\appendices

% you can choose not to have a title for an appendix
% if you want by leaving the argument blank

\section{Proof of Lemma \ref{lem:typical_noise}}
\label{sec:proof_typical_noise}
\begin{IEEEproof}
  It is known that, since $X_i \sim \NN(0,\sigma^2)$, $i=1,\ldots,n$, then $X_i^2$ follows a gamma
  distribution and $\mathbb{E}[X_i^2] = \sigma^2$, $\var(X_i^2)=2\sigma^4$. Consequently, by the independence of the $X_i$,
  \begin{equation*}
    \mathbb{E}[\rho^2] = n\sigma^2,\ \var(\rho^2)=2n\sigma^4.
  \end{equation*}
  %Lemma \ref{lem:chebyshev_inequality} with $Y=\rho^2$ and $\tau = \kappa \sqrt{2n} \sigma^2$ 
  Chebyshev's inequality states that, for any $\kappa>0$,
  \begin{equation*}
    \prob \left\{|\rho^2 - n\sigma^2| > \kappa\sqrt{2n}\sigma^2 \right\} \leq \frac{1}{\kappa^2}.
  \end{equation*}
  If we choose $\kappa = \kappa(n)$ such that $\lim_{n \to \infty}\kappa = +\infty$, then
  \begin{equation}
    \label{eq:noise_squared_norm}
    \lim_{n \to \infty} \prob \left\{|\rho^2 - n\sigma^2| \leq \kappa\sqrt{2n}\sigma^2 \right\} = 1.
  \end{equation}
  As a consequence,
  \begin{equation*}
    \lim_{n \to \infty} \prob \left\{\rho^2 \leq \sigma^2n \left( 1 + \kappa\sqrt{\frac{2}{n}} \right) \right\} = 1.
  \end{equation*}
  Taking for example $\kappa = \log_2 n$, we have that $\lim_{n \to \infty}\kappa\sqrt{2/n} = 0$. This implies that for $n$ big enough and for every $\varepsilon > 0$
  \begin{equation*}
    \sqrt{1 + \kappa\sqrt{\frac{2}{n}}} < 1 + \varepsilon
  \end{equation*}
  and
  \begin{equation*}
    \prob \left\{\rho \leq \sigma\sqrt{n} \left( \sqrt{1 + \kappa\sqrt{\frac{2}{n}}} \right) \right\}
    \leq \prob \left\{\rho \leq \sigma\sqrt{n} \left( 1 + \varepsilon \right) \right\}.
  \end{equation*}
  This is enough to conclude that
  \begin{equation*}
    \lim_{n \to \infty} \prob \left\{\rho \leq \sigma\sqrt{n} \left( 1 + \varepsilon \right) \right\} = 1,
  \end{equation*}
  which proves the statement restricted to the second inequality. But notice that \eqref{eq:noise_squared_norm} also implies that 
  \begin{equation*}
    \lim_{n \to \infty} \prob \left\{\rho^2 \geq \sigma^2n \left( 1 - \kappa\sqrt{\frac{2}{n}} \right) \right\} = 1.
  \end{equation*}
  This leads to the conclusion that
  \begin{equation*}
    \lim_{n \to \infty} \prob \left\{\rho \geq \sigma\sqrt{n} \left( 1 - \varepsilon \right) \right\} = 1,
  \end{equation*}
  too, and the lemma is proved.
\end{IEEEproof}

\section{Proof of Lemma \ref{lem:integer_sphere_points}}
\label{sec:proof_sphere_points}
\begin{IEEEproof}
  Consider, for every $\mathbf{z} \in \Zn$, the cube $\mathcal{C}_{\mathbf{z}}$ centered at $\mathbf{z}$ of edge (and volume) equal to $1$.
  Let
  \begin{equation*}
    U = \bigcup_{\mathbf{z} \in \Zn \cap B_{\mathbf{c},n}(\rho)}\mathcal{C}_{\mathbf{z}}.
  \end{equation*}
  Now, let $S_1$ be the sphere inscribed in $U$, and $S_2$ the one circumscribed to $U$, both of them centered at $\mathbf{c}$. The definition of $U$ and the fact that the length of the diagonal of any $\mathcal{C}_{\mathbf{z}}$ is $\sqrt{n}$ imply that the radius of $S_1$ is at least $\rho - \sqrt{n}/2$, while the one of $S_2$ is at most $\rho + \sqrt{n}/2$. Therefore,
  \begin{equation*}
    \vol \left( B_{\mathbf{c},n}(\rho)\right) \left(1 - \frac{\sqrt{n}}{2\rho} \right)^n = \vol \left( B_{\mathbf{c},n} \left( \rho - \frac{\sqrt{n}}{2} \right) \right) \leq \vol(S_1) \leq \vol(U)
  \end{equation*}
  and
  \begin{equation*}
    \vol(U) \leq \vol(S_2) \leq \vol \left( B_{\mathbf{c},n} \left( \rho + \frac{\sqrt{n}}{2} \right) \right) = \vol \left( B_{\mathbf{c},n}(\rho)\right) \left(1 + \frac{\sqrt{n}}{2\rho} \right)^n.
  \end{equation*}
  Since $|\Zn \cap B_{\mathbf{c},n}(\rho)| = \vol(U)$, these two inequalities yield the wanted result.
\end{IEEEproof}

\section{Proof of Lemma \ref{lem:equivalent_points}}
\label{sec:proof_equivalent_points}
\begin{IEEEproof}
  Let us start with the case $\mu = 1$, that outlines the strategy for a more general $\mu$. If $\mathbf{z} \equiv \mathbf{x} \bmod p$, then $\mathbf{x} - \mathbf{z} \in p\Zn$. Hence, $x_i - z_i = k_ip$, for some $k_i \in \Z$. Let us call $N = \sum_{i=1}^n|k_i|$; we have
  \begin{equation*}
    \Vert\mathbf{x} - \mathbf{z}\Vert^2 = \sum_{i=1}^n(x_i - z_i)^2 = \sum_{i=1}^n k_i^2p^2 \geq p^2\sum_{i=1}^n |k_i| = p^2N.
  \end{equation*}
  This, together with the fact that both $\mathbf{x}$ and $\mathbf{z}$ lie in $\B$, gives the necessary condition
  \begin{equation*}
    p^2N \leq \Vert\mathbf{x} - \mathbf{z}\Vert^2 \leq 4\rho^2
  \end{equation*}
  or, equivalently,
  \begin{equation*}
    N \leq \frac{4\rho^2}{p^2}.
  \end{equation*}
  Then, the number of $\mathbf{z}$ equivalent to $\mathbf{x}$ in $\B$ is bounded by the number $L$ of different vectors $(k_1,k_2,\ldots,k_n) \in \Zn$ such that $\sum_{i=1}^n|k_i| \leq 4\rho^2/p^2$. One of this vectors is simply $\zero \in \Zn$. Hence, $L$ itself is bounded by $1$ plus the number of possible ways of:
  \begin{enumerate}
  \item fixing $m$ coordinates among $n$ (with $1 \leq m \leq \lfloor 4\rho^2/p^2 \rfloor$; $m=0$ corresponds to $k_i=0$ for every $i$, i.e., to the ``1 plus'');
  \item for each one of the $m$ fixed coordinates, deciding if $k_i$ will be positive or negative (and, for now, fix $k_i = 0$);
  \item choosing for $\lfloor 4\rho^2/p^2 \rfloor$ times to increment one of the $m$ coordinates $k_i$ of $\pm 1$, according to the sign fixed at step 2.
  \end{enumerate}
  As a consequence,
  \begin{align*}
    |\{\mathbf{z} \in \B \cap \Zn : \mathbf{z} \equiv \mathbf{x} \bmod p\}| & \leq L \\
    %\label{eq:binom_approx}
    & \leq 1 + \sum_{m=1}^{\lfloor 4\rho^2/p^2 \rfloor} \binom{n}{m} 2^m m^{\lfloor 4\rho^2/p^2 \rfloor} \\
    & \leq 1 + \sum_{m=1}^{\lfloor 4\rho^2/p^2 \rfloor} n^m 2^m m^{\lfloor 4\rho^2/p^2 \rfloor} \\
    & \leq 1 + \frac{4\rho^2}{p^2}n^{4\rho^2/p^2}2^{4\rho^2/p^2} \left( \frac{4\rho^2}{p^2} \right)^{4\rho^2/p^2} \\
    & = 1 + \frac{4\rho^2}{p^2}\left( \frac{8n\rho^2}{p^2} \right)^{4\rho^2/p^2}, 
  \end{align*}

  The lemma is proved for $\mu = 1$. Now, let us consider the case in which $\mu$ takes another value and let $\mathbf{z}'$ be any point inside the sphere such that $\mathbf{z}' \equiv \mu\mathbf{x} \bmod p$. Then
  \begin{equation*}
    |\{\mathbf{z} \in \B \cap \Zn : \mathbf{z} \equiv \mu \mathbf{x} \bmod p\}| = |\{\mathbf{z} \in \B \cap \Zn : \mathbf{z} \equiv \mathbf{z}' \bmod p\}|
  \end{equation*}
  and the proof works exactly in the same way as before, with $\mathbf{z}'$ instead of $\mathbf{x}$.
\end{IEEEproof}

\section{Proof of Lemma \ref{lem:orthogonal_noise}}
\label{sec:proof_orthogonal_noise}
\begin{IEEEproof}
  If $\mathbf{x} = \zero$, the statement is trivially true. So, suppose from now on that $\mathbf{x} \neq \zero$.
  The scalar product $\mathbf{x}\mathbf{w}^T = \sum_{i=1}^n x_iw_i$ is a sum of i.i.d.\ Gaussian random variables, weighted by the $x_i$, then it is well known that it follows a Gaussian distribution, too. More precisely, $\E[\mathbf{x}\mathbf{w}^T] = 0$ and
  \begin{equation*}
    \var \left(\sum_{i=1}^n x_iw_i\right) = \sum_{i=1}^n x_i^2\var(w_i) = \sigma^2 \Vert\mathbf{x}\Vert^2. 
  \end{equation*}
  Consider $Q(\cdot)$, the tail probability of the standard normal distribution:
  \begin{equation*}
    Q(y) = \frac{1}{\sqrt{2 \pi}} \int_{y}^{\infty} \exp \left(-\frac{u^2}{2}\right) \, \mathrm{d}u.
  \end{equation*}
  For positive $y$, the Chernoff bound states that
  \begin{equation*}
    Q(y) \leq \frac{1}{2}e^{-\frac{y^2}{2}}.
  \end{equation*}
  We apply this bound to our probability and we have
  \begin{align*}
    \prob \{ |\mathbf{x}\mathbf{w}^T| > f(n)\sigma\Vert\mathbf{x}\Vert\} & = 2Q\left(\frac{f(n)\sigma\Vert\mathbf{x}\Vert}{\sigma\Vert\mathbf{x}\Vert}\right) \\
    & \leq \exp\left( - \frac{f(n)^2}{2} \right),
  \end{align*}
  which tends to $0$ because of the choice of $f(n)$ by hypothesis. Hence, 
  \begin{equation*}
    \lim_{n \to \infty} \prob \{ |\mathbf{x}\mathbf{w}^T| \leq f(n)\sigma \Vert\mathbf{x}\Vert\} = 1.
  \end{equation*}
\end{IEEEproof}

\section{Proof of Lemma \ref{lem:good_points}}
\label{sec:proof_good_points}
\begin{IEEEproof}
  If $z \in \Z$ we denote by $\overline{z} \in \Z$ the element of the class of $z$ modulo $p$ with the smallest absolute value; that is, $z \equiv \overline{z} \bmod p$ and $\overline{z}$ is the class representative lying in $\{-(p-1)/2, -(p-3)/2, \ldots, (p-1)/2\}$. The notation adapts to integer vectors, too.
  
  First of all, notice that $\mathbf{z} \equiv \mu \mathbf{x} \bmod p$ means $\mathbf{z} = \mu \mathbf{x} + p\mathbf{k}$, for some $\mathbf{k} \in \Zn$. Hence, if we call $\nu = 1-\mu$,
  \begin{align*}
    \Vert\mathbf{x} - \mathbf{z}\Vert^2 & = \Vert\mathbf{x} - \mu \mathbf{x} - p\mathbf{k}\Vert^2 \\
    & \geq \Vert\overline{(1-\mu)\mathbf{x} - p\mathbf{k}}\Vert^2 \\
    & = \Vert\overline{(1-\mu)\mathbf{x}}\Vert^2 \\
    & = \Vert\overline{\nu \mathbf{x}}\Vert^2.
  \end{align*}
  If $\Vert\overline{\nu \mathbf{x}}\Vert^2 > 4 \rho^2$, then $\Vert\mathbf{x} - \mathbf{z}\Vert^2 > 4 \rho^2$, too. In other words, $\mathbf{z}$ lies outside $B_{\mathbf{x},n}(2\rho)$ and $\mathbf{x}$ does not have to be counted among the ones contributing to $N(\mu)$. That is,
  \begin{equation*}
    N(\mu) \leq |\{\mathbf{x} \in \Zn \cap \B_{\eff} : \Vert\overline{\nu \mathbf{x}}\Vert^2 \leq 4 \rho^2 \}| = N'(\mu).
  \end{equation*}

  Now, let $C \geq 3$ be a fixed constant and, given $\mathbf{x}$, consider 
  \begin{equation*}
    J = \left\{i \in \{1,2,\ldots,n\} : |x_i| < \frac{\sqrt{p}}{C^2} \right\};
  \end{equation*}
  Let us prove that if $n$ is large enough, then $|J| \geq n-n^\gamma$ for every constant $\gamma$ such that $\max\{0,1-\lambda(2R-1)\}<\gamma<1$. Indeed, suppose by contradiction that $|J| < n-n^\gamma$, then we would have at least $n^\gamma$ coordinates $x_i$ of $\mathbf{x}$ such that $|x_i| \geq \sqrt{p}/C^2$. We employ the hypotheses on the ranges of $\gamma$ and $R$ and the relation $p=n^\lambda$ to get:
  \begin{align*}
    \Vert\mathbf{x}\Vert^2 & \geq n^\gamma\frac{p}{C^4} \\
    & > n\frac{p^{2(1-R)}}{2 \pi e}\left(1 + \frac{1}{n^\omega}\right)^2 \\
    & = \rho_{\eff}^2\left(1 + \frac{1}{n^\omega}\right)^2 \\ 
    & \geq \Vert\mathbf{x}\Vert^2,
  \end{align*}
  which is a nonsense (notice that the second - strict - inequality is true for $n$ large enough).

  Before going on, for a given $\mathbf{x} = (x_1,x_2,\ldots,x_n)$ and for a subset of indices $I \subseteq \{1,2,\ldots,n\}$ we define $\mathbf{x}(I)$ to be the vector $(x(I)_1,x(I)_2,\ldots,x(I)_n)$ such that
  \begin{equation*}
    \mathbf{x}(I)_i =
    \begin{cases}
      0, & \text{if } i \in I \\
      x_i, & \text{otherwise}
    \end{cases}.
  \end{equation*}

  {\bf First estimate: $|\nu| \leq \sqrt{p}$.} When $\nu$ is ``small'', denoting $J^c = \{1,2,\ldots,n\} \smallsetminus J$, we have
  \begin{equation*}
    \Vert\overline{\nu \mathbf{x}}\Vert^2 \geq \Vert\overline{\nu \mathbf{x}}(J^c)\Vert^2 = \Vert\nu \mathbf{x}(J^c)\Vert^2 \geq 4\Vert\mathbf{x}(J^c)\Vert^2;
  \end{equation*}
  the equality holds by definition of $J$ and because $|\nu| \leq \sqrt{p}$, whereas the second inequality comes from the hypothesis on the range of $\mu$, that implies $|\nu|\geq 2$.
  Now, if $\Vert\mathbf{x}(J^c)\Vert^2 > \rho^2$, then the previous chain of inequalities gives $\Vert\overline{\nu \mathbf{x}}\Vert^2 > 4\rho^2$. Hence,
  \begin{align*}
    N'(\mu) & \leq |\{\mathbf{x} \in \Zn \cap \B_{\eff} : \Vert \mathbf{x}(J^c)\Vert^2 \leq \rho^2 \}| \\
    & \leq \binom{n}{\lfloor n^\gamma \rfloor} \cdot |\Z^{\lfloor n-n^\gamma \rfloor} \cap B_{\zero,\lfloor n-n^\gamma \rfloor}(\rho)| \cdot |\Z^{\lfloor n^\gamma \rfloor} \cap B_{\zero,\lfloor n^\gamma \rfloor}(\rho_{\eff}(1+1/n^\omega))|.
  \end{align*}
  Notice that the binomial coefficient is upper bounded by the subexponential function $n^{n^\gamma}$ and, similarly,
  \begin{align*}
    |\Z^{\lfloor n^\gamma \rfloor} \cap B_{\zero,\lfloor n^\gamma \rfloor}(\rho_{\eff}(1+1/n^\omega))| & \leq |\Z^{\lfloor n^\gamma \rfloor} \cap [-\rho_{\eff}(1+1/n^\omega), \rho_{\eff}(1+1/n^\omega)]^{\lfloor n^\gamma \rfloor}| \\
    & \leq \left(Dn^{1/2+\lambda(1-R)}\right)^{n^\gamma},
  \end{align*}
  for some constant $D$ and $\gamma < 1$. We can conclude that
  \begin{equation*}
    N'(\mu) \leq f(n) |\Z^{\lfloor n-n^\gamma \rfloor} \cap B_{\zero,\lfloor n-n^\gamma \rfloor}(\rho)|,
  \end{equation*}
  for some subexponential function $f(n)$.

  {\bf Second estimate: $|\nu| > \sqrt{p}$.} Let $\eta$ be a constant such that $0 < \eta < 1$. We say that $\mathbf{x} \in \Zn \cap \B_{\eff}$ is \emph{heavy} if for all $K \subseteq \{1,2,\ldots,n\}$ such that $|K| \leq (1-\eta)n$, we have $\Vert\mathbf{x}(K)\Vert^2 > 4\rho^2/C^2$. Qualitatively speaking, a heavy $\mathbf{x}$ is such that every ``quite small'' subset of coordinates still gives a ``big enough'' contribution to the total norm of $\mathbf{x}$ itself. 

  Now, %let $\nu \in \Fp \smallsetminus \{0\}$ (corresponding to $\mu \neq 1$ and from now on, 
  consider 
  \begin{equation*}
    I = \{i \in \{1,2,\ldots,n\} : |\overline{\nu x_i}| < C|x_i|\}.
  \end{equation*}
  Suppose that $\mathbf{x}$ is heavy, then, if $|I| \leq (1-\eta)n$,
  \begin{equation*}
    \Vert\overline{\nu \mathbf{x}}\Vert^2 \geq \Vert\overline{\nu \mathbf{x}}(I)\Vert^2 \geq C^2\Vert\mathbf{x}(I)\Vert^2 > 4 \rho^2,
  \end{equation*}
  where the second inequality is a direct consequence of the definition of $I$. This means that in this case %$\mathbf{x}$ does not have to be counted and
  \begin{equation*}
    N'(\mu) \leq |\{\mathbf{x} \in \Zn \cap \B_{\eff} : \mathbf{x} \text{ is not heavy} \}| + |\{\mathbf{x} \in \Zn \cap \B_{\eff} : \mathbf{x} \text{ is heavy, } |I| > (1-\eta)n\}|.
  \end{equation*}
  Let us call $N_1(\mu)$ the first addend and $N_2(\mu)$ the second one and estimate them.

  {\bf Estimation of $N_1(\mu)$.} If $\mathbf{x}$ is not heavy, there exists $K \subseteq \{1,2,\ldots,n\}$ such that $|K| \leq (1-\eta)n$ and $\Vert\mathbf{x}(K)\Vert^2 \leq 4 \rho^2/C^2$. 
  Notice that if this is true for $K=\emptyset$, then the same property holds \emph{a fortiori} for a bigger $K$. Then, if $h(\cdot)$ is the binary entropy funcion and supposing without loss of generality that $(1-\eta)n$ is integer,
  \begin{align*}
    N_1(\mu) & \leq |\{\mathbf{x} \in \Zn \cap \B_{\eff} : \exists K \subseteq \{1,2,\ldots,n\} \text{ with } |K| \leq (1-\eta)n, \Vert\mathbf{x}(K)\Vert^2 \leq 4 \rho^2/C^2 \}| \\
    & \leq \binom{n}{\eta n} \cdot |\Z^{(1-\eta)n} \cap B_{\zero,(1-\eta)n}(\rho_{\eff}(1+1/n^\omega))| \cdot |\Z^{\eta n} \cap B_{\zero,\eta n}(2\rho/C)|\\
    & \lesssim 2^{nh(\eta)} \vol\left(B_{\zero,(1-\eta)n}(\rho_{\eff}(1+1/n^\omega))\right)\left(1+\frac{\sqrt{(1-\eta)n}}{2\rho_{\eff}(1+1/n^\omega)}\right)^{(1-\eta)n} \\
    & \ \ \ \ \ \cdot \vol\left(B_{\zero,\eta n}(2\rho/C)\right)\left(1+\frac{C \sqrt{\eta n}}{4\rho}\right)^{\eta n}\\
    & \lesssim \left( \frac{2^{h(\eta)}}{\sqrt{(1-\eta)^{1-\eta}}\sqrt{\eta^{\eta}}} \left(\frac{2(1+\varepsilon)}{C}\right)^\eta \right)^n p^{(1-\eta)n(1-R) + \eta n(1-R_f)} g(n)\\
    & = \left( \frac{2^{h(\eta)}}{\sqrt{(1-\eta)^{1-\eta}}\sqrt{\eta^{\eta}}} \left(\frac{2(1+\varepsilon)}{C}\right)^\eta \right)^n p^{n(1-R) - n \eta(R_f-R)} g(n),
  \end{align*}
  for some subexponential function $g(n)$. Notice that, for every choice of $\eta$ and $\varepsilon$, we can choose $C$ to be large enough to make the whole quantity in the big parenthesis less than $1$. Thus,
  \begin{equation*}
    N_1(\mu) \lesssim p^{n(1-R) - n \eta(R_f-R)} g(n).
  \end{equation*}
   
  {\bf Estimation of $N_2(\mu)$.} Let $\mathbf{x}$ be heavy and suppose that $|I| > (1-\eta)n$. 
  Let us define $I' = I \cap J$. Notice that $|J|$ is asymptotic to $n$ and, for $n$ big enough, $I' \neq \emptyset$. Then,
  \begin{equation*}
    |I'| \geq |J| - |I^c| \geq n (1 - \eta -n^{\gamma-1}) \sim n (1 - \eta).
  \end{equation*}
  Let $S \subseteq \B_{\eff}$ be the set of integer points whose cardinality is $N_2(\mu)$ and that we have to estimate. We will create a relation $\phi: S \to \B_{\eff}$ (a ``function'' with more than one image per point), as follows: if $\mathbf{x} \in S$, fix $|I'|/2$ coordinates of $I'$ and add to each of them $1$ or $-1$, in such a way that the new point is still inside $\B_{\eff}$. The set of images of $\mathbf{x}$ is made of all the $\binom{|I'|}{|I'|/2}$ new points that we obtain with the $\binom{|I'|}{|I'|/2}$ different choices of coordinates to modify. We denote it by $\phi(\mathbf{x}) \subseteq \B_{\eff}$. We have implicitly supposed that $|I'|/2$ is integer, but nothing would substantially change if $|I'|$ were odd. Observe that the number of images of each single $\mathbf{x}$ is bounded from below by 
  \begin{equation*}
    \binom{|I'|}{|I'|/2} \geq \frac{1}{\sqrt{2|I'|}} 2^{|I'|} \geq \frac{1}{\sqrt{2n}} 2^{n (1 - \eta -n^{\gamma-1})},
  \end{equation*}
  independently from $I'$. We have used Lemma \ref{lem:binomial_coeff} to approximate the binomial coefficient.
  
  Now, let 
  \begin{equation*}
    S' = \{ \mathbf{x'} \in \B_{\eff} : \mathbf{x}' \in \phi (\mathbf{x}), \exists \mathbf{x} \in \B_{\eff} \} \subseteq \B_{\eff}.
  \end{equation*}
  It is possible that a certain $ \mathbf{x}' \in S'$ has more than one counterimage in $S$. We would like to estimate how many they can be. In order to count them, pay attention to the following facts: given an $\mathbf{x}$ in $S$, for all $i \in I'=I \cap J$ we have that
  \begin{itemize}
    \item $|x_i| < \sqrt{p}/C^2$ (by definition of $J$),
    \item $|\overline{\nu x_i}| < C|x_i|$ (by definition of $I$).
  \end{itemize}
  The two conditions together say that $|\overline{\nu x_i}| < \sqrt{p}/C$, whereas $|\nu| > \sqrt{p}$ by hypothesis. Then, if we suppose that $x_i$ is positive, we have
  \begin{equation*}
    |\overline{\nu (x_i - 1)}| = |\overline{\nu x_i - \nu}| > \sqrt{p}\left( 1-\frac{1}{C}\right) \geq \frac{\sqrt{p}}{C} > C|x_i| > C|x_i - 1|.
  \end{equation*}
  With the same argument, we also have that for a negative $x_i$,
  \begin{equation*}
    |\overline{\nu (x_i + 1)}| > C|x_i + 1|.
  \end{equation*}

  Now, consider $\mathbf{x}' \in \phi (\mathbf{x})$ for some $\mathbf{x} \in S$; %Lemma \ref{lem:p_absolute_value} 
  what we have just shown implies that all the coordinates $x_i'$ of $\mathbf{x}'$ that are equal to a coordinate of $\mathbf{x}$ plus or minus $1$ (i.e., all the ``modified'' coordinates of $\mathbf{x}$), are such that $|\overline{\nu x_i'}| \geq C|x_i'|$.  As a consequence and by definition of $I$, every $\mathbf{x}' \in S'$ has 
  %at most
  %\begin{equation}
  %  \max_{|I^c|} \left \{ |I'|/2 + |I^c| \right \} = |I'|/2 + \lfloor \eta n \rfloor
  %\end{equation}
  between $|I'|/2 $ and $|I'|/2 + \lfloor \eta n \rfloor$ coordinates such that $|\overline{\nu x_i'}| \geq C|x_i'|$. On the other hand, every $\mathbf{x} \in S$ has between $0$ and $\lfloor \eta n \rfloor$ of them. This means that an upper bound of the number $M$ of counterimages of $\mathbf{x}' \in S'$ is given by the number of possible modifications of plus or minus $1$ (now only towards the surface of $\B_{\eff}$, since $\phi$ always ``pushes'' a point towards the inner region) of $|I'|/2$ coordinates chosen among the at most $|I'|/2 + \eta n$ such that $|\overline{\nu x_i'}| \geq C |x_i'|$; in formulae,
  \begin{align*}
    M & \leq \sum_{k=|I'|/2}^{|I'|/2 + \lfloor \eta n \rfloor}\binom{|I'|/2 + \lfloor \eta n \rfloor}{k} \\
    & \leq \eta n 2^{|I'|/2 + \lfloor \eta n \rfloor} \\
    & \leq \eta n 2^{\frac{n}{2} + \eta n}.
  \end{align*}

  Summarizing, we have created a relation $\phi$ that associates every point in $S$ with at least $2^{n(1 - \eta - n^{\gamma - 1})}/\sqrt{2n}$ points in $S'$ and every point in $S'$ with at most $\eta n 2^{(1/2 + \eta)n}$ counterimages in $S$. In other terms,
  \begin{equation*}
    N_2(\mu) \frac{1}{\sqrt{2n}} 2^{n(1 - \eta - n^{\gamma - 1})} \leq |S'| \eta n 2^{\frac{n}{2} + \eta n} \leq |\B_{\eff} \cap \Zn| \eta n 2^{(\frac{1}{2} + \eta)n}
  \end{equation*}
  and
  \begin{equation*}
    N_2(\mu) \leq \sqrt{2} \eta n^{\frac{3}{2}} 2^{(2\eta - \frac{1}{2} + n^{\gamma-1})n} |\B_{\eff} \cap \Zn|.
  \end{equation*}
  Putting together the estimation of $N_1(\mu)$ and $N_2(\mu)$, we get
  \begin{equation*}
    N'(\mu) \lesssim p^{n(1-R) - n \eta(R_f-R)} g(n) + \sqrt{2} \eta n^{\frac{3}{2}} 2^{(2\eta - \frac{1}{2} + n^{\gamma-1})n} |\B_{\eff} \cap \Zn|.
  \end{equation*}

  {\bf Conclusion.} We have shown that for every value of $\nu$ (hence of $\mu$), it is true that
  \begin{align*}
    N(\mu) \leq  N'(\mu) & \lesssim \max \bigg\{ f(n) |\Z^{\lfloor n-n^\gamma \rfloor} \cap B_{\zero,\lfloor n-n^\gamma \rfloor}(\rho)|, \\
    & \ \ \ \ \ \ p^{n(1-R) - n \eta(R_f-R)} g(n) + \sqrt{2} \eta n^{\frac{3}{2}} 2^{(2\eta - \frac{1}{2} + n^{\gamma-1})n} |\B_{\eff} \cap \Zn|\bigg\}.
  \end{align*}
  Since the number of different $\mu$ is bounded by $p$, we can multiply by $p$ the previous bound and get
  \begin{align*}
    N = \sum_{\mu \in \Fp \smallsetminus \{0,1,2\}}N(\mu) & \lesssim \max \bigg\{ p f(n) |\Z^{\lfloor n-n^\gamma \rfloor} \cap B_{\zero,\lfloor n-n^\gamma \rfloor}(\rho)|, \\
    & \ \ \ \ \ \ p^{n(1-R) - n \eta(R_f-R)+1} g(n) + p\sqrt{2} \eta n^{\frac{3}{2}} 2^{(2\eta - \frac{1}{2} + n^{-(1-\gamma)})n} |\B_{\eff} \cap \Zn|\bigg\}.
  \end{align*}
  Recall that the goal of this lemma is to prove that $N = o(p^{n(1-R)}/t(n))$ for every subexponential function $t(n)$. Let us consider the previous terms separately. 
  First of all, using \eqref{eq:rates_conditions} in the second inequality:
  \begin{equation*}
    \frac{p f(n) t(n) |\Z^{\lfloor n-n^\gamma \rfloor} \cap B_{\zero,\lfloor n-n^\gamma \rfloor}(\rho)|}{p^{n(1-R)}} \lesssim \frac{r(n)}{p^{n(R_f-R)}},
  \end{equation*}
  for some subexponential function $r(n)$. Thanks to the hypothesis $p^{R_f-R}=\Omega>1$, the whole quantity decreases at least exponentially to $0$, as wanted. Similarly,
  \begin{equation*}
    \frac{p^{n(1-R) - n \eta(R_f-R)+1} g(n)t(n)}{p^{n(1-R)}} = \frac{g(n)t(n)p}{p^{\eta n(R_f-R)}} \rightarrow 0.
  \end{equation*}

  Finally,
  \begin{equation*}
    \frac{p\sqrt{2} \eta n^{\frac{3}{2}} 2^{(2\eta - \frac{1}{2} + n^{-(1-\gamma)})n} |\B_{\eff} \cap \Zn|t(n)}{p^{n(1-R)}} \lesssim 2^{(2\eta - \frac{1}{2} + n^{-(1-\gamma)})n}s(n),
  \end{equation*}
  for some subexponential function $s(n)$. The whole quantity tends to $0$ because the dominating term is exponential and $\eta$ can be chosen in such a way that the exponent is negative. This ends the proof.
\end{IEEEproof}

\section{Proof of Lemma \ref{lem:good_graphs}}
\label{sec:proof_good_graphs}
\begin{IEEEproof}
  First of all, let us order the set $V_L$ (putting it in bijection with $\{1,2,\ldots,n\}$) and the set $V_R$ (in bijection with
  $\{1,2,\ldots,fn\}$); let us also order the set $E$ of edges and call $e_1,e_2,\ldots,e_{f\Delta}$ 
  the edges linked to the first element of $V_L$, $e_{f\Delta + 1},e_{f\Delta + 2},\ldots,e_{2f\Delta}$ the edges linked
  to the second element of $V_L$, and so on. At the same time, call $f_1,f_2,\ldots,f_{\Delta}$ the edges linked to
  the first element of $V_R$, $f_{\Delta + 1},f_{\Delta + 2},\ldots,f_{2\Delta}$ the edges linked to the second element
  of $V_R$, and so on. Then, a graph is determined by a permutation of $\{1,2,\ldots,\Delta fn\}$ that
  assigns to every $e_m$ exactly one of the $f_l$.

  By definition,
  \begin{equation*}
    %\label{eq:prob_1}
    \prob\{\text{$\mathcal{G}$ is not $D$-good from left to right}\} \leq \prob\{\text{$\mathcal{G}$ does not satisfy \eqref{eq:condition_1}}\}.
  \end{equation*}
  Evaluating this probability corresponds to counting the number of possible permutations of $\{1,2,\ldots,\Delta f n\}$ that do not guarantee the expansion property.
  \begin{align}
    \prob & \{\text{$\mathcal{G}$ does not satisfy \eqref{eq:condition_1}}\} \nonumber \\
    & = \prob\{ \exists S \subseteq V_L : |S| \leq \frac{n}{D+1} \text{ and } |N(S)| < fD|S|\} \nonumber \\
    & \leq \sum_{\substack{S \subseteq V_L \\ 1 \leq |S| \leq \lfloor \frac{n}{D+1} \rfloor}}{\prob\{|N(S)| < fD|S|\}} \nonumber \\
    & \leq \sum_{\substack{S \subseteq V_L \\ 1 \leq |S| \leq \lfloor \frac{n}{D+1} \rfloor}} \sum_{\substack{T \subseteq V_R \\ |T|=\lfloor fD|S| \rfloor}} \prob\{N(S) \subseteq T\} \nonumber \\
    & = \sum_{\substack{S \subseteq V_L \\ 1 \leq |S| \leq \lfloor \frac{n}{D+1} \rfloor}} \binom{fn}{\lfloor fD|S| \rfloor} \binom{\lfloor fD|S| \rfloor \Delta}{f\Delta |S|}\Big/ \binom{\Delta f n}{\Delta f |S|} \nonumber \\
    & = \sum_{s=1}^{\lfloor \frac{n}{D+1} \rfloor} \binom{n}{s} \binom{fn}{\lfloor fDs \rfloor} \binom{\lfloor fDs \rfloor \Delta}{f \Delta s}\Big/ \binom{\Delta f n}{\Delta f s} \nonumber \\
    \label{eq:binomial_product}
    & \leq an^{b(1+fD-f\Delta)} + \sum_{s=c}^{\lfloor \frac{n}{D+1} \rfloor} \binom{n}{s} \binom{fn}{\lfloor fDs \rfloor} \binom{\lfloor fDs \rfloor \Delta}{f \Delta s}\Big/ \binom{\Delta f n}{\Delta f s},
  \end{align}
  for some constants $a,b\geq 0$ and for every $c\in \mathbb{N}\smallsetminus \{0\}$.
  Now, let $s = \xi n$; by Lemma \ref{lem:binomial_coeff}, the sum in \eqref{eq:binomial_product} is upper bounded by
  \begin{align*}
    & \sum_{s=c}^{\lfloor \frac{n}{D+1} \rfloor} 2^{n\left(-(f \Delta-1)h(\xi) + fh\left(\frac{\lfloor fD\xi n\rfloor}{fn}\right) + \frac{\lfloor fD\xi n\rfloor}{n}\Delta h\left(\frac{f\xi n}{\lfloor fD\xi n\rfloor} \right) \right)}\\
    & \sim \sum_{s=c}^{\lfloor \frac{n}{D+1} \rfloor} 2^{n\left(-(f \Delta-1)h(\xi) + fh(D\xi) + D\xi f \Delta h\left(\frac{1}{D} \right) \right)}.
  \end{align*}
  Let us study the function
  \begin{equation*}
    \gamma(\xi) = -(f \Delta-1)h(\xi) + fh(D\xi) + D\xi f \Delta h\left(\frac{1}{D} \right).
  \end{equation*}
  Its second derivative is:
  \begin{equation*}
    \gamma''(\xi) = \frac{f \Delta -1}{\xi(1-\xi)} - \frac{fD}{\xi(1-D\xi)}.
  \end{equation*}
  Recalling that $\xi \in \left[\frac{c}{n},\frac{1}{D+1} \right]$, it is easy to show that $\gamma''(\xi) > 0$ under the condition
  \begin{equation*}
    \Delta > D^2 + \frac{1}{f},
  \end{equation*}
  that is assumed in \eqref{eq:Delta_expansion}. Thus, $\gamma$ is convex and
  \begin{equation*}
    \max_{c/n \leq \xi \leq 1/(D+1)} \gamma(\xi) = \max \left\{\gamma \left(\frac{c}{n}\right), \gamma \left(\frac{1}{D+1}\right) \right\}.
  \end{equation*}
  Now, \eqref{eq:Delta_expansion} also implies that $\gamma(1/(D+1))$ is constant and negative, whereas it is clear that $\gamma(c/n)$ tends to $0$ when $n$ tends to infinity. Hence, for $n$ big enough,
  \begin{equation*}
    \max_{c/n \leq \xi \leq 1/(D+1)} \gamma(\xi) = \gamma \left(\frac{c}{n}\right)
  \end{equation*}
  and, for some other positive constants $u$ and $v$, using again Lemma \ref{lem:binomial_coeff} we obtain:
  \begin{align}
    \eqref{eq:binomial_product} & \lesssim an^{b(1+fD-f\Delta)} + \sum_{s=c}^{\lfloor \frac{n}{D+1} \rfloor} 2^{n\gamma\left(\frac{c}{n}\right)} \nonumber \\
    & \leq an^{b(1+fD-f\Delta)} + un \binom{n}{c} \binom{fn}{\lfloor fDc \rfloor} \binom{\lfloor fDc \rfloor \Delta}{f \Delta c}\Big/ \binom{\Delta f n}{\Delta f c} \nonumber \\
    \label{eq:final_polynomial}
    & \leq an^{b(1+fD-f\Delta)} + vn^{1+c(1+fD-f\Delta)}.
  \end{align}
  \eqref{eq:Delta_expansion} implies that $f \Delta > fD +1$ and we can choose $c$ such that $1+c(1+fD-f\Delta)<0$, therefore \eqref{eq:final_polynomial} is vanishing when $n$ grows. This concludes the proof.
\end{IEEEproof}

\section{Proof of Lemma \ref{lem:volume_ratio}}
\label{sec:proof_volume_ratio}
\begin{IEEEproof}
  The proof of the lemma is a simple application of Lemma \ref{lem:integer_sphere_points} and Lemma \ref{lem:volume_sphere}:
  \begin{align*}
    \frac{|\Z^{n-m} \cap B_{\mathbf{c}',n-m}(\rho)|}{|\Zn \cap B_{\mathbf{c},n}(\rho)|} & \leq \frac{\vol\left(B_{\mathbf{c}',n-m}\left(\rho + \frac{\sqrt{n-m}}{2}\right)\right)}{\vol\left(B_{\mathbf{c},n}\left(\rho - \frac{\sqrt{n}}{2}\right)\right)} \\
    & \leq \frac{\vol\left(B_{\mathbf{c}',n-m}\left(\rho\right)\right)}{\vol\left(B_{\mathbf{c},n}\left(\rho\right)\right)} \frac{\left(1 + \frac{\sqrt{n}}{2\rho} \right)^n}{\left(1 - \frac{\sqrt{n}}{2\rho} \right)^n} \\
    & \sim \frac{(\sqrt{n})^{n+1}}{(\sqrt{n-m})^{n-m+1}} \left(\sqrt{2\pi e}\right)^{-m} \left( \frac{2\rho + \sqrt{n}}{2\rho - \sqrt{n}}\right)^n \rho^{-m} \\
    & = \frac{(\sqrt{n})^{n+1}}{(\sqrt{n-m})^{n-m+1}} \left(\sqrt{2\pi e}\right)^{-m} \left(1+\frac{2\sqrt{n}}{2\rho - \sqrt{n}} \right)^n \rho^{-m}. 
  \end{align*}
\end{IEEEproof}

% use section* for acknowledgment
%\section*{Acknowledgment}
%The research work presented in this paper on LDA lattices
%is supported by QNRF, a member of Qatar Foundation, 
%under NPRP project 5-597-2-241.

% Can use something like this to put references on a page
% by themselves when using endfloat and the captionsoff option.
\ifCLASSOPTIONcaptionsoff
  \newpage
\fi

\end{document}